%% file: main.tex
\documentclass[sigconf]{acmart}

\def\BibTeX{{\rm B\kern-.05em{\sc i\kern-.025em b}\kern-.08emT\kern-.1667em\lower.7ex\hbox{E}\kern-.125emX}}

\renewcommand\footnotetextcopyrightpermission[1]{} 
\usepackage{lineno,hyperref}
\usepackage{tikz,tikz-qtree}
\usepackage{pgfplots}
\usetikzlibrary{patterns,shapes}

\usepackage[linesnumbered,ruled,vlined]{algorithm2e}
\usepackage[noend]{algpseudocode}

\usepackage{caption}
\usepackage[inline]{enumitem}
\usepackage[normalem]{ulem}
\usepackage{mathtools}
\usepackage{multirow}
\usepackage{subfigure}
\usepackage{color}
\usepackage{amsmath,bm,amssymb}
\usepackage{url}
\usepackage{comment}
\usepackage{diagbox}

\usepackage{enumitem}
\newlist{tabitem}{itemize}{1}
\setlist[tabitem]{wide=0pt, nosep, leftmargin= * ,label=\textbullet,after=\vspace{-\baselineskip},before=\vspace{-0.6\baselineskip}}

\usepackage{multibib}
\usepackage{adjustbox}
\usepackage{graphicx}
\usepackage{rotating}
\usepackage{tabularx}
\usepackage{ltablex}
\usepackage{booktabs}
\usepackage{soul}

\newcommand{\citeSubject}[1]{\citeauthor{#1} (\citeyear{#1})}
\newcommand{\citeSub}[1]{\citeauthor{#1} \citeyear{#1}}

\usepackage{pdfrender}

\begin{document}

\title{Research Commentary on Recommendations with Side Information: A Survey and Research Directions}

\author{Zhu Sun$^1$, Qing Guo$^1$, Jie Yang$^2$, Hui Fang$^3$, Guibing Guo$^4$, Jie Zhang$^1$, Robin Burke$^5$}
\affiliation{
\institution{$^1$Nanyang Technological University, Singapore; $^2$Amazon Research, USA} 
\institution{$^3$Shanghai University of Finance and Economics, China}
\institution{$^4$Northeastern University, China; $^5$University of Colorado, USA}
}
\authornote{Hui Fang is the corresponding author. Email addresses of all authors are as follows: \{zhu.sun,qguo006,zhangj\}@ntu.edu.sg; jiy@amazon.com; fang.hui@mail.shufe.edu.cn; guogb@swc.neu.edu.cn; robin.burke@colorado.edu
}

\renewcommand{\shortauthors}{}

\begin{abstract}
Recommender systems have become an essential tool to help resolve the \textit{information overload} problem in recent decades. Traditional recommender systems, however, suffer from \textit{data sparsity} and \textit{cold start} problems. To address these issues, a great number of recommendation algorithms have been proposed to leverage side information of users or items (e.g., social network and item category), demonstrating a high degree of effectiveness in improving recommendation performance. This Research Commentary aims to provide a comprehensive and systematic survey of the recent research on recommender systems with side information. Specifically, we provide an overview of state-of-the-art recommendation algorithms with side information from two orthogonal perspectives. One involves the different methodologies of recommendation: the memory-based methods, latent factor, representation learning and deep learning models.  The others cover different representations of side information, including structural data (flat, network, and hierarchical features, and knowledge graphs); and non-structural data (text, image and video features). 
Finally, we discuss challenges and provide new potential directions in recommendation, along with the conclusion of this survey.   
\end{abstract}


\fancyhead{}
\settopmatter{printacmref=false,printfolios=false}

\maketitle

\input{section/intro}

\input{section/related}
\input{section/conventional.tex}

\input{section/deeplearning.tex}

\input{section/future.tex}

\input{section/conclusion}

\section{Acknowledgements}
This work was partly conducted within the Delta-NTU Corporate Lab for Cyber-Physical Systems with funding support from Delta Electronics Inc. and the National Research Foundation (NRF) Singapore under the Corp Lab@University Scheme, and also supported by the funding awarded to Dr. Jie Zhang by the BMW Tech Office Singapore.
We also gratefully acknowledge the support of National Natural Science Foundation of China (Grant No. 71601104, 71601116, 71771141 and 61702084) and the support of the
Fundamental Research Funds for the Central Universities in China
under Grant No. N181705007.

\bibliographystyle{ACM-Reference-Format}
\bibliography{reference}

\end{document}

%% file: section/intro.tex
\section{Introduction}
With the advent of the era of big data, the volume of information on the web has increased in an exponential fashion. Users are submerged in the flood of countless products, news, movies, etc. Aiming to provide personalized recommendation services for users based on their historical interaction data, recommender systems have become a vital and indispensable tool to help tackle the \textit{information overload} problem (\citeSub{ricci2015recommender}; \citeSub{desrosiers2011comprehensive}).
\begin{figure}[!t]
    \centering
    \includegraphics[scale=.4]{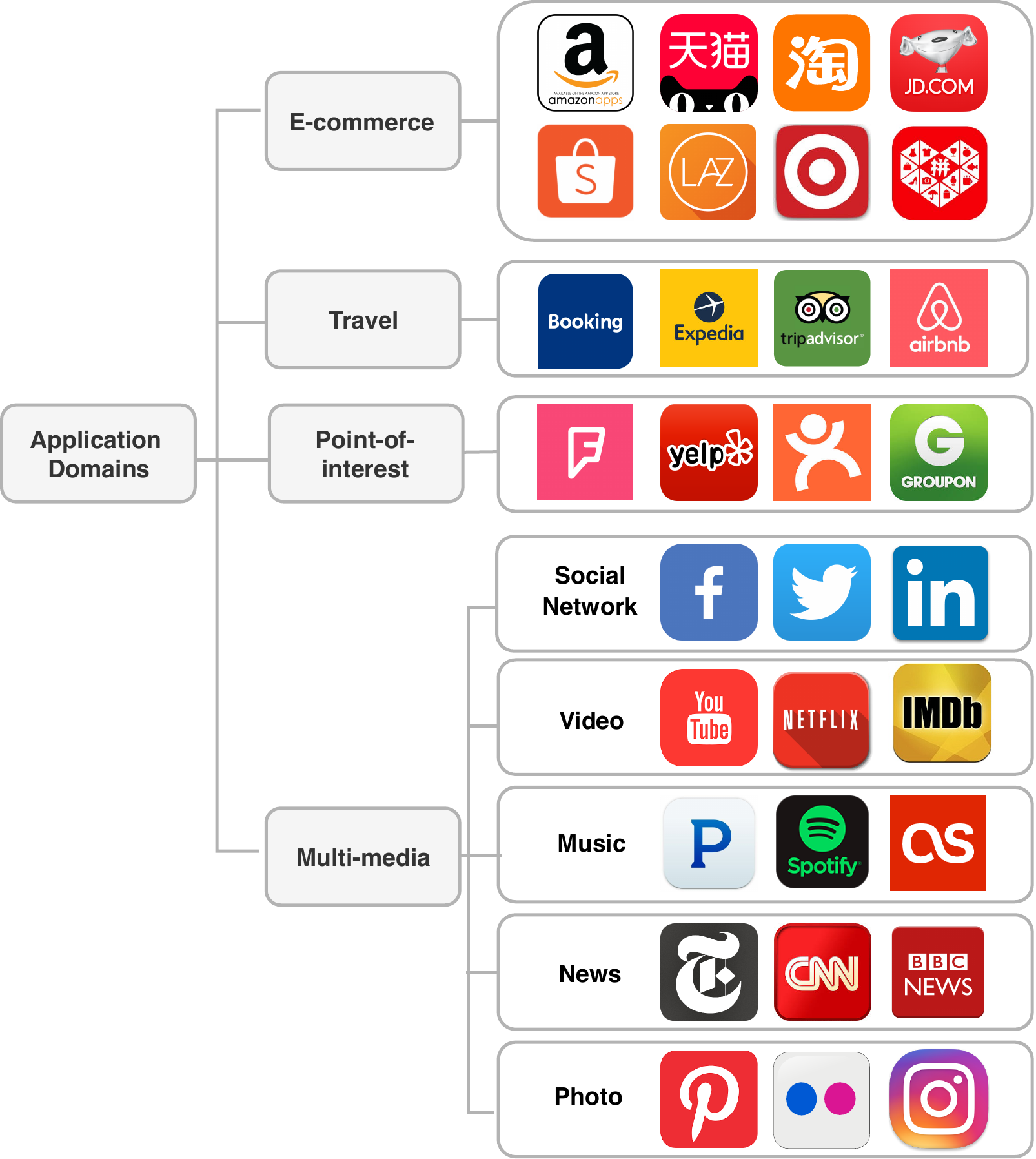}
    \caption{Popular apps that are utilized in various domains with recommender systems.}\label{fig:apps}
    \vspace{-0.2in}
\end{figure}
Empirical studies have demonstrated the effectiveness in facilitating decision-making process and boosting business across various domains
(\citeSub{zhang2017deep}; \citeSub{song2012survey}; \citeSub{adomavicius2005toward}), such as e-commerce (Amazon, Target, Taobao), point-of-interest (Foursquare, Yelp, Groupon), and multi-media (Youtube, Pinterest, Spotify),
to name a few. 
Fig. \ref{fig:apps} summarizes popular online services in various domains where recommender systems have been launched to improve the user experience. 
The great blossoming of recommender systems in practical applications has been promoted, to a large extent, by the flourishing research on recommendation. 
For example, recommender systems have become an important topic in a number of top tier research conferences and journals\footnote{Some of the key conferences and journals include NIPS (Neural Information Processing Systems), ICML (International Conference on Machine Learning), WWW (World Wide Web Conference), WSDM (Conference on Web Search and Data Mining), KDD (Conference on Knowledge Discovery and Data Mining), SIGIR (Conference on Research and Development in Information Retrieval), CIKM (International Conference on Information and Knowledge Management), IJCAI (International Joint Conferences on Artificial Intelligence), AAAI (Conference on Artificial Intelligence), UAI (Conference on Uncertainty in Artificial Intelligence), RecSys (Conference on Recommender Systems), ICLR (International Conference on Learning Representations), and TKDE (IEEE Transactions on Knowledge and Data Engineering), TOIS (ACM Transactions on Information Systems), CSUR (ACM Computing Surveys), etc.}.

The number of publications on recommender systems has increased dramatically in the last few years. 
RecSys (\url{recsys.acm.org}), the leading international conference on recommender systems, has continously attracted a tremendous amount of interest from both academia and industry
(\citeSub{cheng2016wide}; \citeSub{covington2016deep}; \citeSub{davidson2010youtube}; \citeSub{gomez2016netflix};
\citeSub{okura2017embedding}).
Among the different recommender systems, most of them are based on \emph{collaborative filtering} (CF), which is one of the most successful techniques for recommendation
(\citeSub{schafer2007collaborative}; \citeSub{ekstrand2011collaborative}; \citeSub{bobadilla2013recommender}; \citeSub{shi2014collaborative}). 
Traditional CF-based methods rely on user-item interaction matrices for making recommendations, assuming that a user's preference can be inferred by aggregating the tastes of similar users. They have been widely investigated
(\citeSub{linden2003amazon}; \citeSub{adomavicius2005toward}; \citeSub{ekstrand2011collaborative}; \citeSub{koren2009matrix}; \citeSub{mnih2008probabilistic}; \citeSub{rendle2009bpr}), 
with various variants of CF-based methods developed
(\citeSub{adomavicius2005toward}; \citeSub{ekstrand2011collaborative}; \citeSub{koren2009matrix}; \citeSub{mnih2008probabilistic}; \citeSub{rendle2009bpr}).
Despite that, traditional CF-based methods are confronted with two fundamental issues when only the user-item interaction matrices are taken into consideration:
\begin{itemize}[leftmargin=*]
    \item \textit{Data sparsity.} Usually, users face an extremely large amount of items to choose from. Even the most active users only rate a small set of items and most items have a very limited amount of feedback from users. This sparsity issue makes it hard for recommender systems to learn users' preferences. 
    \item \textit{Cold start}. It is a critical issue for both new users and items. Without historical data, it is difficult to generate decent recommendations. As a common solution, popular items might be recommended to new users, which will fail to create personalized recommendations.
\end{itemize}

\begin{figure}[t]
    \centering
    \includegraphics[scale=.34]{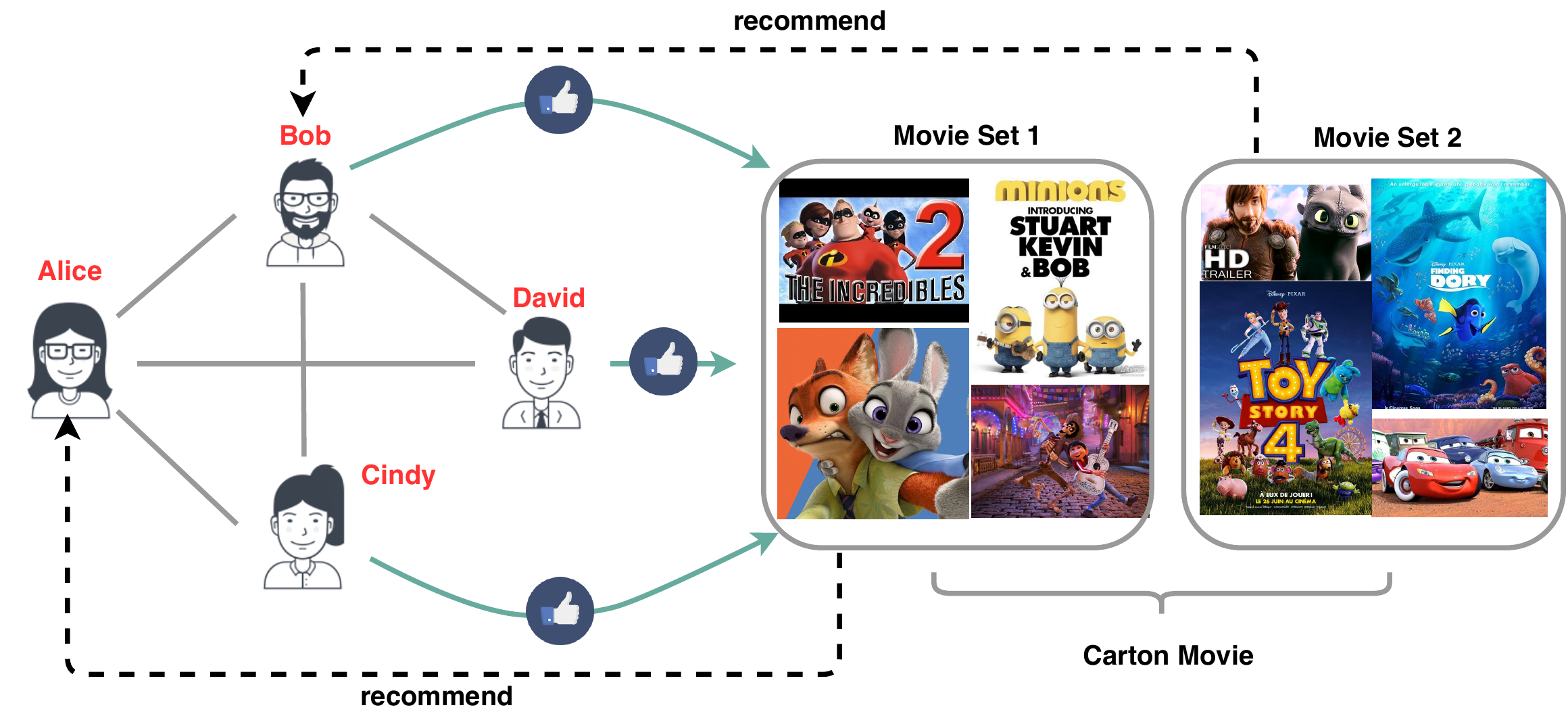}
    \caption{A toy example on leveraging user and item side information (social networks and movie genres) for more accurate recommendations. For instance, Alice has social connections with her friends, Bob, Cindy and David. As all her friends liked movies in Movie Set 1 (e.g., Zootopia and Coco), Alice would also be more likely to favor these movies in Movie Set 1. Besides, as the movies in both Movie Set 1 and Movie Set 2 belong to the genre of Carton movies, Bob would also prefer movies in Movie Set 2 (e.g., Toy Story and Cars), given that he liked movies in Movie Set 1. }\label{fig:example}
    \vspace{-0.2in}
\end{figure}

To address the two issues, different types of side information, such as social networks, user profiles and item descriptions, have been utilized for recommender systems in various domains (\citeSub{guo2019exploiting}) (see Fig. \ref{fig:apps}).
For instance, due to the emergence of social networks, a number of trust-aware recommendation algorithms
(\citeSub{ma2009learning}; \citeSub{jamali2010matrix}; \citeSub{ma2011recommender};
\citeSub{yang2012circle}; \citeSub{guo2012simple}; \citeSub{liusocial}; \citeSub{bao2014leveraging}; \citeSub{guo2015trustsvd}) have been proposed based on the assumption that users share similar preferences with their trusted friends. For example, for restaurants, users often have meals with their trusted friends
(\citeSub{ye2010location}; \citeSub{yang2013sentiment}). Besides social information, the side information for items (e.g., categories, genres, locations and brands) provides an in-depth understanding of both item properties and user preferences.
Many recommendation approaches
(\citeSub{kim2003recommendation}; \citeSub{shi2011tags}; \citeSub{koenigstein2011yahoo}; \citeSub{kanagal2012supercharging}; \citeSub{hu2014your}; \citeSub{sun2017mrlr}; \citeSub{sun2018rkge}) have been proposed by exploiting that kind of item information.
Fig. \ref{fig:example} depicts an example of how side information facilitates the generation of more accurate recommendations for users.  
Regardless of either user or item side information, the evolution of recommendation approaches with side information -- especially with the emergence and rapid development of deep learning based approaches, which have superior scalability and flexibility to accommodate arbitrary side information --   
has proven to be able to achieve great success with resolving the \textit{data sparsity} and \textit{cold start} problems, thus boosting recommendation performance. 

\medskip\noindent\textbf{Differences between this research commentary and other surveys.}
Due to the effectiveness of side information for recommender systems, the number of recent research studies have exploded in this field. And, there are also quite a few survey papers published on recommender systems. 
For instance, earlier works endeavored to conduct literature reviews on collaborative filtering techniques
(\citeSub{sarwar2001item}; \citeSub{breese1998empirical}; \citeSub{adomavicius2005toward}; \citeSub{schafer2007collaborative}; \citeSub{su2009survey};
\citeSub{desrosiers2011comprehensive}; \citeSub{ekstrand2011collaborative};
\citeSub{bobadilla2013recommender}; \citeSub{ricci2015recommender}). 
\citeSubject{lops2011content} provided a review on the state-of-the-arts and trends in content-based recommender systems. \citeSubject{burke2002hybrid} presented a survey on hybrid recommender systems. \citeSubject{bellogin2013empirical} introduced
an empirical comparison of social, collaborative filtering, and hybrid recommenders.
\citeSubject{gomez2016netflix} and \citeSubject{song2012survey} discussed various algorithms in recommending movie (Netflix) and music, respectively.
\citeSubject{zhang2019deep} proposed a comprehensive review on how deep learning based algorithms are applied for recommender systems. And finally, \citeSubject{shi2014collaborative} provided a systematic review on how side information is employed in collaborative filtering based approaches. 

Existing survey papers have mainly focused on a single perspective, instead of conducting a thorough investigation. In other words, they either discussed the general methodologies for recommender systems (e.g., \citeSub{zhang2019deep}; \citeSub{gomez2016netflix}) or side information per se (e.g., \citeSub{shi2014collaborative}), but ignored to explore the inherent dependency between them that together leads to high-quality recommendations. 
As a matter of fact, on the one hand, there are plenty of recent research efforts on dealing with the complexity of side information for realizing its full potential for better recommendations. Throughout our investigation, we discovered that existing research studies have been exploring more sophisticated structures to represent various kind of side information, including flat, network, hierarchical features and knowledge graphs. The different structures encode important relationships among the side information.
For example, category hierarchies can reflect the affiliation among categories, whereas the flat structure of the category does not have such a property. 
Such a relationship can be of high value for improving recommendation performance.

\begin{table*}[t]
\footnotesize
\centering
\caption{Classifications of recommender systems from different perspectives.}\label{tab:rs_classification}
\vspace{-0.1in}
\begin{tabular}{p{1.8cm}p{3.7cm}p{2cm}p{2.9cm}}
\specialrule{.15em}{.05em}{.05em}
\multicolumn{1}{l}{\textbf{Perspective}} &\multicolumn{1}{l}{Strategies}& \multicolumn{1}{l}{Tasks}  &\multicolumn{1}{l}{Outputs} \\
\specialrule{.05em}{.05em}{.05em}
\specialrule{.05em}{.05em}{.05em}
\multirow{3}{*}{\textbf{Category}}
&
\begin{tabitem}
  \item Content-based filtering
  \item Collaborative filtering
  \item Hybrid methods
\end{tabitem}
& 
\begin{tabitem}
    \item General
    \item Temporal
    \item Sequential
\end{tabitem}
& 
\begin{tabitem}
    \item Rating Prediction
    \item Item Ranking 
\end{tabitem}
\tabularnewline
\specialrule{.15em}{.05em}{.05em}
\end{tabular}

\end{table*}

On the other hand, many research studies have proposed more advanced recommendation methodologies to accommodate the diverse side information, evolving from memory-based methods to latent factor, representation learning and deep learning models. Based on our literature review, recommendation performance depends on both the structures representing the rich side information and the fundamental recommendation methodologies of employing them. The more complex representation of side information often needs to be coupled with more advanced methodologies to fully exploit the value of side information. In other words, it is often impossible to disentangle the useful side information from the fundamental methodologies for better recommendations.

This survey seeks to provide the research community a comprehensive and systematic overview of current progress in the recommendation area by considering both the representation of side information and the fundamental recommendation methodologies. It should not only focus on some cutting-edge techniques (e.g., knowledge graphs and deep learning models), but also other conventional ones (e.g., social networks and latent factor models) which have been the cornerstone in the development of recommender systems with side information. In this way, this Research Commentary provides a complete picture for both researchers and practitioners in this area.

\medskip\noindent\textbf{Article collection.}
To cover recent studies, we collected hundreds of papers published in prestigious international conferences and journals related to recommender systems, including NIPS, ICML, UAI, KDD, WWW, WSDM, IJCAI, AAAI, SIGIR, RecSys, CIKM, ICLR, and TKDE, TOIS, CSUR, etc. Google Scholar was primarily used to searching for papers while other academic search engines were also adopted, such as ACM Digital Library (\url{dl.acm.org}), IEEE Xplore (\url{ieeexplore.ieee.org}), Web of Science (\url{www.webofknowledge.com}), and Springer (\url{www.springer.com}). 
A number of keywords and their combinations were utilized to search for related papers, including recommender systems, recommendations, side information, auxiliary information, social networks, feature hierarchies, knowledge graphs, collaborative filtering, factorization, representation learning, deep learning, neural networks, etc. 

\medskip\noindent\textbf{Contributions.}
This survey aims to provide a thorough literature review on the approaches of exploiting side information for recommender systems. It is expected to help both academic researchers or industrial practitioners who are interested in recommender systems gain an in-depth understanding of how to improve recommendation performance with the usage of different types of side information. In summary, we make the following key contributions: 
(1) we conduct a systematic review for recommendation approaches with the incorporation of side information from two orthogonal perspectives. That is, different fundamental methodologies and various representations of side information; 
(2) we propose a novel taxonomy to classify existing recommendation approaches, which clearly demonstrates the evolution process of recent research studies; 
(3) we provide a comprehensive literature review of state-of-the-art studies by providing insightful comparison and analysis; and (4) we identify future directions and potential trends in this research area to shed light and promote further investigation on side information for more effective recommendations.

%% file: section/related.tex
\section{Evolution of recommenders with side information}

Prior to diving into state-of-the-art approaches on exploiting side information, 
we first introduce the relevant concepts and provide an overview of the evolution of research focusing on both recommendation methodologies and side information. 

\subsection{Overview of recommender systems} 
Generally, recommender systems predict users' preferences on items to assist users for making easier decisions. This section provides an overview of recommender systems from different perspectives. Specifically, recommender systems can be classified based on the strategies, tasks and outputs, as shown in Table \ref{tab:rs_classification}.

\medskip\noindent\textbf{Classification by strategies.}
Recommendation strategies can usually be classified into three categories: (1) content-based filtering, (2) collaborative filtering and (3) hybrid methods. The first two are relevant to our review as the content-based filtering methods provide us a vital clue on the various side information as well as ways to use it for recommendation, and the collaborative filtering methods give us a complete picture on the development of fundamental recommendation methodologies that are then studied to incorporate side information for better recommendations. That being said, the hybrid methods are the main focus of our investigation as they inherit and develop both content-based and collaborative filtering strategies.  
More detailed descriptions of the three types of strategies are presented as follows:
\begin{itemize}[leftmargin=0.4cm]
    \item \textbf{Content-based filtering.} It mainly utilizes user profiles and item descriptions to infer users' preferences towards items. The basic process is to build the profile of a user based on her personal attributes or descriptions of historical items that she has purchased or liked. The recommendations are created by matching the content of items with user profiles. In particular, a range of auxiliary data, such as categories, tags, brands, and images, can be utilized to construct descriptive features of an item. As these methods mainly rely on the rich content features of users and items, they are capable of handling the data sparsity and cold-start problems better. Meanwhile, they enable us to gain a deep understanding of how side information is exploited by state-of-the-art algorithms.
    \item \textbf{Collaborative filtering (CF).} This technique aims to predict users' preferences towards items by learning from user-item historical interactions, either in the form of explicit feedback (e.g., ratings and reviews) or implicit feedback (e.g., click and view). Generally, there are two types of CF-based techniques: memory- and model-based methods. 
    The former methods (\citeSub{hwang2012using}; \citeSub{guo2012simple}) usually exploit original user-item interaction data (e.g., rating matrices) to predict unobserved ratings by aggregating the preferences of similar users or similar items. The latter assume that the preference of a user or the characteristic of an item can be represented by a low-dimensional latent vector. More specifically, model-based methods learn the latent feature vectors of users and items from user-item matrices, and predict the recommendations by calculating the dot product of the latent vectors of the user and item (\citeSub{koren2009matrix}; \citeSub{mnih2008probabilistic}). Empirical studies have proven that model-based methods outperform memory-based ones in most cases. However, the \textit{data sparsity} and \textit{cold start} issues inherently hinder the effectiveness of CF-based methods when user-item interaction data are very sparse. 
    As the most successful technique in recommendation, these methods enable us to have a comprehensive understanding on the evolution of fundamental methodologies in this area. 
    \item \textbf{Hybrid methods.} They take advantage of both CF- and content-based approaches so as to remedy their shortcomings. There are two types of techniques for blending different recommendation models: {early fusion} and {late fusion}. The former refers to combining both explicit contents (e.g., visual, textual, and knowledge-aware features) and historical user-item interaction data, and then feeding them into some CF-based methods to boost recommendation performance (\citeSub{zhang2016collaborative}; \citeSub{tuan20173d}). On the other hand, late fusion methods build separate recommender systems that are specialized to each kind of information, and then combine the predictions of these systems (\citeSub{park2006naive}; \citeSub{melville2002content}; \citeSub{pero2013opinion}). Hybrid recommendation methods are known to empirically outperform the pure CF- or content-based methods, especially for solving the {data sparsity} and {cold start} problems. Our investigation mainly focuses on state-of-the-art hybrid recommendation methods. The vast majority of them were developed in the recent $10$ years.
    In total, around $95\%$ of the papers were published in $2010-2019$, and more than $60\%$ of the papers were published in the recent five years. 
\end{itemize}

\medskip\noindent\textbf{Classification by tasks.}
In terms of whether to consider time information (e.g., the order of historical interactions), recommender systems can be categorized by general, temporal and sequential recommendation tasks.  
\begin{itemize}[leftmargin=0.4cm]
    \item \textbf{General recommendation}. It normally leverages global user-item interaction data to recommend the top-N items for users. 
    The algorithms, such as matrix factorization (\citeSub{koren2009matrix})  and its derived models (e.g., \citeSub{singh2008relational}; \citeSub{chen2012svdfeature}; \citeSub{rendle2012factorization}), are able to effectively model user preferences, thus providing a static list of recommendations reflecting long-term interests of each user.
    \item \textbf{Temporal recommendation.} It usually captures user preferences given a timestamp or a time period. More specifically, some methods (e.g., TimeSVD++ (\citeSub{koren2009collaborative})) split time into several segments, and model the user-item interactions in each segment. To build an effective temporal recommender system, the key is to model the dynamics of user preferences that exhibit significant (short- or long-term) temporal drift (e.g., `what users prefer to have for lunch' or `which places users want to visit on weekends?') (\citeSub{koren2009collaborative};
    \citeSub{xiong2010temporal}; \citeSub{wu2017recurrent};  \citeSub{hosseini2018recurrent}).
    \item \textbf{Sequential recommendation (or next-item recommendation).} It is different from the above tasks, as sequential recommendation predicts users' \textit{next preferences} based on their most recent activities (\citeSub{rendle2010factorizing}; \citeSub{hidasi2015session}; \citeSub{wang2015learning}; \citeSub{yu2016dynamic}; \citeSub{jing2017neural}; \citeSub{tang2018personalized}; \citeSub{kang2018self}; \citeSub{pasricha2018translation}; \citeSub{zhang2018next}).
    In other words, sequential recommendation seeks to model sequential patterns among successive items, and generate well-timed recommendations for users. Therefore, it is more difficult than the other two types of recommendation tasks mentioned above. 
\end{itemize}

\medskip\noindent\textbf{Classification by outputs.}
Another categorization is based on the form of outputs, and there generally are two types of tasks: rating- and ranking-based item recommendation tasks
(\citeSub{sun2015exploiting}). Rating-based recommendation (rating prediction) predicts users' explicit preference scores towards items, which is usually considered as a regression task. In contrast, ranking-based recommendation (item ranking) focuses on the (relative) ranking positions of items and usually generates a top-$N$ item recommendation list to each user. 

\medskip\noindent\textbf{Discussion.}
In summary, there can be different ways to categorize recommender systems from various perspectives. Existing classification taxonomies, however, cannot help deliver a complete picture of the research studies in recommendation with side information.
In this view, we create a new taxonomy to classify the literature based on two aspects: the representation of side information and the fundamental recommendation methodologies. 
The proposed taxonomy mainly focuses on the hybrid recommendation methods, sweeping recent state-of-the-art algorithms in various tasks (general, temporal and sequential) with different types of outputs (rating prediction and item ranking).
More importantly, it allows the research community to capture a comprehensive understanding of how side information is leveraged for effective recommendations. Detailed discussions of the relevant literature will be presented in Sections \ref{sec:conventional} and \ref{sec:deep}.
\begin{figure}[t]
\centering
	\subfigure[Evolution of fundamental methodologies that are applied into recommendation with the progressive timeline marked by the red dots.]{
		\includegraphics[width=0.44\textwidth]{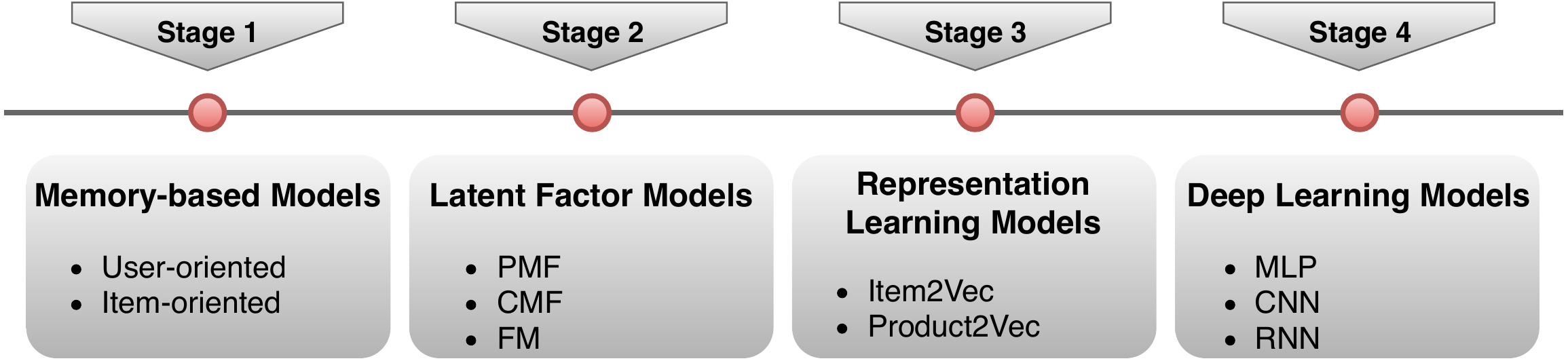}
	}
	\subfigure[Evolution of side information that are exploited for recommendation with the progressive timeline marked by the red dots.]{
		\includegraphics[width=0.4\textwidth]{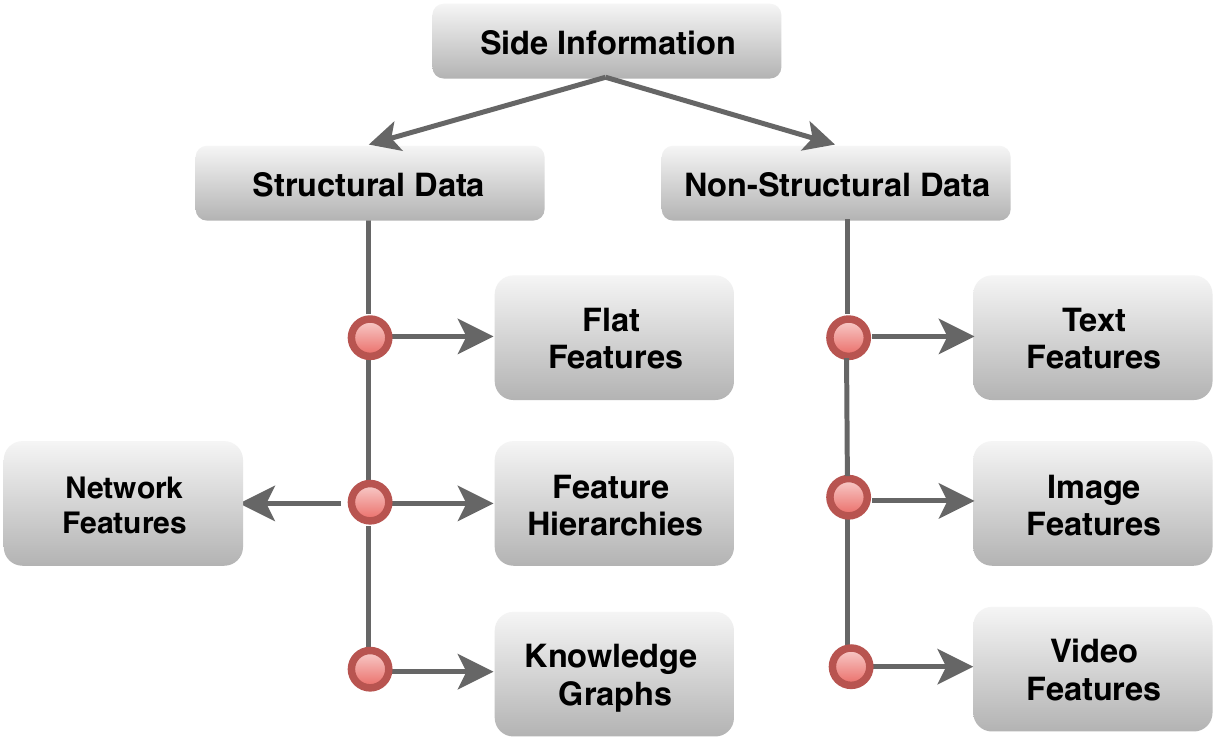}
	}
	\vspace{-0.1in}
	\caption{Evolution of fundamental methodologies and side information that are exploited in recommender systems with the progressive timeline marked by the red dots. }\label{fig:evolution}
\end{figure}

\subsection{Evolution of fundamental methodologies for recommendation}
In terms of the fundamental recommendation methodologies, we mainly focus on CF-based approaches as most of the advances fall into this category
(\citeSub{koren2008factorization}; \citeSub{rendle2009bpr}; \citeSub{koren2009matrix}; \citeSub{sedhain2015autorec}; \citeSub{he2017neural}; \citeSub{wu2017recurrent}). Before diving deep into the specific methods that employ side information, we illustrate the evolution of CF-based recommendation techniques with the progressive timeline shown in Fig. \ref{fig:evolution}a. 
Generally, two types of CF-based approaches are widely investigated, namely memory-based and model-based (e.g., latent factor models) approaches.

\medskip\noindent\textbf{Memory-based approaches.}
Memory-based approaches are also referred to as \textit{neighborhood-based collaborating filtering algorithms}. They are among the earliest techniques that aggregate the interests of neighbors for recommendation.
Specifically, memory-based approaches exploit user-user or item-item similarity derived from the user-item historical interaction matrix to make recommendations. User- and item-oriented methods are two kinds of typical memory-based approaches. User-oriented approaches identify like-minded users who can complement each other's ratings. The ratings of a target user are predicted based on the ratings of similar users found in a system. In contrast, item-oriented approaches evaluate a user's preference for an item based on the ratings of similar items rated by the same user. Although memory-based approaches have been adopted in real-world applications such as CiteULike, Youtube, and Last.fm, they are ineffective for large-scale datasets as searching for similar users or items can be time-consuming in large user or item space.

\medskip\noindent\textbf{Model-based approaches.}
Model-based approaches aim to build predictive models by adopting data mining or machine learning techniques on user-item rating matrices to uncover complex user behavior patterns. The learned models are then used to predict users' ratings of unknown items. Besides the user-item rating matrix, side information can serve as additionally valuable features that are fed into the predictive models, and thus assist in resolving the data sparsity and cold start issues. Model-based approaches can better adapt and scale up to large-scale datasets with significant performance improvements when compared with 
memory-based ones. Typically, successful model-based recommendation approaches fall into three categories: latent factor models, representation learning models and deep learning models. 

\begin{itemize}[leftmargin=0.4cm]
    \item \textbf{Latent factor models (LFMs)}. They decompose the high dimensional user-item rating matrix into low-dimensional user and item latent matrices. Due to high efficiency, state-of-the-art recommendation methods are dominated by LFMs (\citeSub{shi2014collaborative}). The basic idea of LFMs is that both users and items can be characterized by a few latent features, and thus the prediction can be computed as the inner product of user-feature and item-feature vectors. Many effective approaches fall into this category, such as \textit{matrix factorization} (MF) (\citeSub{koren2009matrix}), \textit{non-negative matrix factorization} (NMF) (\citeSub{zhang2006learning}), \textit{tensor factorization} (TensorF) (\citeSub{bhargava2015and}), \textit{factorization machine} (FM) (\citeauthor{rendle2010factorization} \citeyear{rendle2010factorization,rendle2012factorization}), SVD++ (\citeSub{koren2008factorization}), \textit{collective matrix factorization} (CMF) (\citeSub{singh2008relational}) and SVDFeature (\citeSub{chen2012svdfeature}). 
    \item \textbf{Representation learning models (RLMs)}. They have been proven to be effective in capturing local item relationships by modeling item co-occurrence in an individual user's interaction records. RLMs were originally inspired by word embedding techniques, which can be traced back to the classical neural network language model (\citeSub{bengio2003neural}), and the recent breakthroughs of Word2Vec techniques, including CBOW and Skip-gram (\citeSub{mikolov2013efficient}). Many Item2Vec (\citeSub{barkan2016item2vec}) based recommendation approaches, which are analogous with the Word2Vec technique, have been proposed to date
    (\citeSub{wang2015learning}; \citeSub{grbovic2015commerce}; \citeSub{liang2016factorization}; \citeSub{feng2017poi2vec}).
    \item \textbf{Deep learning models (DLMs)}. They have brought significant breakthroughs in various domains, such as computer vision, speech recognition, and natural language processing
    (\citeSub{lecun1995convolutional}; \citeSub{socher2011parsing}; \citeSub{krizhevsky2012imagenet}; \citeSub{luong2015effective}; \citeSub{wang2016recursive}), with recommender systems being no exception. 
    In contrast to LFMs and RLMs, DLMs (e.g., AutoRec (\citeSub{sedhain2015autorec}), NCF (\citeSub{he2017neural}) and DMF (\citeSub{xue2017deep})) can learn nonlinear latent representations via various types of activation functions (e.g., sigmoid, ReLU (\citeSub{nair2010rectified})). For instance, recurrent neural network (RNN) based approaches
    (\citeSub{hidasi2015session}; \citeSub{jing2017neural}; \citeSub{wu2017recurrent}; \citeSub{hosseini2018recurrent}) have shown powerful capabilities for sequential recommendation due to the ability of preserving historical information over time. Convolutional neural network (CNN) based approaches
    (\citeSub{zhang2016collaborative}; \citeSub{he2016sherlock}; \citeSub{he2016vista}) are capable of extracting local features so as to capture more contextual influences. In summary, DLMs possess essential advantages, and have promoted active and advanced studies in recommendation.
\end{itemize}

\medskip\noindent\textbf{Discussion.}
In essence, both the LFMs (e.g., matrix factorization) and RLMs (e.g., item2vec) can be considered as a special case of DLMs, that is, the shallow neural networks (\citeSub{he2017neural}). For instance, matrix factorization can be regarded as a one-layer neural network which transforms one-hot user and item vectors to dense representations with a linear inner product of these vectors for prediction. 
Although DLMs achieve superior performance against other model-based recommendation methods, the investigation into how to efficiently incorporate diverse side information into DLMs has not reached its full potential. 
In contrast, such research issues have been well studied for LFMs and RLMs in the recent decades, which could provide inspiration for the development of DLMs with side information.
On the other hand, in comparison with DLMs, which involve more computational cost but often only achieve small performance increments, traditional model-based methods (e.g., LFMs and RLMs) have the potential to be further developed to produce better recommendation accuracy. Trading-off between the recommendation accuracy and the computational cost is, therefore, an important direction for future research that requires a comprehensive review of the different types of recommendation methodologies. To this end, we conduct a systematic and comprehensive review on state-of-the-art algorithms along with the evolution of fundamental methodologies, so as to deliver a complete picture in this field.

\subsection{Evolution of side information for recommendation}
In order to resolve the data sparsity and cold-start issues, recent CF-based recommendation techniques focus more on exploiting different kinds of side information such as social networks and item categories. Such side information can be used to estimate users' preferences even with insufficient user-item historical interaction data. 
For example, the emergence of social networks help us to indirectly infer a user's preference by aggregating her friends' preferences. Other side information (e.g., item tags or categories) can be directly used for understanding a user's interests, such as the categories of movies or music albums reflect what types of movies or music she enjoys. 
To achieve a systematic understanding, we propose a new taxonomy to categorize the side information by the presence of their intrinsic structures, including structural data (i.e., flat features, network features, hierarchical features and knowledge graphs) and non-structural data (i.e., text features, image features and video features). The taxonomy is shown in Fig. \ref{fig:evolution}b.

\begin{figure}[t]
    \centering
    \includegraphics[scale=.50]{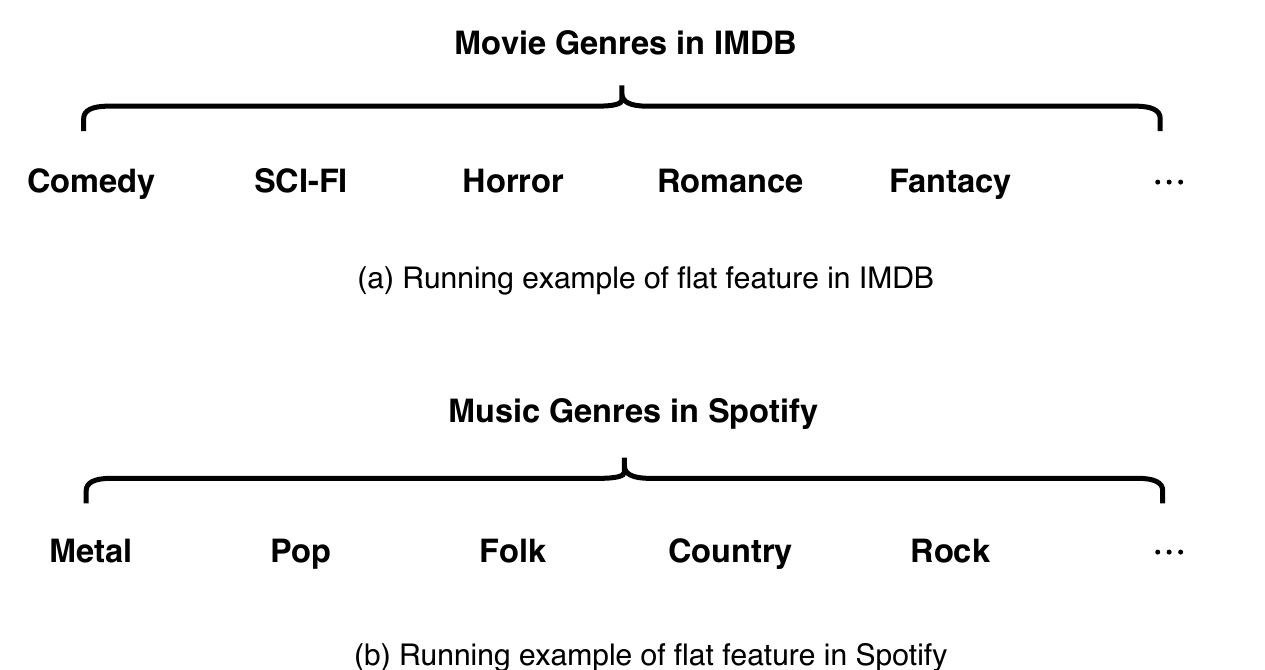}
    \caption{Examples of flat features, where all features are independently orgainzed at the same layer. Both movies in IMDB and music in Spotify are classified by genres.}
    \label{fig:flat_feature}
    \vspace{-0.2in}
\end{figure}

\medskip\noindent\textbf{Flat features (FFs).}
Early studies 
(\citeSub{lippert2008relation}; \citeSub{sharma2011improving}; \citeSub{hwang2012using}; \citeSub{yang2012circle}; \citeSub{liu2013personalized}; \citeSub{ji2014two}; \citeSub{hu2014your}; \citeSub{vasile2016meta}) mainly focused on integrating flat features (FFs), where the features are organized independently at the same layer. 
Fig. \ref{fig:flat_feature} illustrates an example of flat features in IMDB and Spotify to organize movies or music by genres. 
Assume that if a user prefers one movie/song under a certain genre, she is more likely to favor other movies/songs under this genre. 
Such side information has been widely leveraged for better movie or music recommendations
(\citeSub{koenigstein2011yahoo}; \citeSub{pei2017interacting}; \citeSub{sun2017unified}).

\begin{figure}[t]
    \centering
    \includegraphics[scale=.35]{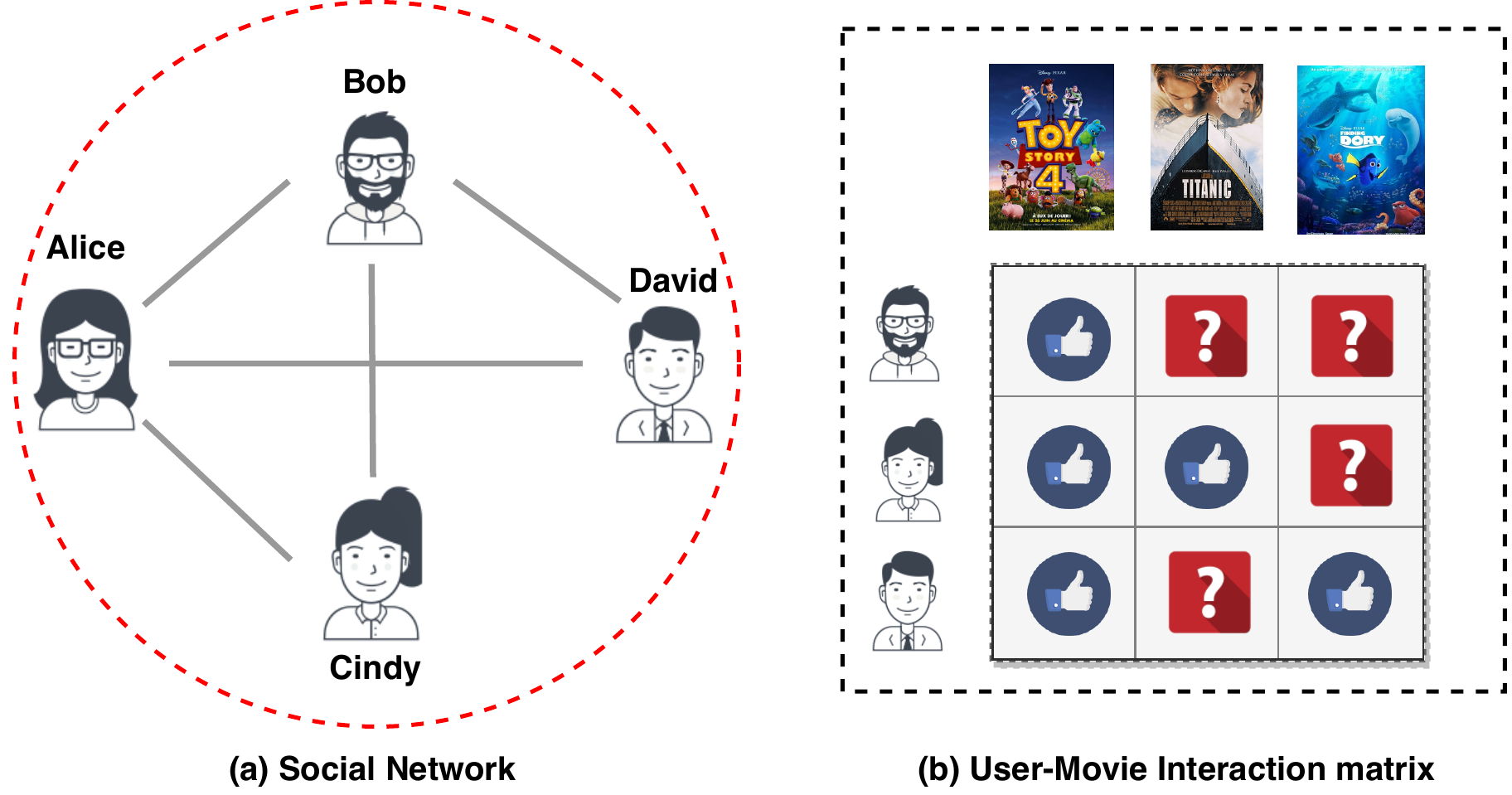}
    \caption{An example of social networks. (a) shows the social network where Bob, Cindy and David are friends of Alice; and (b) presents the user-movie interactions.}
    \label{fig:network}
    \vspace{-0.2in}
\end{figure}

\medskip\noindent\textbf{Network features (NFs).}
The advent of social networks has promoted active research on trust-aware recommender systems
(\citeSub{guo2012simple}; \citeSub{bao2014leveraging}; \citeSub{guo2015leveraging}). As a kind of homogeneous graph with single type of entity (user) and entity relation (friendship), social networks provide an alternative view of user preferences other than item ratings. The intuition is that social friends may share similar preferences and influence each other by recommending items. 
It has been proven that the fusion of social networks can yield significant performance enhancements
(\citeSub{jamali2010matrix}; \citeSub{ma2011recommender}; \citeSub{forsati2014matrix}; \citeSub{guo2015trustsvd}; \citeSub{ding2017baydnn}).
Fig. \ref{fig:network} illustrates an example of social networks to help resolve the cold start issue of recommender systems. Alice, as a newly enrolled user, can also get movie recommendations (Toy Story 4), as all of her friends (Bob, Cindy and David) favor this movie.

\begin{figure}[t]
    \centering
    \includegraphics[scale=.6]{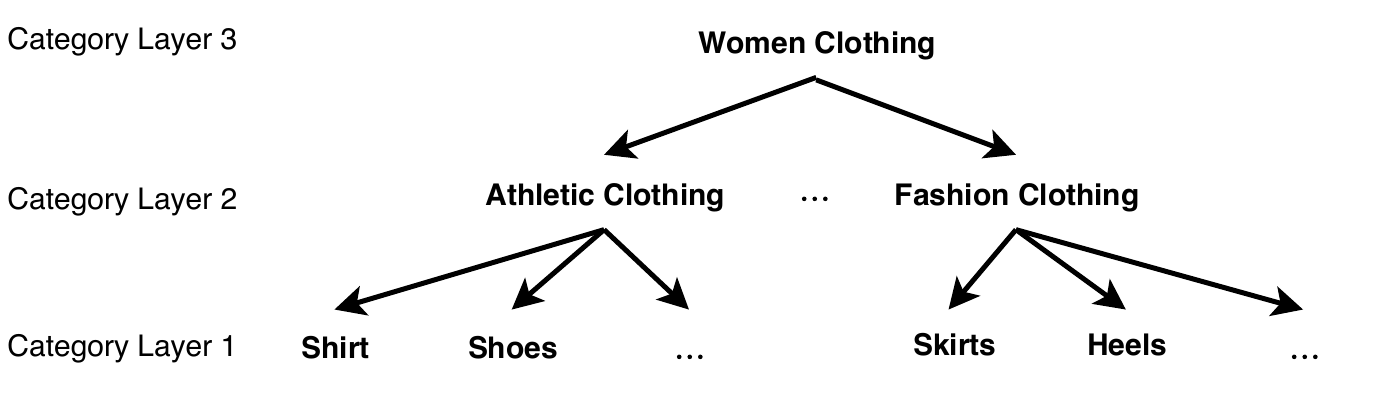}
    \caption{An example of the FH in Amazon Women's Clothing, where Women Clothing is first classified into several general categories (e.g., Athletic Clothing), and then is divided into more specific sub-categories (e.g., Shirts).}
    \label{fig:hierarchy}
    \vspace{-0.3in}
\end{figure}

\medskip\noindent\textbf{Feature hierarchies (FHs).}
More recently, researchers have attempted to investigate user / item features with a more complicated structure, a \textit{feature hierarchy} (FH), to further enhance recommendation performance.
A FH is  a natural yet powerful structure for human knowledge, and it provides a machine- and human-readable description of a set of features and their \textit{affiliatedTo} relations.
The benefits brought by explicitly modeling
feature relations through FHs have been studied in a broad spectrum of disciplines,
from machine learning (\citeSub{jenatton2010proximal}; \citeSub{kim2010tree}) to natural language processing (\citeSub{hu2015entity}). In the context of recommender systems, FHs have been proven to be more effective in generating high-quality recommendations than FFs (\citeSub{ziegler2004taxonomy}; \citeSub{weng2008exploiting}; \citeSub{menon2011response}; \citeSub{koenigstein2011yahoo}; \citeSub{mnih2011taxonomy}; \citeSub{kanagal2012supercharging}; \citeSub{he2016sherlock}; \citeSub{he2016sherlock}; \citeSub{yang2016learning}; \citeSub{sun2017exploiting}).
Typical examples of FHs include online products hierarchies (e.g., the Amazon web store (\citeSub{mcauley2015image})) and food hierarchies (e.g., Gowalla (\citeSub{liu2013personalized})). Fig. \ref{fig:hierarchy}
offers an example of a 3-layer FH for Women's Clothing in Amazon. If a customer prefers skirts, she may possibly like heels to match her skirt instead of athletic shoes. This is due to both Skirts and Heels belonging to a higher layer category -- Fashion Clothing, and they inherit similar characteristics of fashion style. By considering the \textit{affiliatedTo} relations among features in FHs, recommendations can be generated in a more accurate and diverse manner.

\begin{figure}[t]
    \centering
    \includegraphics[scale=0.21]{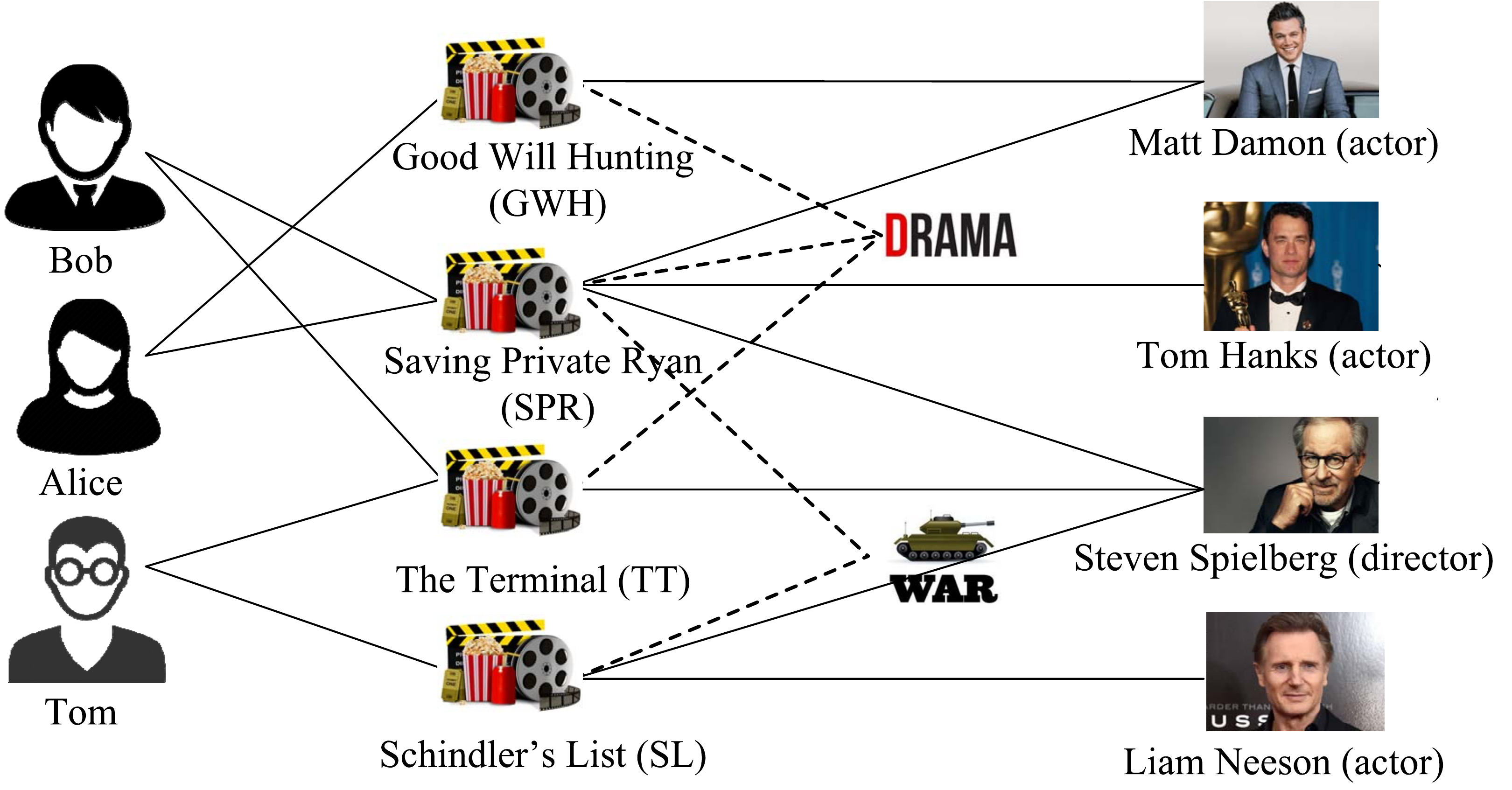}
    \caption{An example of a KG in movie domain, which contains users, movies, actors, directors and genres as entities; rating, categorizing, acting, and directing as entity relations.}
    \label{fig:kg}
    \vspace{-0.3in}
\end{figure}

\medskip\noindent\textbf{Knowledge graphs (KGs).}
Recently, with the development of semantic web, knowledge graphs (KGs) 
(\citeSub{yan2007graph}; \citeSub{lin2015learning}; \citeSub{wang2017knowledge}; \citeSub{cai2018comprehensive}) as an auxiliary data source have attracted extensive interest in the community of recommender systems. In contrast with FHs, which are generally limited to describing features with the child-parent (i.e., \textit{affiliatedTo}) relationship, KGs connect various types of features related to users (e.g., demographics and social networks) or items (e.g., the genre, director and actor of a movie), in a unified global representation space (See Fig. \ref{fig:kg}).  Leveraging the heterogeneous connected information from
KGs helps with the inference of subtler user or item relationships from different angles, which are difficult to be uncovered merely with homogeneous information (e.g., genre). The recommendation accuracy can, therefore, be further boosted with the incorporation of KGs (\citeSub{yu2013collaborative}; \citeSub{yu2013recommendation}; \citeSub{luo2014hete}; \citeSub{shi2015semantic}; \citeSub{grad2015recommendations}; \citeSub{catherine2016personalized}; \citeSub{shi2016integrating}; \citeSub{zhang2016collaborative}; \citeSub{zheng2017recommendation}; \citeSub{wang2017flickr}; \citeSub{zhang2017joint}; \citeSub{sun2018rkge}; \citeSub{wang2018explainable}).

\medskip\noindent\textbf{Non-structural data.}
All the aforementioned side information, including FFs, FHs, NFs and KGs, is structural knowledge. Apart from that, some non-structural data (e.g., text, image and video content) has also been widely utilized for generating high-quality recommendations.
For instance, reviews posted by users have been adopted for evaluating their experience (e.g., online shopping, POI check-in). Compared with ratings, reviews can better reflect different aspects of users' preferences
(\citeSub{yin2013lcars}; \citeSub{he2015trirank}; \citeSub{gao2015content}; \citeSub{wang2017location}).

Suppose a user, Sarah, posted a review for a restaurant -- ``\textit{The \textcolor{blue}{staff} was super friendly and \textcolor{blue}{food} was nicely cooked! will visit again}". From this we may infer that Sarah is quite satisfied with the ``food'' and ``service'' of the restaurant.  
Hence, reviews can serve as complementary information to explain the ratings and model users' preferences in a finer granularity (\citeSub{wu2016joint}; \citeSub{catherine2017transnets}; \citeSub{zheng2017joint}; \citeSub{seo2017interpretable}; \citeSub{tay2018multi}; \citeSub{lu2018like}).
Moreover, image has also been taken into account for better visual recommendations (\citeSub{lei2016comparative}; \citeSub{liu2017deepstyle}; \citeSub{niu2018neural}) and general recommendations
(\citeSub{mcauley2015image}; \citeSub{zhou2016applying}; \citeSub{wang2017your}; \citeSub{alashkar2017examples}; \citeSub{yu2018aesthetic}), as the visual features related to items (e.g., movie poster, book covers, hotel/food/clothing photos) play an important role to attract users and further affect their decision-making process (\citeSub{zhang2016collaborative}; \citeSub{he2016vbpr}; \citeSub{he2016ups};  \citeSub{chen2017attentive}; \citeSub{chu2017hybrid}).

\begin{table}[t]
\footnotesize
\centering
\addtolength{\tabcolsep}{-1.2mm}
\caption{Comparison of different data structure w.r.t. entity types and entity relations. }\label{tab:structure}
\vspace{-0.1in}
\begin{tabular}{|l|c|c|c|c|}
    \specialrule{.15em}{.05em}{.05em}
    \textbf{Data}&{Flat features}& {Network features}& {Feature hierarchies}& {Knowledge graphs}\\
    \specialrule{.05em}{.05em}{.05em}
    \specialrule{.05em}{.05em}{.05em}
    \textbf{Types} &1&1&$>=1$&$>1$\\
    \textbf{Relations} &0 &1 & 1 & $>1$\\
    \specialrule{.15em}{.05em}{.05em}
\end{tabular}
\end{table}

\medskip\noindent\textbf{Discussion.}
For the structural information, from flat features to network features and feature hierarchies, and to knowledge graphs, the structure becomes more and more complex, evolving from a homogeneous structure to a heterogeneous one, with increasing entity types and entity relations, as summarized in Table \ref{tab:structure}. For instance, in the flat features, there is only one type of entity (genres of movie) and no entity relation; while in the social network, besides one entity type (users), there is only one type of entity relation (friendship). In terms of the knowledge graph, it contains multiple types of entities and entity relations in a unified space. 
The more sophisticated the side information is, the more knowledge and information are encoded. Therefore, it is a necessity to develop more advanced fundamental methodologies to efficiently accommodate such information. When it comes to the non-structural side information (e.g., text, images, videos), we need to utilize the deep learning advances to help extract the hidden features. In sum, it is often impossible to disentangle various useful side information from the fundamental methodologies for better recommendations: they are mutually enhanced by each other in a cooperative fashion.

To sum up, Fig. \ref{fig:evolution}a and b depict the overall scheme of the proposed new taxonomies to categorize the fundamental methodologies and diverse side information for recommendation. 
Specifically, we propose a novel taxonomy to categorize: (1) the fundamental recommendation methodologies from memory-based methods, latent factor models and representation learning models towards deep learning models; and (2) the side information by their intrinsic data types, including structural data (flat features, network features, feature hierarchies and knowledge graphs), and non-structural data (text, images and videos). 
Based on this, we conducted a systematic, comprehensive, and insightful analysis on state-of-the-art hybrid recommendation approaches with side information. 
Table \ref{tab:summary} summarizes the statistics of all representative algorithms that we selectd for coverage (164-28=136 in total) from the above two perspectives. 
Around 95\% of the papers were published in recent 10 years. 
For ease of exposition, we will present and analyze all the conventional models (i.e., memory based methods, latent factor models and representation learning models) with various types of side information in Section \ref{sec:conventional}. Following this, Section \ref{sec:deep} introduces deep learning models with diverse side information.  

\begin{table}[t]
\centering
\addtolength{\tabcolsep}{-0.3mm}
\footnotesize
\renewcommand{\arraystretch}{0.75}
\caption{Summary of representative state-of-the-art recommendation algorithms with side information, where `FFs, NFs, FHs, KGs' denote the structural side information, namely {flat features, network features, feature hierarchies,} and {knowledge graphs}, respectively; `MMs, LFMs, RLMs, DLMs' represent {memory-based, latent factor, representation learning} and {deep learning} models, respectively. Note that they have the same meanings for all the following tables.
Besides, in this table we also include the `Basic' methods without incorporating side information for each type of methodology. }\label{tab:summary}
\vspace{-0.1in}
\begin{tabular}{|l|c|cccc|ccc|c|}
    \specialrule{.15em}{.05em}{.05em}
    \multirow{2}{*}{\textbf{No.}}&\multirow{2}{*}{\textbf{Basic}}&\multicolumn{4}{c|}{\textbf{Structural Data}}
    &\multicolumn{3}{c|}{\textbf{Non-Structural Data}} &\multirow{2}{*}{{Total}}
    \\\cline{3-9}
     & &\textbf{FFs}&\textbf{NFs}&\textbf{FHs}&\textbf{KGs}& \textbf{Text}&\textbf{Images}&\textbf{Videos}&\\
    \specialrule{.05em}{.05em}{.05em}
    \specialrule{.05em}{.05em}{.05em}
    \textbf{MMs} &2 &2 &5 &3 &-- &2 &-- &--&14 \\
    \textbf{LFMs} &8 &17 &15 &10 &6 &10 &9 &--&75\\
    \textbf{RLMs} &6 &4&--&--&--&--&--&--& 10\\
    \textbf{DLMs}&12 &7 &5 &2 &14 &18 &6 &1 &65\\
    {Total}&28 &30 &25 &15 &20 &30 &15 &1 &164\\
    \specialrule{.15em}{.05em}{.05em}
\end{tabular}
\end{table}

%% file: section/conventional.tex
\section{Conventional models with side information}\label{sec:conventional}

In this section, we present and analyze the exploitation of various side information for conventional recommendation models, including memory-based methods, and latent factor models, as well as representation learning models.

\subsection{Memory-based methods with side information}

Early recommendation approaches with side information were mainly built upon memory-based methods (MMs)
(\citeSub{schafer2007collaborative}; \citeSub{adomavicius2005toward}; \citeSub{desrosiers2011comprehensive}). Typical research includes approaches either with item side information (e.g., item categories (\citeSub{sharma2011improving}; \citeSub{hwang2012using})) or user side information (e.g., social networks (\citeSub{guo2012simple})).

\medskip\noindent\textbf{MMs+FFs.} 
Many MMs consider \textit{flat features} (FFs) for recommendation with pre- or post-filtering manner, based on the assumption that users may have similar interests with other users who are affiliated to the same features. For instance, \citeSubject{hwang2012using} introduced the notion of {category experts}, and predicted unknown ratings for the target user by aggregating the ratings of category experts instead of traditional similar users.
It is equivalent to leveraging the flat categories to cluster (i.e., pre-filter) users into different groups. 
\citeSubject{davidson2010youtube} proposed a Youtube video recommender, where flat categories are used to post-filter videos, to further ensure the diversity of the final recommended videos. 

\medskip\noindent\textbf{MMs+NFs.} Later, the advent of social networks has promoted active research in the area of trust-aware recommender systems.  
A number of works leverage social networks, that is, the \textit{network features} (NFs), for effective recommendations
(\citeSub{guo2012simple}; \citeSub{guo2012resolving}; \citeSub{guo2013integrating}; \citeSub{guo2014merging}; \citeSub{guo2015leveraging}). These methods posit that social friends may share similar interests. Specifically, they estimate the unknown ratings for the target user by merging the ratings of her trusted friends.

\medskip\noindent\textbf{MMs+FHs.} Several researchers also attempted to fuse \textit{feature hierarchies} (FHs) into MMs by exploiting the user- and product-taxonomy distributions. For example, \citeSubject{ziegler2004taxonomy} devised a user-based taxonomy-driven product recommendation method. In particular, they first represented each product by a taxonomy distribution vector, where elements denote the scores of the product's affiliation to the respective topics in the taxonomy. Then, the user taxonomy vector is obtained by summarizing the vectors of products that the user has interacted with. It discovers the user neighbors by calculating the similarity of the corresponding user-taxonomy vectors. Following this, \citeSubject{weng2008exploiting} proposed an item-based approach named HTR with the incorporation of both the user-item preference and user-taxonomic preference. 
Besides, category hierarchies give a precise description about functions and properties of products. They are utilized to estimate user preferences at different category levels for recommending POIs to users who visit a new city (\citeSub{bao2012location}).

\medskip\noindent\textbf{MMs+TFs.}
Some researchers adopted text features (e.g. reviews, comments) via either word-level text similarity or extracted sentiment. 
For instance, \citeSubject{terzi2014text}
proposed TextKNN to measure the similarity between users based on the similarity of text reviews instead of ratings. \citeSubject{pappas2013sentiment} developed a \textit{sentiment-aware nearest neighbor model} (SANN) for recommendations over TED talks. It adapts the estimated ratings by making use of the sentiment scores extracted from user comments. 

\medskip\noindent\textbf{Discussion.}
Memory-based methods (\citeSub{sarwar2001item};\citeSub{koren2008factorization}), however, are widely recognized as being less effective than model-based ones in large-scale datasets due to the time-consuming search in the user or item space. In a nutshell, the weak scalability of MMs limits their exploitation of the knowledge encoded in various side information, and even hinders them to encode side information with more complicated structural data (e.g., knowledge graphs) and non-structural data (e.g., images and videos). On the other hand, the underlying principles of fusing side information still provide valuable guidance for model-based methods.

\subsection{Latent factor models with side information}
Due to the high efficiency, state-of-the-art recommendation methods are mainly dominated by latent factor models (LFMs) (\citeSub{shi2014collaborative}), including \textit{matrix factorization} (MF)
(\citeSub{mnih2008probabilistic}; \citeSub{koren2009matrix}; \citeSub{wang2015exploring}),
\textit{weighted non-negative matrix factorization} (WNMF) (\citeSub{zhang2006learning}), \textit{Bayesian personalized ranking} (BPR) (\citeSub{rendle2009bpr}), 
\textit{tensor factorization} (TensorF) (\citeSub{karatzoglou2010multiverse}), \textit{factorization machine} (FM) (\citeSub{rendle2010factorization},  \citeyear{rendle2012factorization}), SVD++ (\citeSub{koren2008factorization}), timeSVD++ (\citeSub{koren2009collaborative}) and RFSS (\citeSub{zhao2017recommendation}). As discussed, they typically learn and model users' behavior (e.g., ratings, purchases) patterns by employing the global statistical information of historical user-item interaction data.
Specifically, they usually decompose the high-dimensional user-item rating matrices into low-rank user and item latent matrices. 
The basic idea is that both users and items can be characterized by a number of latent features, and thus the prediction can be computed as the inner product of corresponding user and item latent vectors. Many effective recommendation methods with side information fall into this category
(\citeSub{shi2011tags}; \citeSub{yang2012circle}; \citeSub{chen2012svdfeature}; \citeSub{hu2014your}; \citeSub{sun2017unified}). 

\begin{table}[t]
\footnotesize
\addtolength{\tabcolsep}{-1.2mm}
\renewcommand{\arraystretch}{0.65}
\centering
\caption{Summary of state-of-the-art latent factor model based recommendation algorithms with side information, where `FFs, NFs, FHs, KGs' represent structural side information, namely {flat features, network features, feature hierarchies} and {knowledge graphs}; `TFs, IFs' denote the non-structural side information, namely {text features} and {image features}. }\label{tab:lfm} 
\vspace{-0.1in}
\begin{tabular}{|lll|cccc|cc|l|}
\specialrule{.15em}{.05em}{.05em}
    \multirow{2}{*}{\textbf{Algorithm}} &\multirow{2}{*}{\textbf{Venue}} &\multirow{2}{*}{\textbf{Year}}  
    &\multicolumn{4}{c|}{\textbf{Structural}}
    &\multicolumn{2}{c|}{\textbf{Non-Str.}}
    &\multirow{2}{*}{\textbf{Reference}}
    \\\cline{4-9}
    &&&\textbf{FFs}&\textbf{NFs}&\textbf{FHs}&\textbf{KGs}&\textbf{TFs}&\textbf{IFs}&\\
    \specialrule{.05em}{.05em}{.05em}
    \specialrule{.05em}{.05em}{.05em}  
    CMF &KDD&2008&$\checkmark$&--&--&--&--&--&\citeauthor{singh2008relational}\\
    TensorF &RecSys &2010&$\checkmark$&--&--&--&--&--&\citeauthor{karatzoglou2010multiverse}\\
    HOSVD &TKDE & 2010  &$\checkmark$&--&--&--&--&--&\citeauthor{symeonidis2010unified} \\
    FPMC &WWW & 2010  &$\checkmark$&--&--&--&--&--&\citeauthor{rendle2010factorizing} \\
    TagCDCF &UMAP &2011 &$\checkmark$&--&--&--&--&--&\citeauthor{shi2011tags}\\
    CircleCon &KDD&2012&$\checkmark$&$\checkmark$&--&--&--&--&\citeauthor{yang2012circle}\\
    FM &TIST&2012&$\checkmark$&--&--&--&--&--&\citeauthor{rendle2012factorization,rendle2010factorization}\\
    SVDFeature&JMLR&2012&$\checkmark$&--&--&--&--&--&\citeauthor{chen2012svdfeature}\\
    NCRP-MF &SIGIR &2014   &$\checkmark$&--&--&--&$\checkmark$&--&\citeauthor{hu2014your}\\
    GeoMF &KDD &2014&$\checkmark$&--&--&--
    &--&--&\citeauthor{lian2014geomf}\\
    CAPRF &AAAI&2015&$\checkmark$&--&--&--&$\checkmark$&&\citeauthor{gao2015content}\\
    ARMF &KDD &2016&$\checkmark$&$\checkmark$
    &--&--&--&--&\citeauthor{li2016point} \\
    ICLF &UMAP&2017&$\checkmark$&--&--&--&--&--&\citeauthor{sun2017unified}\\
    TransFM &RecSys &2018 &$\checkmark$&--&--&--&--&--&\citeauthor{pasricha2018translation}\\
    TRec &ECRA &2019&$\checkmark$&--&--&--&$\checkmark$&--&\citeauthor{veloso2019online}\\
    \specialrule{.05em}{.05em}{.05em}
    SoRec &CIKM &2008
    &--&$\checkmark$&--&--&--&--&\citeauthor{ma2008sorec}\\
    RSTE &SIGIR &2009&--
    &$\checkmark$&--&--&--&--&\citeauthor{ma2009learning}\\
    RWT &RecSys &2009&--
    &$\checkmark$&--&--&--&--&\citeauthor{ma2009distrust}\\
    SocialMF &RecSys &2010 &--&$\checkmark$&--&--&--&--&\citeauthor{jamali2010matrix}\\
    SoReg &WSDM &2011 &--&$\checkmark$&--&--&--&--&\citeauthor{ma2011recommender}\\
    RSTE &TIST&2011&--&$\checkmark$&--&--&--&--
    &\citeauthor{ma2011learning}\\
    TrustMF &IJCAI &2013 &--&$\checkmark$&--&--&--&--&\citeauthor{liusocial}\\
    SR &SIGIR &2013&--&$\checkmark$&--&--&--&--&\citeauthor{ma2013experimental}\\
    DTrust &AAAI &2014 &--&$\checkmark$&--&--&--&--&\citeauthor{bao2014leveraging}\\
    MFTD &TOIS &2014 &--&$\checkmark$&--&--&--&--&\citeauthor{forsati2014matrix}\\
    TrustSVD &AAAI &2015 &--&$\checkmark$&--&--&--&--&\citeauthor{guo2015trustsvd}\\
    \specialrule{.05em}{.05em}{.05em}
    MF-Tax &RecSys& 2011 &--&--&$\checkmark$&--&--&--&\citeauthor{koenigstein2011yahoo}\\
    TaxLF &JMLR & 2011 &--&--&$\checkmark$&--&--&--&\citeauthor{mnih2011taxonomy}\\
    H+LR++ &KDD &2011 &--&--&$\checkmark$&--&--&--&\citeauthor{menon2011response}\\
    BMF &NIPS &2012 &--&--&$\checkmark$&--&--&--&\citeauthor{mnih2012learning}\\
    Tran-Cate &CIKM &2013 &--&--&$\checkmark$&--&--&--&\citeauthor{liu2013personalized}\\
    TaxF &VLDB &2013&--&--&$\checkmark$&--&--&--&\citeauthor{kanagal2012supercharging}\\
    ReMF &RecSys &2016 &--&--&$\checkmark$&--&--&--&\citeauthor{yang2016learning}\\
    CHMF &UMAP &2016 &--&--&$\checkmark$&--&--&--&\citeauthor{sun2016effective}\\
    Sherlock &IJCAI &2016 &--&--&$\checkmark$&--&--&$\checkmark$&\citeauthor{he2016sherlock}\\
    HieVH &AAAI &2017 &--&--&$\checkmark$&--&--&--&\citeauthor{sun2017exploiting}\\
    \specialrule{.05em}{.05em}{.05em}
    HeteMF &IJCAI &2013&--&--&--&$\checkmark$&--&--&\citeauthor{yu2013collaborative}\\
    HeteRec &RecSys &2013 &--&--&--&$\checkmark$&--&--&\citeauthor{yu2013recommendation}\\
    HeteRec\_p &WSDM &2014&--&--&--&$\checkmark$&--&--&\citeauthor{yu2014personalized}\\
    HeteCF &ICDM &2014  &--&$\checkmark$&--&$\checkmark$&--&--&\citeauthor{luo2014hete}\\
    SemRec &CIKM &2015  &--&$\checkmark$&--&$\checkmark$&--&--&\citeauthor{shi2015semantic}\\
    GraphLF &RecSys&2016&--&--&--&$\checkmark$&--&--&\citeauthor{catherine2016personalized}\\
    \specialrule{.05em}{.05em}{.05em}
    HFT &RecSys &2013 &--&--&--&--
    &$\checkmark$&--&\citeauthor{mcauley2013hidden}\\
    O\_Rec &UMAP &2013 &--&--&--&--
    &$\checkmark$&--&\citeauthor{pero2013opinion}\\
    EFM &SIGIR &2014&--&--&--&--
    &$\checkmark$&--&\citeauthor{zhang2014explicit}\\
    TopicMF &AAAI &2014 &--&--&--&--
    &$\checkmark$&--&\citeauthor{bao2014topicmf}\\
    EnFM &WWW &2017 &--&--&--&--&$\checkmark$&
    $\checkmark$&\citeauthor{chu2017hybrid}\\
    EBR &ECRA &2017&--&--&--&--&$\checkmark$&--
    &\citeauthor{pourgholamali2017embedding}\\
    AFV &ECRA &2018&--&--&--&--&$\checkmark$
    &--&\citeauthor{xu2018adjective}\\
    \specialrule{.05em}{.05em}{.05em}
    IRec &SIGIR&2015&--&--&--&--&--
    &$\checkmark$&\citeauthor{mcauley2015image}\\
    Vista &RecSys&2016&--&--&--&--&--&$\checkmark$&\citeauthor{he2016vista}\\
    VBPR &AAAI &2016&--&--&--&--&--&$\checkmark$
    &\citeauthor{he2016vbpr}\\
    TVBPR &WWW &2016 &--&--&--&-- &--&$\checkmark$&\citeauthor{he2016ups}\\
    VPOI &WWW &2017&--&--&--&--&--&$\checkmark$
    &\citeauthor{wang2017your}\\
    DeepStyle &SIGIR &2017 &$\checkmark$&--&--&--&--&$\checkmark$&\citeauthor{liu2017deepstyle}\\
    DCFA &WWW &2018&--&--&--&--&--&$\checkmark$
    &\citeauthor{yu2018aesthetic}\\
    \specialrule{.15em}{.05em}{.05em}
\end{tabular}
\end{table}

\begin{table}[t]
\footnotesize
\addtolength{\tabcolsep}{-1.5mm}
\centering
\caption{Classifications of state-of-the-arts w.r.t. LFMs+FFs.}\label{tab:lfm+ff}
\vspace{-0.1in}
\begin{tabular}{l|l}
\specialrule{.15em}{.05em}{.05em}
\textbf{Type} &\textbf{Representative Method}\\
\specialrule{.05em}{.05em}{.05em}
\specialrule{.05em}{.05em}{.05em}
\multirow{2}{*}{CMF} & (1) CMF (\citeSub{singh2008relational}); (2) MRMF (\citeSub{lippert2008relation}); \\
&(3) TagCDCF (\citeSub{shi2011tags}); (4) CAPRF (\citeSub{gao2015content})\\
\specialrule{.05em}{.05em}{.05em}
\multirow{2}{*}{SVDFeature} &(1) SVDFeature (\citeSub{chen2012svdfeature}); (2) NCRP-MF (\citeSub{hu2014your}); \\
&(3) TRec (\citeSub{veloso2019online})\\
\specialrule{.05em}{.05em}{.05em}
TensorF & (1) TensorF (\citeSub{karatzoglou2010multiverse}); (2) HOSVD (\citeSub{krizhevsky2012imagenet})\\
\specialrule{.05em}{.05em}{.05em}
FM &(1) FM (\citeauthor{rendle2012factorization} \citeyear{rendle2010factorization,rendle2012factorization}); (2) TransFM (\citeSub{pasricha2018translation})\\
\specialrule{.05em}{.05em}{.05em}
Others &(1) CircleCon (\citeSub{yang2012circle}); (2) ICLF (\citeSub{sun2017unified}); \\
& (3) ARMF (\citeSub{li2016point})\\
\specialrule{.15em}{.05em}{.05em}
\end{tabular}
\vspace{-0.2in}
\end{table}

\medskip\noindent\textbf{LFMs+FFs.} 
Early LFMs (See Table~\ref{tab:lfm}) incorporate \textit{flat features} (FFs) to help learn better user and item latent representations\footnote{In this survey, the following words `embedding', `representation', `latent vector' and `latent feature' are interexchangablely used.}. As further summarized in Table \ref{tab:lfm+ff},
several generic feature-based methods have been proposed. For instance, \citeSubject{singh2008relational} proposed \textit{collective matrix factorization} (\textit{CMF}) by simultaneously decomposing the user-item and user-feature/item-feature matrices.
Then, \citeSubject{chen2012svdfeature} designed \textit{SVDFeature}, which assumes that the representations of users or items can be influenced by those of their affiliated features.
\citeSubject{karatzoglou2010multiverse} proposed \textit{tensor factorization} (\textit{TensorF}), which is a generalization of MF that allows for a flexible and generic integration of features by modeling the data as a \textit{user-item-feature} N-dimensional tensor instead of the traditional 2D user-item matrix.
\citeauthor{rendle2010factorization} (\citeyear{rendle2010factorization}, \citeyear{rendle2012factorization}) devised \textit{factorization machine} (\textit{FM}) algorithm to model the pairwise interactions between all variables using factorized parameters.

Most of the state-of-the-art LFMs+FFs methods are built upon the four types of generic feature-based methods mentioned above: (1) based on CMF, \citeSubject{shi2014collaborative} introduced TagCDCF by factorizing the user-item and cross-domain tag-based user and item similarity matrices. \citeSubject{lippert2008relation} proposed a prediction model -- MRMF by jointly factorizing the user-item and user-feature (e.g., gender) 
as well as item-feature (e.g., genre) 
matrices. \citeSubject{gao2015content} proposed a location recommender -- CAPRF, which jointly decomposes the user-location interaction and location-tag affinity matrices; 
(2) based on SVDFeature, \citeSubject{hu2014your} proposed a rating prediction approach called NCRP-MF, which learns the embeddings of items by adding their affiliated categories. \citeSubject{veloso2019online}
proposed a hotel recommender -- TRec -- with the incorporation of hotel themes.
They argue that the embedding of a hotel should be reflected by those themes that the hotel belongs to; (3) based on TensorF, \citeSubject{symeonidis2010unified} proposed a unified recommendation model (HOSVD) via tensor factorization for user-tag-item triplet data; and (4) based on FM, \citeSubject{pasricha2018translation} proposed a sequential recommendation model -- TransFM, which adopts FM to fuse user and item flat features, such as user gender and item category. 

In addition to the aforementioned ones, there are still other related works. For instance, \citeSubject{yang2012circle} 
leveraged FFs to do pre-filtering. They designed CircleCon to infer the category-specific social trust circle for recommendation by assuming that a user may trust different subsets of friends regarding different categories. Given the assumption that users (items) have different preferences (characteristics) on different categories,  
\citeSubject{sun2017unified} proposed a category-aware model -- ICLF, which estimates a user's preference to an item by multiplying the inner product of the user and category latent vectors, and that of item and category latent vectors, where the category is the one that the item belongs to. Similarly, \citeSubject{li2016point} proposed ARMF to predict a user's taste over an item by multiplying the inner product of user and item latent vectors, and the user's preference to the affiliated categories of the item. 

\medskip\noindent\underline{Summary of LFMs+FFs.}
Table \ref{tab:lfm+ff} summarizes all the methods that belong to LFMs+FFs category. First, significant improvements have been achieved with these methods in comparison with the plain LFMs without considering FFs, which strongly verifies the usefulness of FFs for more effective recommendations. 
Second, comparable performance can be obtained by CMF, SVDFeature and TensorF based methods, while the time complexity of TensorF based methods far exceeds the other two types of methods. 
Third, extensive empirical studies have demonstrated the superiority of the FM based approaches among all the counterparts, as they explicitly consider the pair-wise interactions between users and items as well as their flat features.  
\begin{table}[t]
\footnotesize
\addtolength{\tabcolsep}{-1.5mm}
\centering
\caption{Classifications of state-of-the-arts w.r.t. LFMs+NFs.}\label{tab:lfm+nf}
\vspace{-0.1in}
\begin{tabular}{l|l}
\specialrule{.15em}{.05em}{.05em}
\textbf{Type} &\textbf{Representative Method}\\
\specialrule{.05em}{.05em}{.05em}
\specialrule{.05em}{.05em}{.05em}
\multirow{2}{*}{CMF} & (1) SoRec (\citeSub{ma2008sorec}); (2) DTrust (\citeSub{bao2014leveraging}); \\
&(3) TrustMF (\citeSub{liusocial}); (4) TrustSVD (\citeSub{guo2015trustsvd})\\
\specialrule{.05em}{.05em}{.05em}
Regularization & (1) RSTE (\citeauthor{ma2009learning}\citeyear{ma2009learning,ma2011learning})\\
\specialrule{.05em}{.05em}{.05em}
\multirow{3}{*}{SVDFeature} &(1) SocialMF (\citeSub{jamali2010matrix}); (2) SoReg (\citeSub{ma2011recommender}); \\
&(3) CircleCon (\citeSub{yang2012circle}); (4) SR (\citeSub{ma2013experimental})\\
&(5) MFTD (\citeSub{forsati2014matrix}); (6) RWT/RWD (\citeSub{ma2009distrust})\\
\specialrule{.15em}{.05em}{.05em}
\end{tabular}
\end{table}

\medskip\noindent\textbf{LFMs+NFs.}
Many studies integrated social networks into LFMs for achieving better recommendation performance. The underlying rationale is that users could share similar interests with their trusted friends.
Three types of representative methods, including CMF-based, SVDFeature-based, and regularization-based ones, are discussed in detail as follows.

\medskip\noindent\textit{(1) CMF based methods.}
One line of research is mainly based on \textit{collective matrix factorization} (CMF) (\citeSub{singh2008relational}), which jointly decomposes both the user-item interaction matrix and the user-user trust matrix.  
For example, \citeSubject{ma2008sorec} proposed SoRec to better learn the user embeddings by simultaneously factorizing the user-item and user-trust matrices. 
\citeSubject{bao2014leveraging} proposed DTrust, which decomposes trust into several aspects (e.g.,  benevolence, integrity) and further employs the support vector regression technique to incorporate them into the matrix factorization model for rating prediction. \citeSubject{liusocial} presented TrustMF, which leverages truster and trustee models to properly catch on a twofold influence of trust propagation on the user-item interactions. \citeSubject{guo2015trustsvd} devised TrustSVD, which inherently involves the explicit and implicit influence of rated items, and thus further incorporates both the explicit and implicit influence of trusted users.

\medskip\noindent\textit{(2) SVDFeature based methods.}
Another line of research mainly follows the idea of SVDFeature, which supposes that the representation of a user will be affected by that of her trusted friends. For example, \citeauthor{ma2009learning} (\citeyear{ma2009learning,ma2011learning}) proposed RSTE, which represents the embedding of a user by adding those of her trusted friends.

\medskip\noindent\textit{(3) Regularization based methods.}
The third line of research adopted the regularization technique (\citeSub{smola2003kernels}) to constrain the distance of embeddings between a user and her trusted friends. For example, SocialMF  (\citeSub{jamali2010matrix}) and CircleCon  (\citeSub{yang2012circle}) are designed on the assumption that a user and her trusted friends should be close to each other in their embedding space.
\citeauthor{ma2011recommender} (\citeyear{ma2011recommender,ma2013experimental}) proposed SoReg and SR to minimize the embedding difference of a user and their trusted friends. Later, \citeSubject{forsati2014matrix} and \citeSubject{ma2009distrust} respectively introduced MFTD and RWT/RWD to further employ distrust information to maximize the distance of embeddings between a user and her distrusted users.

\medskip\noindent\underline{Summary of LFMs+NFs.}
Table \ref{tab:lfm+nf} summarizes the three types of LFMs+NFs. To conclude, first, the effectiveness of NFs for more accurate recommendation has been empirically validated, when comparing with the plain LFMs.
Second, \textit{regularization} is generally a quite straightforward and time-efficient way to incorporate social influence, which naturally allows trust propagation among indirect social friends. For instance, suppose that users $u_j, u_k$ are friends of user $u_i$. By regularizing the distances of ($u_i, u_j$) and ($u_i, u_k$) respectively, the distance of ($u_j, u_k$) is indirectly constrained. Third, CMF based methods usually achieve the best performance. For instance, DTrust and TrustSVD outperform most of the other trust-aware approaches. Finally, the methods fusing both trust and distrust information perform better than those merely considering single aspect, suggesting the usefulness of distrust for recommendation; and this is further confirmed by the fact that the distrust-based methods perform almost as well as the trust-based methods (\citeSub{ma2009distrust}), which proves that the distrust information among users is as important as the trust information (\citeSub{fang2015multi}). 

\begin{table}[t]
\footnotesize
\addtolength{\tabcolsep}{-1mm}
\centering
\caption{Classifications of state-of-the-arts w.r.t. LFMs+FHs.}\label{tab:lfm+fh}
\vspace{-0.1in}
\begin{tabular}{l|l}
\specialrule{.15em}{.05em}{.05em}
\textbf{Type} &\textbf{Representative Method}\\
\specialrule{.05em}{.05em}{.05em}
\specialrule{.05em}{.05em}{.05em}
\multirow{2}{*}{{SVDFeature}} &(1) MF-Tax  (\citeSub{koenigstein2011yahoo}); (2) TaxF (\citeSub{kanagal2012supercharging}); \\
&(3) Sherlock (\citeSub{he2016sherlock}); (4) TaxLF (\citeSub{mnih2011taxonomy}); \\
& (5) CHLF (\citeSub{sun2017unified}); (6) HieVH (\citeSub{sun2017exploiting})\\
\specialrule{.05em}{.05em}{.05em}
{Regularization} & (1) H+LR++ (\citeSub{menon2011response}); (2) ReMF (\citeSub{yang2016learning})\\
\specialrule{.15em}{.05em}{.05em}
\end{tabular}
\end{table}

\medskip\noindent\textbf{LFMs+FHs.}
As summarized in Table \ref{tab:lfm+fh}, the first type of algorithms are based on the basic idea of SVDFeature. For instance, both MF-Tax (\citeSub{koenigstein2011yahoo}) and TaxF (\citeSub{kanagal2012supercharging}) model the embedding of an item by equally adding those of its ancestor features in the hierarchy.
Later, \citeSubject{he2016sherlock} proposed Sherlock, which manually defines the various influence of categories at different layers of the hierarchy. In contrast, TaxLF (\citeSub{mnih2011taxonomy}), CHLF (\citeSub{sun2017unified}) and HieVH (\citeSub{sun2017exploiting}) strive to automatically learn the different influences. 
The second type utilizes the regularization technique. For instance, \citeSubject{menon2011response} proposed an ad-click prediction method that regularizes the embeddings of features in the hierarchy via the child-parent relation. However, it assumes that an ad is conditionally independent from all higher layer features.
\citeSubject{yang2016learning} proposed ReMF to automatically learn the impacts of category hierarchies by parameterizing regularization traversing from the root to leaf categories.

\medskip\noindent\underline{Summary of LFMs+FHs.}
All representative LFMs+FHs methods are summarized in Table \ref{tab:lfm+fh}. 
Compared with FFs where features are independently organized at the same layer, the FHs provide human- and machine-readable descriptions of a set of features, and their parent-child relations. The richer knowledge encoded in FHs enables a more accurate and diverse recommendation.
Regardless of SVDFeature or regularization based methods, they all indicate that the categories at different layers of the hierarchy play different roles in characterizing the user-item interactions. The type of methods being able to automatically identify the different saliency of the hierarchical categories can achieve a better exploitation of FHs, so as to generate much more high-quality recommendations.

\medskip\noindent\textbf{LFMs+KGs.}
Most of the LFMs+KGs methods generally first extract meta paths (\citeSub{sun2011pathsim}) from KGs, and these paths are then fed into LFMs {for high-quality recommendations}.
Some of these methods adopt the regularization technique to incorporate the influence of the extracted meta paths. 
For instance, \citeSubject{yu2013collaborative} extracted paths connecting item pairs, and leveraged the path-based item similarity as the regularization coefficient of the pairwise item embeddings.
Another type of methods employs the path-based similarity to
learn the user preference diffusion. For example,  \citeSubject{yu2013recommendation} developed HeteRec
to learn the user preference diffusion to the unrated items that are connected with her rated items via meta paths.
It was further extended to HeteRec\_p for incorporating personalization via clustering users based on their interests. Similarly, \citeSubject{luo2014hete} proposed HeteCF, which leverages the path-based similarity to model user preference diffusion to unrated items. In addition, it also adds pairwise user (item) regularization to constrain the distance of embeddings of users (items) that are connected by meta paths. \citeSubject{shi2015semantic} devised SemRec that predicts the rating of a user to an item via a weighted combination of those of her similar users under different meta paths.  

Besides the meta-path-based approaches, there is another line of research focusing on designing graph-based methods mainly attributed to the underlying technique of random walk. For instance, by combining the strengths of LFMs with graphs, \citeSubject{catherine2016personalized} proposed GraphLF which adopts a general-purpose probabilistic logic system (ProPPR) for recommendation. 

\begin{table}[t]
\footnotesize
\centering
\caption{Classifications of state-of-the-arts w.r.t. LFMs+KGs.}\label{tab:lfm+kg}
\vspace{-0.1in}
\begin{tabular}{p{2.5cm}|p{2.5cm}|p{2cm}}
\specialrule{.15em}{.05em}{.05em}
\multicolumn{2}{c|}{\textbf{Meta-path-based method}} &\multirow{2}{*}{\textbf{Graph Method}} \\\cline{1-2}
\multicolumn{1}{c|}{\textbf{Regularization}}&\multicolumn{1}{c|}{\textbf{Diffusion}}& \\
\specialrule{.05em}{.05em}{.05em}
\specialrule{.05em}{.05em}{.05em}
\begin{tabitem}
  \item HeteMF \cite{yu2013collaborative}
  \item HeteCF \cite{luo2014hete}
\end{tabitem}
& 
\begin{tabitem}
    \item HeteRec \cite{yu2013recommendation}
    \item HeteRec\_p \cite{yu2014personalized}
    \item HeteCF \cite{luo2014hete}
    \item SemRec \cite{shi2015semantic}
\end{tabitem}
& 
\begin{tabitem}
    \item GraphLF \cite{catherine2016personalized}
\end{tabitem}
\tabularnewline
\specialrule{.15em}{.05em}{.05em}
\end{tabular}
\vspace{-0.2in}
\end{table}

\medskip\noindent\underline{Summary of LFMs+KGs.}
Table \ref{tab:lfm+kg} summarizes all the representative recommendation methods under LFMs+KGs. For a more in-depth discussion, first, by simplifying entity types and relation types, the complex KGs can be downgraded to other simple structural side information, such as FFs and NFs. For instance, we can only keep the item-category affinity relations, or user-user friendship relations in KGs to mimic FFs and NFs, respectively. 
From this point of view,
LFMs+KGs can be regarded as the generalized version of feature-based approaches.   
Second, the majority of these methods make use of meta paths (\citeSub{sun2011pathsim}) to extract knowledge from the KG. By incorporating meta paths, the ideas of other recommendation models such as user-/item-oriented CF can be easily
modeled in a generic way. Consider an example where we start with user $u_i$ and follow a meta path:
\begin{equation*}\small
    \text{User}\xrightarrow{\text{isFriendOf}} \text{User} \xrightarrow{\text{watched}} \text{Movie}.
\end{equation*} 
We thus can reach the movies that are watched by the friends of $u_i$. Hence, this meta path underpins the idea of user-oriented CF. To sum up, the usage of meta paths helps deliver an ensemble recommender. 
Third, the success of these methods, nevertheless, heavily relies on the quality and quantity of the handcrafted meta paths, which additionally requires domain knowledge. Besides, the manually designed features are often incomplete to cover all possible entity relations. These issues largely limit the capability of these methods to generate high-quality recommendations. 

\medskip\noindent\textbf{LFMs+TFs.}
Aside from the user-item rating matrices, the
relevant reviews often provide the rationale for users' ratings and identify what aspects of an item they cared most about, and what sentiment they held for the item. 
We summarize four types of methods under LFMs+TFs: word-level, sentiment-level, aspect-level, and topic-level methods. 
They mainly focus on extracting useful information encoded in the text features, such as reviews, tips, comments, content and descriptions, to further boost recommendation accuracy. 

\medskip\noindent\textit{(1) Word-level methods.}
Word-level methods usually directly extract the words from textual information. For instance, \citeSubject{hu2014your} proposed a SVDFeature based method (NCRP-MF) that models the embedding of a business by adding those of words extracted from its relevant reviews. 
\citeSubject{pourgholamali2017embedding} proposed a feature-based matrix factorization method (EBR) which first extracts words from product descriptions and user review texts, and then employs the word embedding technique (\citeSub{mikolov2013efficient}) to learn semantic product and user representations. 
These representations are ultimately incorporated into the matrix factorization model for better recommendations.   
\citeSubject{chu2017hybrid} proposed EnFM, which extracts important words from textual reviews via \textit{term frequency and inverse document frequency} (TF-IDF) technique (\citeSub{ramos2003using}).
They enhanced the \textit{factorization machine} (FM) by fusing the extracted words as features of users and items.  

\medskip\noindent\textit{(2) Sentiment-level methods.}
The second type is the sentiment-level, that is, analyzing the sentiment expressed by the textual information. Some studies leverage the extracted sentiment to do pre- or post-filtering. For instance, \citeSubject{pero2013opinion} proposed O\_pre to pre-process the user-item interaction matrix to generate the user-item opinion matrix, where the opinion matrix is obtained based on textual reviews. They also devise O\_post to post-process the predicted ratings by adding the estimated opinion score. \citeSubject{veloso2019online} proposed TRec to utilize binary sentiment score extracted from reviews to post-filter low-quality items from the final item ranking list. 
Other studies leverage the extracted sentiment from reviews as the corresponding confidence for the factorization, which indicates the importance of each user-item interaction pair. 
O\_model (\citeSub{pero2013opinion}) and CAPRF (\citeSub{gao2015content}) both adopted the user-item sentiment matrix as a confidence matrix to constrain the factorization process.
Recently, \citeSubject{xu2018adjective} proposed AFV based on SVD (\citeSub{paterek2007improving}), which employs the adjective features extracted from user reviews to reflect users' perceptions on items.
It automatically learns user and item representations under these features for more accurate and explainable item recommendations.  

\medskip\noindent\textit{(3) Aspect-level methods.}
To minimize the reliance on sentiment analysis accuracy, the third type is based on the aspect-level, which extracts aspects (the specific properties of items) from textual information. 
For instance, \citeSubject{he2015trirank} proposed Trirank that accommodates users, items and aspects into a heterogeneous graph. They adopted graph regularization technique (\citeSub{smola2003kernels}) to constrain the distance of user-item, user-aspect and item-aspect pairs. \citeSubject{guo2017aspect} developed a knowledge graph named as \textit{aspect-aware geo-social influence graph}, which incorporates the geographical, social and aspect information into a unified graph.
\citeSubject{zhang2014explicit} devised EFM
by extracting both aspect and sentiment from the user reviews. It builds user-aspect attention and item-aspect quality matrices based on the phrase-level sentiment analysis, and then simultaneously decomposes these two matrices together with the user-item interaction matrix. 

\medskip\noindent\textit{(4) Topic-level methods.}
This type of methods exploits the topic modeling methods, such as \textit{latent Dirichlet allocation} (LDA) (\citeSub{blei2003latent}), to extract the latent topics in the review texts. For instance, \citeSubject{mcauley2013hidden} proposed the \textit{hidden factors as topics} (HFT) approach, which learns the item latent-topic distribution from all related reviews. The learned distribution is then linked with the corresponding item latent factor via a transformation function. Later, \citeSubject{bao2014topicmf} further extended HFT by proposing TopicMF. It correlates the latent topics of each review with the user and item latent factors simultaneously. 
Also, AFV (\citeSub{xu2018adjective}) adopts LDA to learn the item-topic distribution. Then the Kullback-Leibler (KL) divergence (\citeSub{hershey2007approximating}) is utilized to calculate the review-topic-based neighbors of items.

\begin{table}[t]
\footnotesize
\addtolength{\tabcolsep}{-1mm}
\centering
\caption{Classifications of state-of-the-arts w.r.t. LFMs+TFs.}\label{tab:lfm+tf}
\vspace{-0.1in}
\begin{tabular}{l|l}
\specialrule{.15em}{.05em}{.05em}
\textbf{Type} &\textbf{Representative Method}\\
\specialrule{.05em}{.05em}{.05em}
\specialrule{.05em}{.05em}{.05em}
\multirow{2}{*}{\textbf{Word}} &(1) NCRP-MF (\citeSub{hu2014your}); (2) EBR (\citeSub{pourgholamali2017embedding}) \\
&(3) EnFM (\citeSub{chu2017hybrid}) \\
\specialrule{.05em}{.05em}{.05em}
\multirow{3}{*}{\textbf{Sentiment}} & 
(1) O\_pre, O\_post, O\_model (\citeSub{pero2013opinion});\\
&(2) CAPRF (\citeSub{gao2015content});  (3) AFV (\citeSub{xu2018adjective});\\ &(4) TRec (\citeSub{veloso2019online});  (5) ORec (\citeSub{zhang2015geosoca}) \\
\specialrule{.05em}{.05em}{.05em}
\textbf{Aspect} &(1) Trirank (\citeSub{he2015trirank}); (2) EFM (\citeSub{zhang2014explicit})  \\
\specialrule{.05em}{.05em}{.05em}
\multirow{2}{*}{\textbf{Topic}}
& (1) HFT (\citeSub{mcauley2013hidden}); (2) TopicMF (\citeSub{bao2014topicmf}) \\
& (3) AFV (\citeSub{xu2018adjective}) \\
\specialrule{.15em}{.05em}{.05em}
\end{tabular}
\vspace{-0.1in}
\end{table}

\medskip\noindent\underline{Summary of LFMs+TFs.}
We summarize these state-of-the-art methods in Table \ref{tab:lfm+tf}, where word-level approaches are the most straightforward ones without any understanding about the content of text features, while the aspect- and sentiment-level approaches leverage \textit{natural language processing} (NLP) toolkits to extract useful information from the text features. The NLP technique helps the algorithms explicitly understand what aspects of an item that users care most about, and what opinions they possess for the item. The accuracy of aspect extraction and sentiment analysis, nonetheless, is the bottleneck for further advancements.  
Moreover, the topic-level approaches go deeper to extract the latent topics hidden in the text features. The learned latent topic distribution enables the algorithms to achieve a subtle understanding of the user-item interactions. 
In a nutshell, from word-level to the sentiment- and aspect-level, and ultimately to the topic-level, an increasingly deeper and more subtle understanding on the text features is gradually achieved, which enables the algorithms to model text features from a raw- to fine-grained manner. 
All methods mentioned above, however, ignore the fact that not all reviews written by a user (or written for an item) are equally important for modeling user preferences (or item characteristics), and even not all the words contained in one review contribute identically to represent this review. These issues have been well addressed by the deep learning models with attention mechanisms, which we will introduce later. 

\medskip\noindent\textbf{LFMs+IFs.}
There are also several latent factor models that consider image features, since images play a vital role in domains like fashion, where the visual appearances of products have great influence on user decisions. 
As a type of extremely complicated non-structural data, the fusion of image features for LFMs, however, generally follows two phases: (1) extract visual features from images based on the pre-trained models, such as deep neural networks; and (2) fuse the extracted visual features into LFMs for better recommendations.  

\citeauthor{he2016sherlock} (\citeyear{he2016vbpr,he2016sherlock,he2016ups,he2016vista}) proposed a series of recommendation methods for the fashion domain to exploit visual features that are extracted from product images by pre-trained deep neural networks (\citeSub{jia2013caffe}). They include:
(1) {VBPR} (\citeSub{he2016vbpr}) is an extension of BPR-MF (\citeSub{rendle2009bpr}) that learns an embedding kernel to linearly transform the high-dimensional product raw visual features into a much lower-dimensional `visual rating' space. The low-dimensional product visual features are then fused into BPR-MF for more accurate recommendations;
(2) {Sherlock} (\citeSub{he2016sherlock}) upgrades VBPR, and learns additional product visual vectors by mapping the raw product visual features across hierarchical categories of products. It, thus, accounts for both high-level and subtle product visual characteristics simultaneously; 
(3) {TVBPR} (\citeSub{he2016ups}) advances VBPR by studying the evolving visual factors that customers consider when evaluating products, so as to make better recommendations; 
and (4) {Vista} (\citeSub{he2016vista}) further takes into account visual, temporal and social influences simultaneously for a sequential recommendation in fashion domain. 

Other researchers also endeavored to make use of visual features for more accurate recommendation. For instance, 
\citeSubject{mcauley2015image} proposed an image-based recommender, IRec, which employs the visual features to distinguish alternative and complementary products. 
\citeSubject{chu2017hybrid} devised a restaurant recommder, EnFM, by leveraging two types of visual features, namely \textit{convolutional neural network} (CNN) (\citeSub{lecun1995convolutional}) features and \textit{color name} (CN). It first represents the attribute of a restaurant by the visual features of its related images, and then feeds the attribute into the factorization model to help achieve high-quality recommendations.    
\citeSubject{wang2017your} designed VPOI to incorporate the visual content of restaurants for an enhanced \textit{point-of-interest} (POI) recommendation. For a user-location pair $(u,l)$, it minimizes the difference between user $u$ (location $l$) and her posted images (its relevant images) with regard to their embeddings.
\citeSubject{liu2017deepstyle} proposed DeepStyle for learning {style} features of items and sensing preferences of users. 
It learns item style representations by 
subtracting the corresponding item category representations from the visual features generated via CNN. The learned item style representations are then fused into BPR for personalized recommendations. 
Similarly, \citeSubject{yu2018aesthetic} proposed a tensor factorization model -- DCFA to leverage the aesthetic features rather than the conventional features to represent an image.
They believed that a user' s decision largely depends on whether the product is in line with her aesthetics.

\medskip\noindent\underline{Summary of LFMs+IFs.}
Undoubtedly, the image features have inspired new models for recommender systems, that is, exploiting visual features (e.g., style of an item) to model the user-item interactions, and have greatly boosted recommendation accuracy and attractiveness.  
As we mentioned earlier, due to the intricacy of image features, they can not be directly used by LFMs. To this end, two modules are required to be separately trained: (1) feature extraction module usually adopts deep learning techniques, such as CNN (\citeSub{lecun1995convolutional}), to learn visual representations of items from images; and
(2) the preference learning module, such as matrix factorization, takes the learned visual representations to help adjust the final item representations learning process. 
Such separately learning hinders these methods from achieving optimal performance improvements. In this view, it calls for unified and elegant recommendation models to exploit such kinds of features.  

\medskip\noindent\textbf{Discussion of LFMs with side information.}
First, compared with memory-based methods, LFMs have relatively higher scalability and flexibility, which enables them to incorporate various types of side information, regardless of the structural or non-structural ones, for more accurate recommendations.
However, the investigation of fusing structural data is more prevalent than that of non-structural data, as illustrated by Table \ref{tab:lfm};
Second, most of the LFMs with side information are extended from the generic feature-based methods, such as matrix factorization with side information regularization (\citeSub{jamali2010matrix}), \textit{collective matrix factorization} (CMF) (\citeSub{singh2008relational}), SVDFeature (\citeSub{chen2012svdfeature}), \textit{tensor factorization} (TensorF) (\citeSub{karatzoglou2010multiverse}), and \textit{factorization machine} (FM) (\citeauthor{rendle2010factorization} \citeyear{rendle2010factorization,rendle2012factorization});
Third, generally, the more complex side information has been incorporated, the more high-quality recommendations can be achieved. For instance, LFMs+FHs outperforms LFMs+FFs, and the performance of LFMs+KGs is better than that of other LFMs with structural side information; 
Fourth, although LFMs are capable of accommodating more complex side information (i.e., knowledge graphs, text and image features), the information encoded in such complex data cannot be directly utilized by LFMs.
In this sense, most of these methods are composed of two independent phases, namely \textit{feature extraction} and \textit{preference learning}. 
For instance, 
the fusion of knowledge graphs is based on meta-path (\citeSub{sun2011pathsim}) extraction. The integration of text features relies on aspect extraction and sentiment analysis advances (\citeSub{zhang2014users}), and the utilization of image features highly depends on deep neural networks (\citeSub{jia2013caffe}; \citeSub{lecun1995convolutional}) to help extract visual features of images.  
The independence of these two phases limits further performance increments of LFMs with side information to some degree. This issue has been partially alleviated by deep learning models, as elaborated in the subsequent subsections.  

\begin{figure*}[ht]
\centering
    \subfigure[Auto-Encoder Model]{
    \includegraphics[scale=0.5]{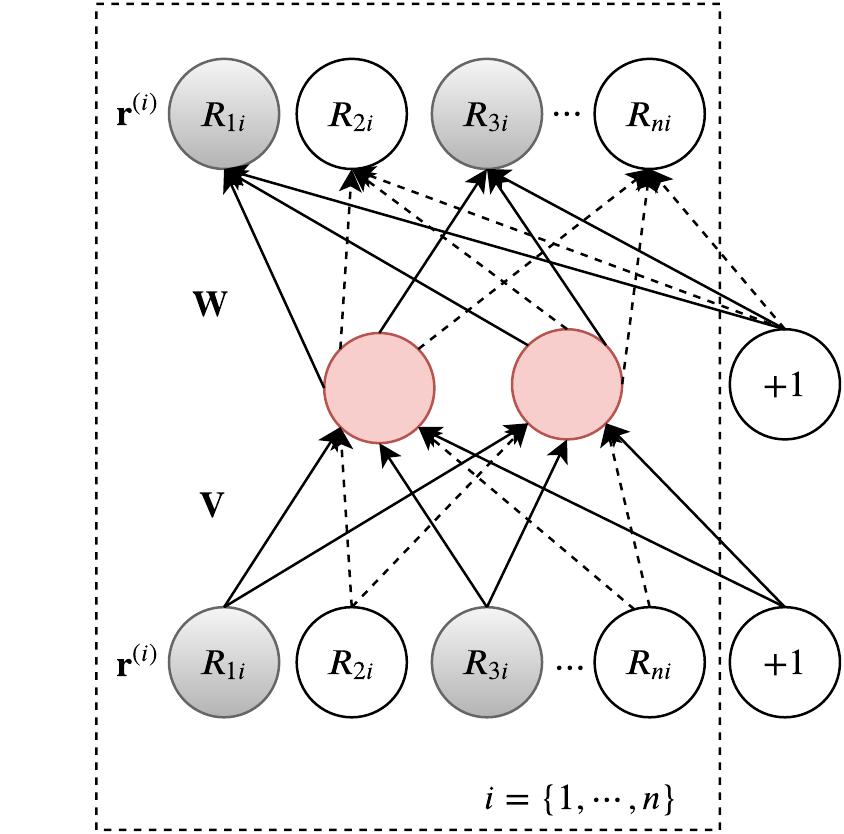}
    }
    \hspace{0.1in}
    \subfigure[NeuMF Model]{
    \includegraphics[scale=0.5]{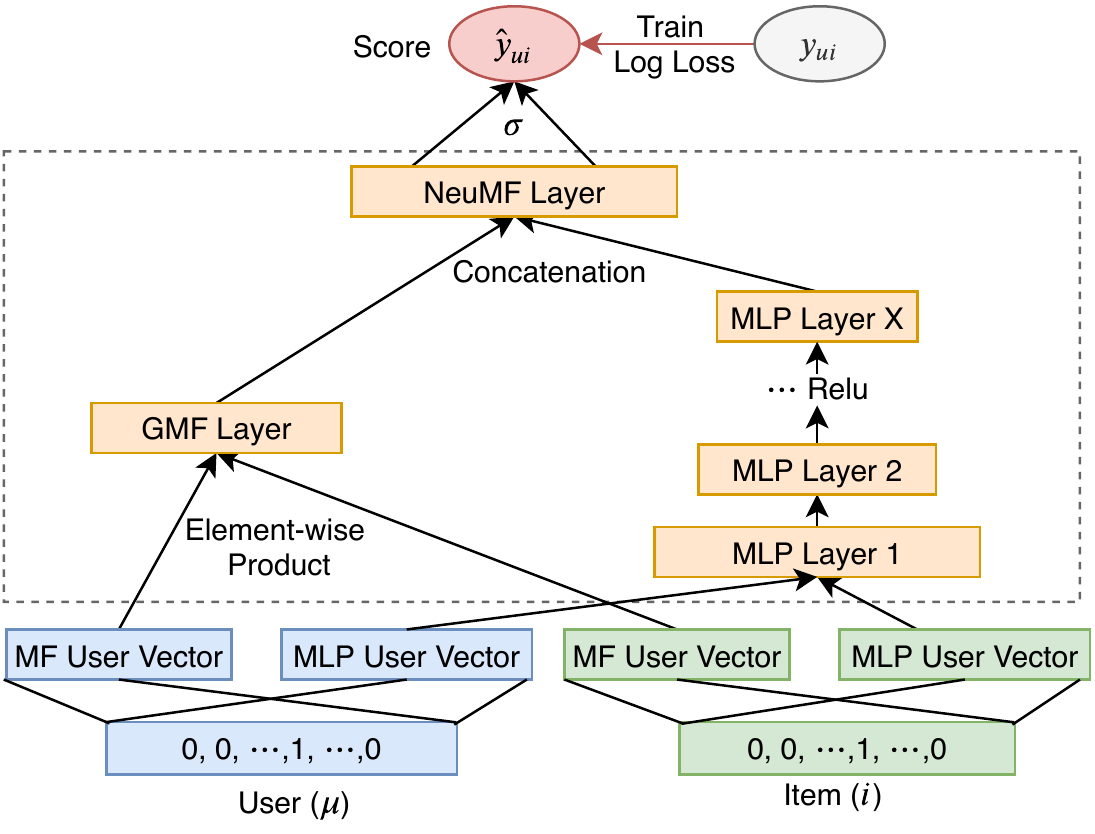}
    }
    \hspace{0.1in}
    \subfigure[DMF Model]{
    \includegraphics[scale=0.5]{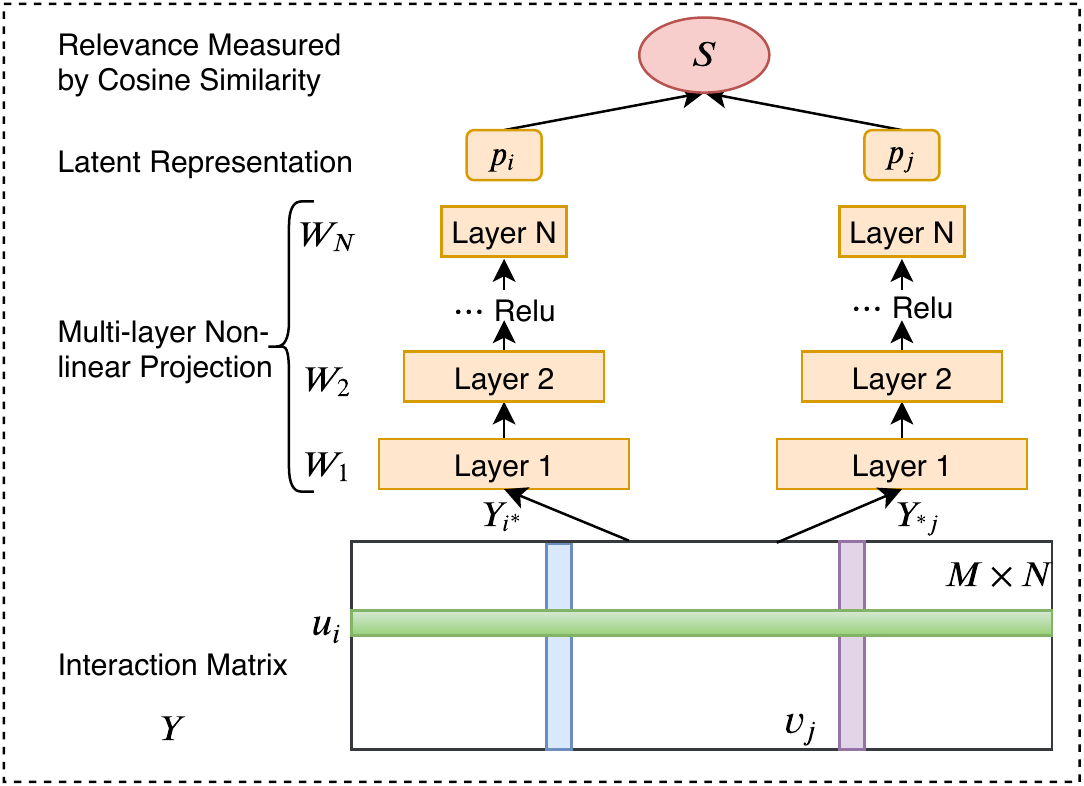}
    }
    \vspace{-0.15in}
    \caption{(a) Item-based AutoRec model, where we use the plate notation to indicate that there are n copies of the neural network (one for each item), where $W, V$ are tied across all copies; $\mathbf{r}^{(i)}$ denotes the observed rating vector for item $i$ (\citeSub{sedhain2015autorec});
    (b) Neural matrix factorization model, which fuses generalized matrix factorization (GMF) on the left side and multi-layer perceptron (MLP) on the right side (\citeSub{he2017neural}); (c) Deep matrix factorization model, which leverages multi-layer non-linear projections to learn user and item representations (\citeSub{xue2017deep}). }
    \label{fig:mlp}
    \vspace{-0.2in}
\end{figure*}

\subsection{Representation learning models with side information}
In contrast to LFMs, \textit{representation learning models} (RLMs) have proven to be effective for recommendation tasks in terms of capturing local item relations by utilizing item embedding techniques.

There are many studies related to RLMs for recommendation. For instance, \citeSubject{barkan2016item2vec} first devised a neural item embedding model (Item2Vec) for collaborative filtering, which is capable of inferring item-to-item relationships. 
Note that the original Item2Vec cannot model user preferences by only learning item representations. Some studies, therefore, extended Item2Vec by taking personalization into account. For instance,  
\citeSubject{grbovic2015commerce} developed three recommendation models: Prod2Vec, BagProd2Vec and User2Vec. Specifically, Prod2Vec learns product representations at product-level over the entire set of receipt logs, whereas BagProd2Vec learns at the receipt-level. User2Vec simultaneously learns representations of products and users by considering the user as a ``global context", motivated by Paragraph2Vec algorithm (\citeSub{le2014distributed}). 
Next, \citeSubject{wang2015learning} proposed a novel \textit{hierarchical representation model} (HRM) to predict what users will buy in the next
basket (sequential recommendation). HRM can capture both sequential behavior and users' general tastes.
\citeSubject{liang2016factorization} proposed CoFactor based upon CMF (\citeSub{singh2008relational}), which synchronously decomposes the user-item interaction matrix and the item-item co-occurrence matrix. 

By taking advantages of RLMs, some researchers attempted to integrate side information (e.g., categories, tags) into RLMs to help learn better user and item embeddings, thus to gain further performance enhancements for recommendation (\citeSub{grbovic2015commerce}; \citeSub{vasile2016meta}; \citeSub{sun2017mrlr}).
For instance, 
\citeSubject{vasile2016meta} extended Item2Vec to a more generic non-personalized model -- MetaProd2Vec, which utilizes item categories to assist in regularizing the learning of item embeddings.
\citeSubject{liu2016exploring} proposed temporal-aware model (CWARP-T) by leveraging the Skip-gram model. It jointly learns the latent representations for a user and a location, so as to respectively capture the user's preference as well as the influence of the context of the location. 
\citeSubject{feng2017poi2vec} designed POI2Vec which incorporates the geographical influence to jointly model user
preferences and POI sequential transitions. 
Recently, \citeSubject{sun2017mrlr} proposed a personalized recommender, MRLR, to jointly learn the user and item representations, where the item representation is regularized by item categories. 

\begin{table}[t]
\footnotesize
\addtolength{\tabcolsep}{-0.8mm}
\centering
\caption{Classifications of state-of-the-arts w.r.t. representation learning models (RLMs), where `Basic' denotes the fundamental RLMs without side information; and `Side' represents the RLMs with side information.}\label{tab:rlm}
\vspace{-0.1in}
\begin{tabular}{l|l|l}
\specialrule{.15em}{.05em}{.05em}
\textbf{Type} &\textbf{Non-Personalized} &\textbf{Personalized} \\
\specialrule{.05em}{.05em}{.05em}
\specialrule{.05em}{.05em}{.05em}
\parbox[t]{2mm}{\multirow{3}{*}{\rotatebox[origin=c]{90}{\textbf{Basic}}}} &Item2Vec (\citeSub{barkan2016item2vec}) &User2Vec (\citeSub{grbovic2015commerce})\\
&Prod2Vec (\citeSub{grbovic2015commerce})& HRM (\citeSub{wang2015learning}) \\
&BagProd2Vec (\citeSub{grbovic2015commerce}) &CoFactor (\citeSub{liang2016factorization}) \\
\specialrule{.05em}{.05em}{.05em} 
\parbox[t]{2mm}{\multirow{3}{*}{\rotatebox[origin=c]{90}{\textbf{Side}}}}
&MetaProd2Vec (\citeSub{vasile2016meta}) & CWAPR-T (\citeSub{liu2016exploring}) \\
&&POI2Vec (\citeSub{feng2017poi2vec})\\
&&MRLR (\citeSub{sun2017mrlr})\\
\specialrule{.15em}{.05em}{.05em}
\end{tabular}
\end{table}

\medskip\noindent\textbf{Discussion of RLMs with side information.} Table \ref{tab:rlm} summarizes all the RLMs based recommendation approaches. 
First, although there is much less work on RLMs with side information than that on LFMs with side information, RLMs provide a different viewpoint to learn item representations: by capturing local item relations in terms of each individual user's interaction data
while LFMs aim to learn user and item representations at the global level.
Second, the objective function of fundamental RLMs (Item2Vec) is actually a softmax layer which has been widely adopted in the attention mechanisms (\citeSub{chen2017attentive}; \citeSub{seo2017interpretable}), or as the output layer of many deep learning models 
(\citeSub{zhang2017next}; \citeSub{yang2017bridging}).
From this viewpoint, RLMs can be considered as transitions from shallow to deep neural networks. 
Third, the fundamental RLMs (Item2Vec) do not consider personalization very well, and should be further extended to accommodate the user's preference. This can be expressed in different manners, such as the averaged representations of items that the user has interacted with, or be treated as the ``global context" via Paragraph2Vec (\citeSub{le2014distributed}). 
Fourth, most of the RLMs with side information focus on incorporating simple flat features, such as item categories (\citeSub{sun2017mrlr}; \citeSub{vasile2016meta}). This is equivalent to adding regularizations on the item embedding learning process. Similarly, other types of data (e.g., FHs) could be considered and adapted to further augment the performance of RLMs based recommendation methods. 
Lastly, the exploitation of Item2Vec stemming from Word2Vec inspires more technique transformations from the NLP domain to recommendation tasks, such as Paragraph2Vec and Document2Vec (\citeSub{le2014distributed}).

%% file: section/deeplearning.tex
\begin{figure*}[htbp]
    \centering
    \hspace{-0.3in}
    \subfigure[Overview of Caser]{
    \includegraphics[scale=0.36]{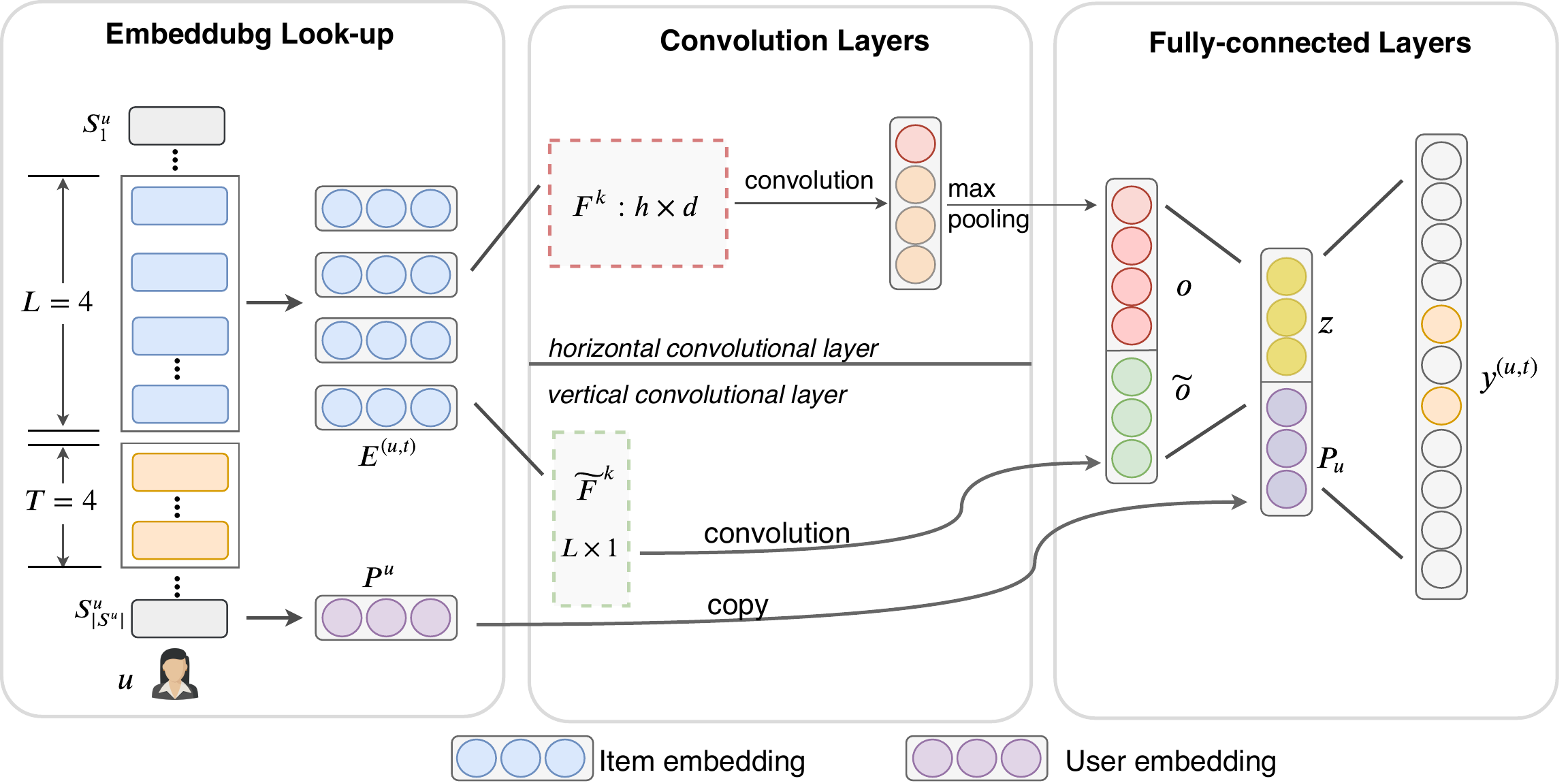}
    }
    \hspace{0.2in}
    \subfigure[Overview of ConvNCF]{
    \includegraphics[width=0.28\textwidth]{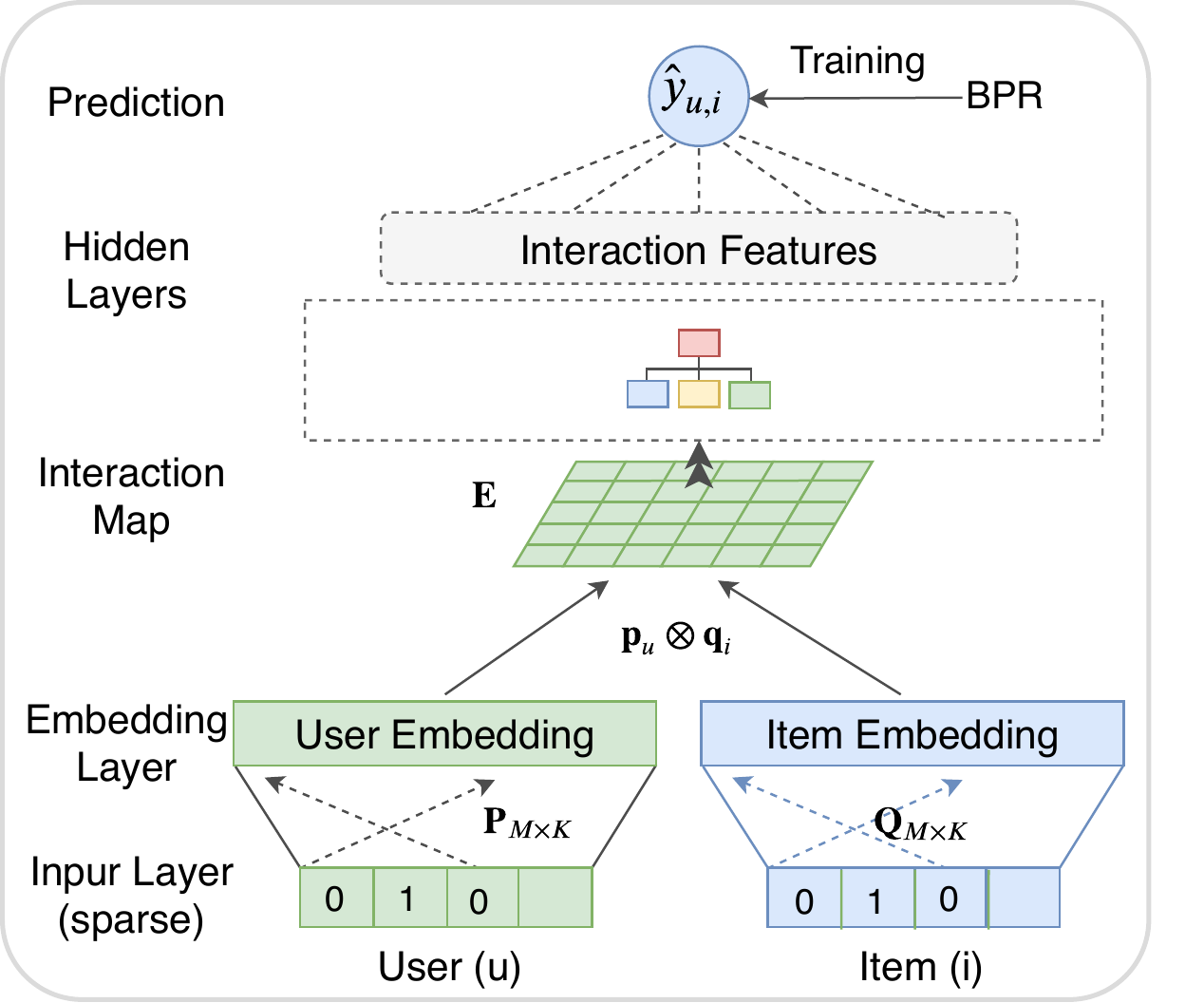}
    }
    \hspace{-0.3in}
    \caption{(a) the architecture of Caser, where the dash rectangular boxes are convolutional filters with different sizes, It uses previous 4 actions (L = 4) to predict which items user $u$ will interact with 
    in next 2 steps (T = 2) (\citeSub{tang2018personalized});
    (b) the overall framework of ConvNCF, where the correlation of user and item embeddings is expressed by the outer product, and then CNN is adopted to learn the high-level abstract correlation (\citeSub{he2018outer}). }
    \label{fig:cnn}
\end{figure*}
\section{Deep learning models with side information}\label{sec:deep}
Deep learning models (DLMs) have gained significant success in various domains, such as computer vision (CV) (\citeSub{krizhevsky2012imagenet}), speech recognition (\citeSub{schmidhuber2015deep}), and natural language processing (NLP) (\citeSub{cho2014learning}). They have also recently attracted tremendous research interest from the recommendation community. In contrast to LFMs and RLMs, DLMs based recommendation approaches (e.g., AutoRec (\citeSub{sedhain2015autorec}), NCF (\citeSub{he2017neural})) can learn nonlinear latent representations through various types of activation functions, such as sigmoid, ReLU (\citeSub{nair2010rectified}).
Thanks to the excellent flexibility of DLMs, side information can be efficiently integrated. Plenty of DLMs employ different kinds of structural side information to help achieve better recommendations, such as item categories (\citeSub{pei2017interacting}), social networks (\citeSub{ding2017baydnn}) and knowledge graphs (\citeSub{sun2018rkge}; \citeSub{wang2018explainable}). Moreover, as DLMs have achieved superior performance in CV and NLP, another important research line focuses on leveraging non-structural side information for more effective recommendations, including visual content (\citeSub{niu2018neural}; \citeSub{liu2017deepstyle}; \citeSub{chen2017attentive}) and textual content (\citeSub{catherine2017transnets}; \citeSub{zheng2017joint}; \citeSub{seo2017interpretable}; \citeSub{tay2018multi}).
Therefore, this section aims to provide in-depth analysis on DLMs with various types of side information.\footnote{Here, for facilitating the presentation, we consider all artificial neural networks as deep learning models, including the ones with one hidden layer (e.g., shallow auto-encoder). } 

\subsection{Basic deep learning models}
We first provide an overview of the basic DLMs without integration of side information. These methods, though, merely take into account the user-item historical interaction data, have achieved significant improvements on the recommendation performance due to the superiority of DLMs. They can be broadly classified into five categories as introduced below. We provide a relatively detailed elaboration of these models as they are the bases for more sophisticated deep learning models with side information.

\medskip\noindent\textbf{Auto-Encoder based methods.}
Auto-Encoder is the simplest neural network with three layers which projects (encodes) the high-dimensional input layer into a low-dimensional hidden layer, and finally re-projects (decodes) the hidden layer to the output layer. The goal is to minimize the reconstruction error, that is, to find the most efficient compact representations for the input data.
One early work was AutoRec proposed by (\citeSub{sedhain2015autorec}), as illustrated by Fig. \ref{fig:cnn}(a). It adopts fully-connected layers to project the partially observed user or item vectors into a low-dimensional hidden space, which is then reconstructed into the output space to predict the missing ratings.

\medskip\noindent\textbf{MLP based methods.}
MLP is short for \textit{multi-layer perceptron} (\citeSub{rumelhart1985learning}), which contains one or more hidden layers with arbitrary activation functions providing levels of abstraction.
Thus, it is a universal network to extract the high-level features for approximating the user-item interactions. 
Based on MLP, \citeSubject{he2017neural} proposed the \textit{neural collaborative filtering} (NCF) framework which tries to integrate generalized matrix factorization (GMF) with MLP: (1) GMF applies a linear kernel to model the user-item interactions in latent space; and (2) MLP uses a non-linear kernel to learn the user-item interaction function from the data. Fig. \ref{fig:mlp}b shows a way of fusing GMF and MLP (called the NeuMF model), where their outputs are concatenated and fed into the NeuMF layer.
\citeSubject{xue2017deep} designed a \textit{deep matrix factorization} model (DMF), which exploits multi-layer non-linear projections to learn the user and item representations by making use of both explicit and implicit feedbacks (See Fig. \ref{fig:mlp}c).

\medskip\noindent\textbf{CNN based methods.}
In essence, \textit{convolutional neural network} (CNN) (\citeSub{lecun1995convolutional}) can be treated as a variant of MLP. It takes input and output with fixed sizes, and its hidden layers typically consist of convolutional layers, pooling layers, and fully connected layers. 
By regarding the input data as an image, CNN can be utilized to help capture the local features. 
For instance, \citeSubject{tang2018personalized} proposed Caser for next item recommendation, as depicted in Fig. \ref{fig:cnn}a. It embeds a sequence of recent interacted items into a latent space which is considered as an image.
Convolutional filters for two directions are then adopted: (1) horizontal filters facilitate to capture union-level patterns with multiple union sizes; and (2) vertical filters help capture point-level sequential patterns through weighted sums over latent representations of previous items.  
\citeSubject{he2018outer} designed ConvNCF, as shown by Fig. \ref{fig:cnn}b. It first utilizes an outer product (interaction map) to explicitly model the pairwise correlations between user and item embeddings, and then employs CNN to learn high-order correlations among embedding dimensions from locally to globally in a hierarchical way.

\begin{figure*}[ht]
    \centering
    \subfigure[DREAM]{
    \includegraphics[width=0.7\textwidth]{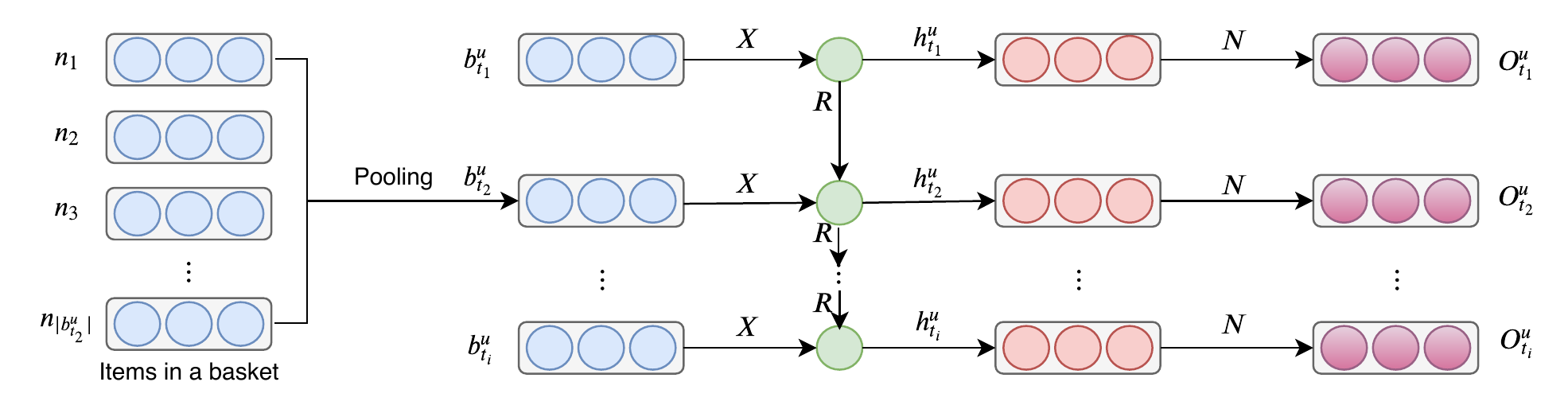}
    }
    \subfigure[ST-RNN]{
    \includegraphics[width=0.3\textwidth]{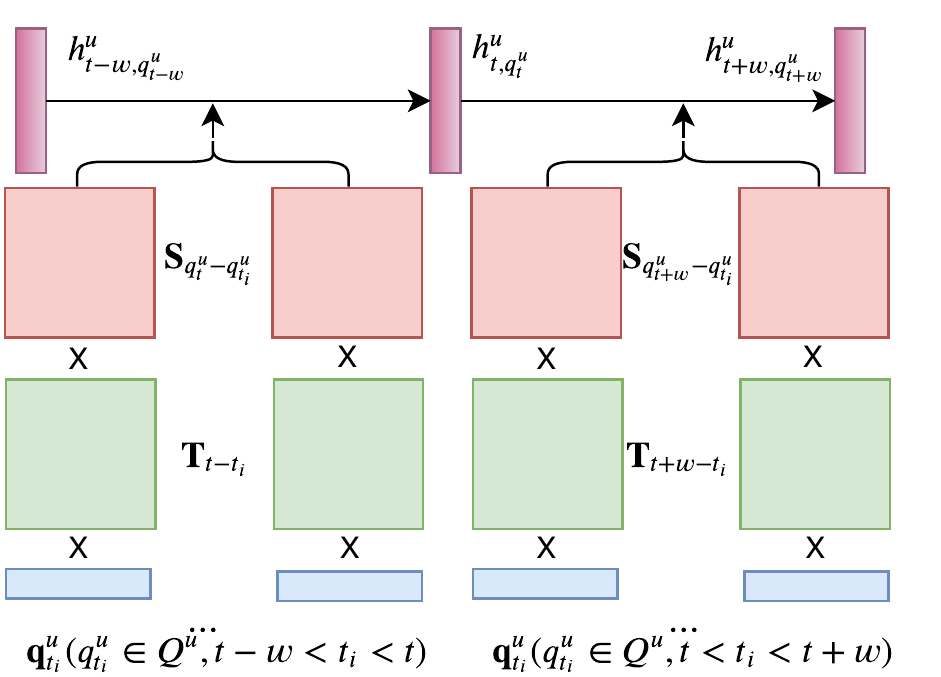}
    }
    \hspace{0.1in}
    \subfigure[RRN]{
    \includegraphics[width=0.3\textwidth]{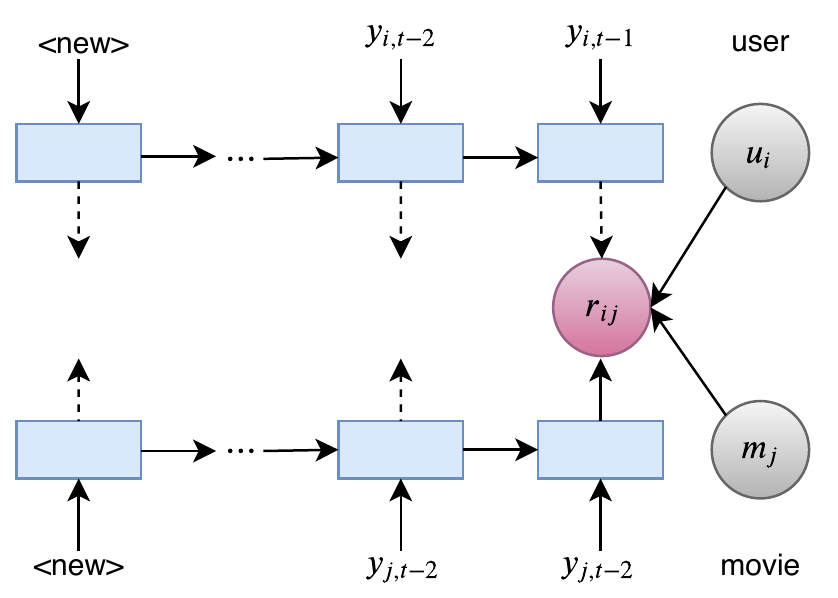}
    }
    \caption{(a) the overall framework of DREAM, where the pooling operation on the items in a basket aims to get the representation
    of the basket. The input layer comprises a series of basket representations of a user. The hidden layer handles the dynamic representation of the user, and the output layer shows scores of this user for all items (\citeSub{yu2016dynamic});
    (b) ST-RNN injects the time- and distance-specific transition matrices to the input embedding (i.e., embedding of location visited by user $u$ at time $t_i$ -- $q_{t_i}^u$) of RNN at each time step (\citeSub{liu2016predicting}); (c) RRN utilizes individual recurrent networks to address the temporal evolution of user and movie state respectively. The state evolution for a user depends on which movies (and how) a user rated previously. Likewise, a movie's parameters are dependent on the users that rated it in the previous time interval and its popularity among them. To capture stationary attributes, it adopts an additional (conventional) set of auxiliary parameters $u_i$ and $m_j$ for users and movies respectively (\citeSub{wu2017recurrent}). 
    }
    \label{fig:rnn}
    \vspace{-0.2in}
\end{figure*}

\medskip\noindent\textbf{RNN based methods.}
\textit{Recurrent neural network} (RNN) (\citeSub{collobert2011natural}) has been introduced in recommendation tasks mainly for temporal recommendation and sequential recommendation (or next item recommendation), as it is capable of memorizing historical information and finding patterns across time. For instance, \citeSubject{hidasi2015session} developed SeRNN for session-based next-item recommendation. It is built upon the \textit{gated recurrent unit} (GRU) (\citeSub{cho2014properties}), a more elaborate model of RNN for dealing with the vanishing gradient problem.
\citeSubject{yu2016dynamic} designed a \textit{dynamic recurrent basket model} (DREAM) based on RNN, which not only learns a dynamic representation
of a user but also captures the global sequential features among the baskets, as illustrated by Fig. \ref{fig:rnn}a.
\citeSubject{jing2017neural} devised NSR based on \textit{long-short term memory} (LSTM) (\citeSub{hochreiter1997long})
to estimate when a user will return to a site and predict her future listening behavior. 
\citeSubject{liu2016predicting} proposed a novel \textit{spatial temporal RNN} (ST-RNN) for next-location recommendation as depicted in Fig. \ref{fig:rnn}b, which models both local temporal and spatial contexts in each layer with time-specific as well as distance-specific transition matrices, respectively.
\citeSubject{wu2017recurrent} proposed RRN to capture both the user and item temporal dynamics by endowing both users and items with a LSTM auto-regressive model, as depicted by Fig. \ref{fig:rnn}c. 

\begin{figure*}[t]
    \centering
    \subfigure[DeepMove]{
    \includegraphics[scale=0.45]
    {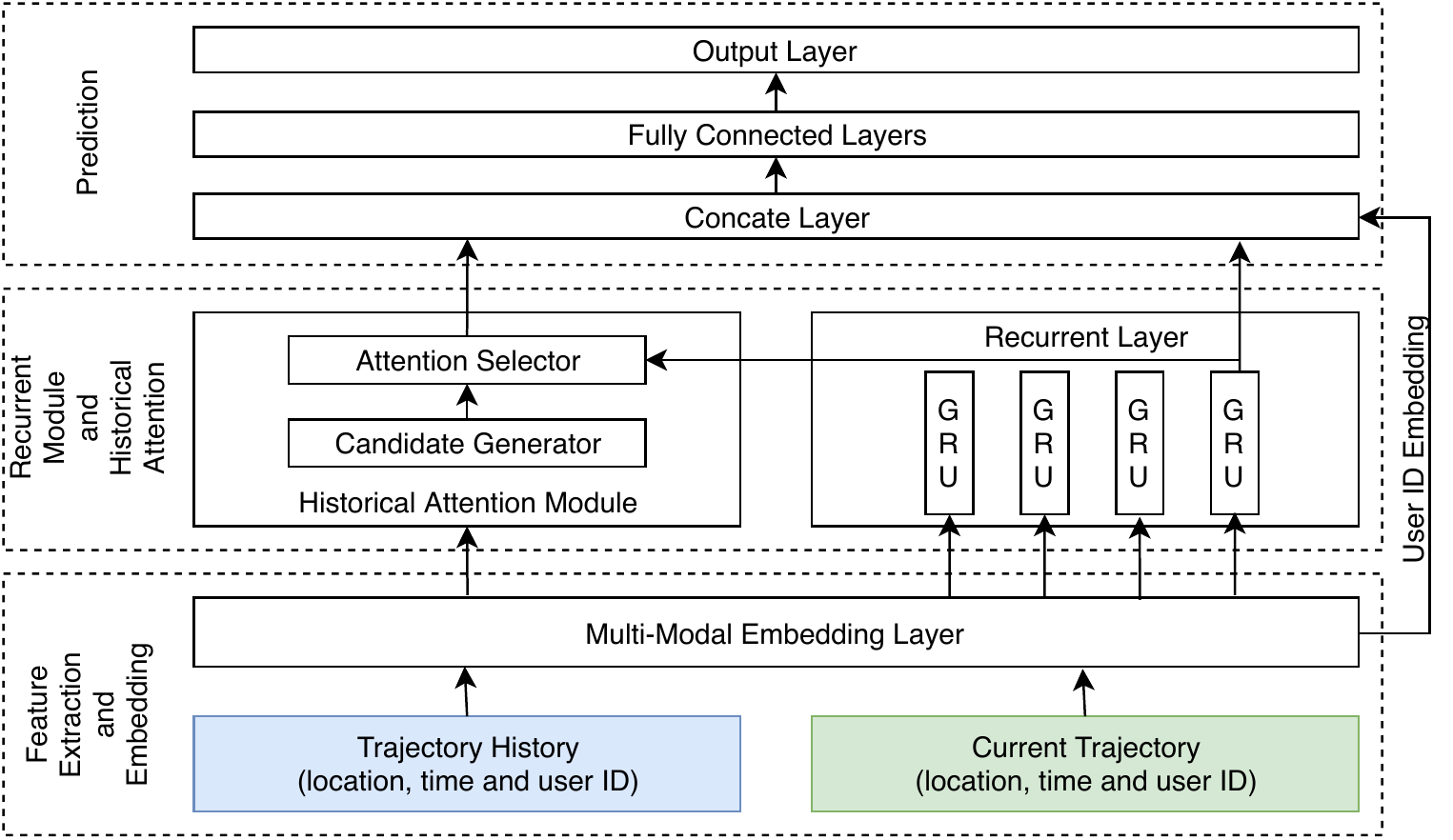}
    }
    \hspace{0.2in}
    \subfigure[AttRec]{
    \includegraphics[scale=0.55]
    {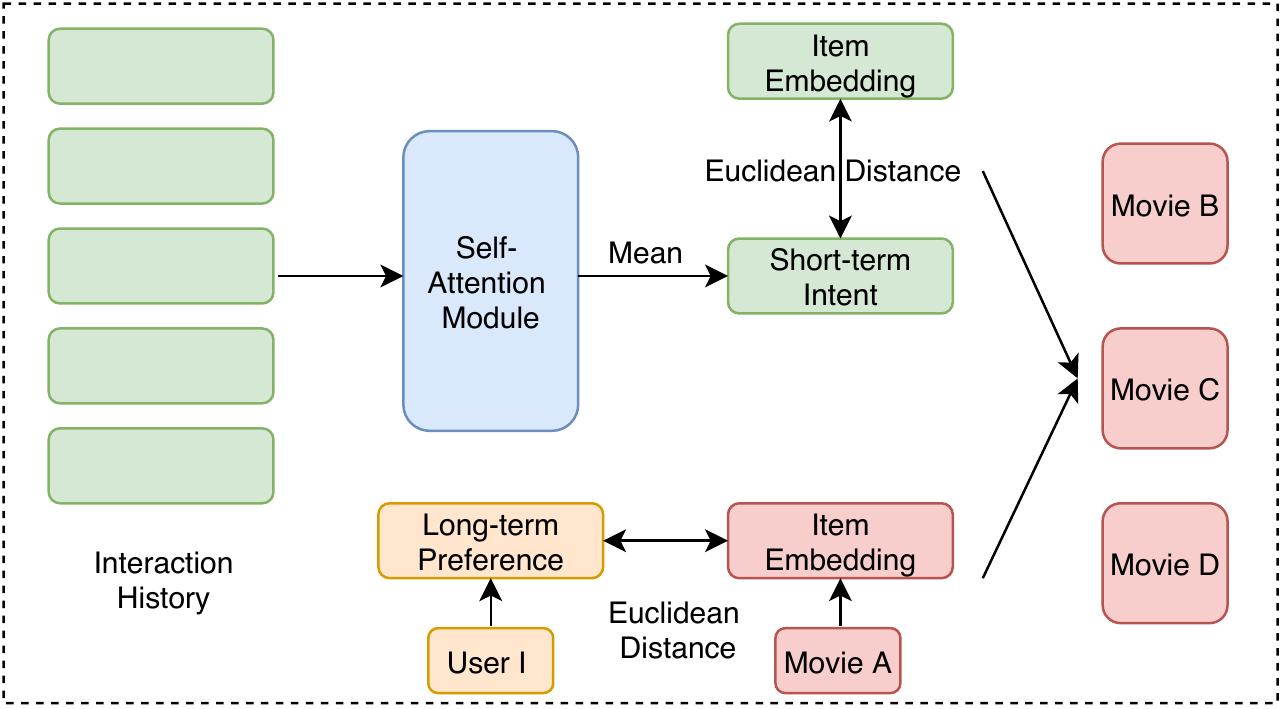}
    }
    \vspace{-0.15in}
    \caption{(a) the overall framework of DeepMove, where the Recurrent Layer captures sequential transitions contained in the current trajectory, and the Historical Attention Module captures the mobility regularity from the historical trajectories. The Concate Layer combines all the features from Attention, Recurrent and Embedding modules into a new vector (\citeSub{feng2018deepmove}); (b) the overview of AttRec, where both short- and long-term user's interests have been considered, and the short-term interest is learned via a self-attention mechanism (\citeSub{zhang2018next}). }
    \label{fig:attention}
    \vspace{-0.15in}
\end{figure*}

\medskip\noindent\textbf{Attention based methods.}
Motivated by the human visual attention nature and attention mechanisms in natural language processing (\citeSub{yang2016hierarchical}) and computer vision (\citeSub{pei2017temporal}; \citeSub{xu2015show}), attention has gained tremendous popularity in the community of recommender systems. 
It mainly aims to cope with the data noisy problem by identifying relevant parts of the input data for modeling the user-item interactions (\citeSub{pei2017interacting}). 
The standard vanilla attention mechanism learns the attention scores for the input data by transforming the representations of input data via fully-connected layers, and then adopting an extra softmax layer to normalize the scores
(\citeSub{pei2017interacting}; \citeSub{chen2017attentive}; \citeSub{wang2018dkn}). Normally, the attention mechanism often cooperates with either RNN to better memorize very long-range dependencies, or CNN to help concentrate on the important parts of the input. 
For instance, \citeSubject{feng2018deepmove} designed DeepMove for user mobility prediction depicted by Fig. \ref{fig:attention}a. It designs RNN to capture the sequential transitions contained in the current trajectory, and meanwhile proposes a historical attention model to capture the mobility regularity from the lengthy historical records.

Recently, self-attention (\citeSub{vaswani2017attention}) has started to gain exposure, as it can replace RNN and CNN in the sequence learning, achieving better accuracy with lower computation complexity. 
It focuses on co-learning and self-matching
of two sequences whereby the attention weights of one sequence are conditioned on the other sequence, and vice verse (\citeSub{zhang2018next}).
Inspired by self-attention, \citeSubject{zhang2018next} proposed a novel sequence-aware recommendation model, AttRec, by considering both short- and long-term user interests shown as Fig. \ref{fig:attention}b. 
It utilizes self-attention mechanism (\citeSub{vaswani2017attention}) to estimate the relative weights of each item in the user's interaction trajectories to learn better representations for the user’s transient interests.
Similarly, \citeSubject{kang2018self} also developed a self-attention based sequential model (SASRec) that can capture the long-term user interests, but makes the predictions based on relatively few actions. It identifies which items are `relevant' from a user's action history with an attention mechanism.

\begin{table}[t]
\footnotesize
\addtolength{\tabcolsep}{2mm}
\centering
\caption{Classifications of basic deep learning models for recommendation, where `AE' denotes auto-encoder; `Attn' refers to attention.}\label{tab:dl+basic}
\vspace{-0.1in}
\begin{tabular}{l|l}
\specialrule{.15em}{.05em}{.05em}
\textbf{Type} &\textbf{Representative Method}\\
\specialrule{.05em}{.05em}{.05em}
\specialrule{.05em}{.05em}{.05em}
\textbf{AE} & (1) AutoRec (\citeSub{sedhain2015autorec}); \\
\specialrule{.05em}{.05em}{.05em}
\textbf{MLP} & (1) NCF (\citeSub{he2017neural}); (2) DMF (\citeSub{xue2017deep}) \\
\specialrule{.05em}{.05em}{.05em}
\textbf{CNN} & (1) Caser (\citeSub{tang2018personalized}); (2) ConvNCF (\citeSub{he2018outer}) \\
\specialrule{.05em}{.05em}{.05em}
\multirow{2}{*}{\textbf{RNN}} & (1) SeRNN (\citeSub{hidasi2015session}); (2) DREAM (\citeSub{yu2016dynamic}) \\
& (3) NSR (\citeSub{jing2017neural}); (4) RRN (\citeSub{wu2017recurrent})\\
\specialrule{.05em}{.05em}{.05em}
\multirow{2}{*}{\textbf{Attn}} & (1) DeepMove (\citeSub{feng2018deepmove}); (2) AttRec (\citeSub{zhang2018next}) \\
& (3) SASRec (\citeSub{kang2018self}) \\
\specialrule{.15em}{.05em}{.05em}
\end{tabular}
\vspace{-0.1in}
\end{table}

\medskip\noindent\underline{Summary of basic DLMs.} 
Table \ref{tab:dl+basic} summarizes the representative basic DLMs that are the essential bases, and can be readily adapted for DLMs with side information. 
First, Auto-Encoder, as the simplest neural network, can be extended to fuse both structural and non-structural side information by learning the contextual representations of items from flat features (e.g., item categories) (\citeSub{dong2017hybrid}), text (e.g., articles, reviews) (\citeSub{okura2017embedding}) or image (e.g., movie posters) (\citeSub{zhang2016collaborative}) features, as we will discuss later.   
Second, as a universal network, MLP helps efficiently extract the high-level user and item representations for better recommendations. Besides, it can be easily extended to fuse structural side information by concatenating flat features with user or item embeddings as the input data
(\citeSub{cheng2016wide}; \citeSub{covington2016deep}; \citeSub{niu2018neural}). 
Third, CNN is extensively exploited to capture the spatial patterns, that is, the local relations among the features in the ``image'' format data with fixed input and output lengths. Thus, it is more capable of coping with non-structural side information, such as texts and images.  
Fourth, RNN is generally employed to capture sequential patterns or temporal dynamics with arbitrary input and output lengths. Hence, it is more suitable for sequential recommendation to predict what next item that users will be interested in (\citeSub{yao2017serm}; \citeSub{zhang2017next}), or explainable recommendation to generate texts (e.g., review, tips) (\citeSub{li2017neural}; \citeSub{lu2018like}). 
Last, the emergence of vanilla attention mechanisms further advance existing neural networks (e.g., CNN, RNN) by explicitly distinguishing the different importance of the input data. Self-attention mechanisms can revolutionize the deep learning structures. 
To sum up, all of the above methods are foundations of DLMs with side information taken into account (as summarized in Table \ref{tab:dl}), and will be elaborated in the following subsections.

\subsection{Deep learning models with flat features (DLMs+FFs)}

Plenty of DLMs have incorporated flat features (e.g., user gender, item categories) for better recommendations. 

\medskip\noindent\textbf{Auto-Encoder based methods.}
\citeSubject{dong2017hybrid} developed a hybrid recommender, HDS, which makes use of both the user-item rating matrix and flat features. It learns deep user and item representations based on two \textit{additional stacked denoising auto-encoders} (aSDAEs) (\citeSub{vincent2010stacked}) with the side information as the input (e.g., user gender, item categories). The aSDAEs are jointly trained with the matrix factorization to minimize the rating estimation error as well as aSDAE reconstruction error. 
\citeSubject{okura2017embedding} presented a news recommender, ENR, which first learns distributed representations of articles based on a variant of a denoising autoencoder (\citeSub{vincent2008extracting}).  The model is trained in a pair-wise manner with a triplet $(a_0, a_1, a_2)$ as input to preserve their categorical similarity, where the articles $a_0, a_1$ are in the same category and $a_0, a_2$ belong to different categories. Then, it generates user representations by using a \textit{recurrent neural network} (RNN) with browsing histories as input sequences, and finally matches and lists articles for users based on inner-product operations. 

\begin{table}[htbp]
\footnotesize
\addtolength{\tabcolsep}{-1.4mm}
\renewcommand{\arraystretch}{0.9}
\caption{Summary of state-of-the-art deep learning based recommendation algorithms with side information, where `FFs, NFs, FHs, KGs' represent structural side information, namely \textit{flat features, network features, feature hierarchies} and \textit{knowledge graphs}; `TFs, IFs, VFs' denote the non-structural side information, namely \textit{text features, image features} and \textit{video features}.}\label{tab:dl}
\vspace{-0.1in}
\begin{tabular}{|lll|cccc|ccc|l|}
\specialrule{.15em}{.05em}{.05em}
    \multirow{2}{*}{\textbf{Algorithm}} &\multirow{2}{*}{\textbf{Venue}} &\multirow{2}{*}{\textbf{Year}}  
    &\multicolumn{4}{c|}{\textbf{Structural}}
    &\multicolumn{3}{c|}{\textbf{Non-Struct.}}
    &\multirow{2}{*}{\textbf{Reference}}
    \\\cline{4-10}
    &&&\textbf{FFs}&\textbf{NFs}&\textbf{FHs}&\textbf{KGs}&\textbf{TFs}&\textbf{IFs}&\textbf{VFs}&\\
    \specialrule{.05em}{.05em}{.05em}
    \specialrule{.05em}{.05em}{.05em}
    Wide\&Deep &RecSys&2016&$\checkmark$&--&--&--&--&--&--&\citeauthor{cheng2016wide}\\
    DNN &RecSys &2016  &$\checkmark$&--&--&--&$\checkmark$&--&--&\citeauthor{covington2016deep}\\
    CDL-Image &CVPR &2016&$\checkmark$&$\checkmark$&--&--&--&$\checkmark$&--&\citeauthor{lei2016comparative}\\
    HDS &AAAI &2017&$\checkmark$&--&--&--&--&--&--&\citeauthor{dong2017hybrid}\\
    IARN &CIKM &2017  &$\checkmark$&--&$\checkmark$&--&--&--&--&\citeauthor{pei2017interacting}\\
    ENR &KDD &2017 &$\checkmark$&--&--&--&$\checkmark$&--&--&\citeauthor{okura2017embedding}\\
    SH-CDL &TKDE &2017&$\checkmark$&--&--&--&$\checkmark$
    &--&--&\citeauthor{yin2017spatial}\\
    NPR &WSDM &2018&$\checkmark$&--&--&--&--&$\checkmark$&--&\citeauthor{niu2018neural}\\
    \specialrule{.05em}{.05em}{.05em}
    3D-CNN &RecSys&2017&--&--&$\checkmark$&--&$\checkmark$&--&--&\citeauthor{tuan20173d}\\
    NEXT &arXiv &2017&--&$\checkmark$&--&--&$\checkmark$&--&--&\citeauthor{zhang2017next}\\
    BayDNN &CIKM&2017&--&$\checkmark$&--&--&--&--&--&\citeauthor{ding2017baydnn}\\
    DeepSoR &AAAI &2018 &--&$\checkmark$&--&--&--&--&--&\citeauthor{fan2018deep}\\
    GraphRec &arXiv&2019&--&$\checkmark$&--&--&--&--&--&\citeauthor{fan2019graph}\\
    \specialrule{.05em}{.05em}{.05em}
    CKE &KDD&2016&--&--&--&$\checkmark$&$\checkmark$&$\checkmark$&--&\citeauthor{zhang2016collaborative}\\
    TransNets & RecSys & 2017&--&--&--&$\checkmark$ &--&--&--&\citeauthor{catherine2017transnets}\\
    PACE & KDD & 2017 &--&--&--&$\checkmark$ &$\checkmark$&--&--&\citeauthor{yang2017bridging}\\
    RKGE &RecSys&2018 &--&--&--&$\checkmark$&--&--&--&\citeauthor{sun2018rkge}\\
    KPRN &AAAI &2018  &--&--&--&$\checkmark$&--&--&--&\citeauthor{wang2018explainable}\\
    DKN &WWW &2018&--&--&--&$\checkmark$&$\checkmark$&--&--&\citeauthor{wang2018dkn}\\
    RippleNet &CIKM &2018&--&--&--
    &$\checkmark$&--&--&--&\citeauthor{wang2018ripplenet}\\
    RippleNet-agg &TOIS &2019&--&--&--
    &$\checkmark$&--&--&--&\citeauthor{wang2019exploring}\\
    RCF &arXiv &2019&--&--&--
    &$\checkmark$&--&--&--&\citeauthor{xin2019relational}\\
    KGCN &arXiv &2019&--&--&--&$\checkmark$&--&--&--&\citeauthor{wang2019knowledgegraph}\\
    KGCN-LS &arXiv &2019&--&--&--
    &$\checkmark$&--&--&--&\citeauthor{wang2019knowledge}\\
    MKR &arXiv &2019&--&--&--
    &$\checkmark$&--&--&--&\citeauthor{wang2019multi}\\
    KTUP &arXiv &2019&--&--&--
    &$\checkmark$&--&--&--&\citeauthor{cao2019unifying}\\
    KGAT &arXiv&2019&--&--&--&$\checkmark$&--&--&--&\citeauthor{wang2019kgat}\\
    \specialrule{.05em}{.05em}{.05em}
    CDL &KDD &2015&--&--&--&--&$\checkmark$&--&--&\citeauthor{wang2015collaborative}\\
    SERM &CIKM &2017&--&--&--&--&$\checkmark$
    &--&--&\citeauthor{yao2017serm}\\
    JRL &CIKM &2017  &--&--&--&--&$\checkmark$ &$\checkmark$&--&\citeauthor{zhang2017joint}\\
    DeepCoNN&WSDM&2017&--&--&--&--&$\checkmark$&--&--&\citeauthor{zheng2017joint}\\
    NRT &SIGIR &2017&--&--&--&--&$\checkmark$&--&--&\citeauthor{li2017neural}\\
    D-Attn &RecSys&2017&--&--&--&--&$\checkmark$&--&--&\citeauthor{seo2017interpretable}\\
    RRN-Text &arXiv &2017  &--&--&--&--&$\checkmark$&--&--&\citeauthor{wu2016joint}\\
    MT &RecSys &2018 &--&--&--&--&$\checkmark$&--&--&\citeauthor{lu2018like}\\
    MPCN &KDD &2018&--&--&--&--&$\checkmark$&--&--&\citeauthor{tay2018multi}\\
    \specialrule{.05em}{.05em}{.05em}
    Exp-Rul &AAAI&2017&--&--&--&--&--&$\checkmark$&--&\citeauthor{alashkar2017examples}\\
    ACF &SIGIR &2017&--&--&--&--&--&$\checkmark$&$\checkmark$&\citeauthor{chen2017attentive}\\
    \specialrule{.15em}{.05em}{.05em}
\end{tabular}
\end{table}

\medskip\noindent\textbf{MLP based methods.}
\citeSubject{cheng2016wide} designed Wide\&Deep to jointly train wide linear regression models and deep neural networks, where the categorical features are converted into low-dimensional embeddings, and then fed into the hidden layers of the deep neural network.
\citeSubject{covington2016deep} introduced DNN for video recommendation on Youtube (\url{www.youtube.com}). It includes: (1) a deep candidate generation model where a user's watch history and side information such as the demographic features of users, are concatenated into a wide layer followed by several fully connected layers with ReLU (\citeSub{nair2010rectified}) to generate video candidates that users are most likely to watch; and (2) a deep ranking model which has similar architecture as the candidate generation model to assign a ranking score to each candidate video.  
\citeSubject{niu2018neural} proposed a pair-wise image recommender, NPR, with the fusion of multiple contextual information including tags, geographic and visual features. 
It adopts one fully-connected layer and element-wise product to learn the user's contextual (i.e., topic, geographical and visual) preference representations, which is concatenated with her general preference representation in the merged layer. Finally, the merged preference representation is connected with the feed-forward network to generate recommendations.

\medskip\noindent\textbf{CNN based methods.}
\citeSubject{lei2016comparative} developed a pair-wise learning method (CDL-Image) for image recommendation. Three sub-networks are involved, wherein two identical sub-networks are used to learn representations of positive and negative images for each user via CNN, and the remaining one is used to learn the representation of user preference via four fully-connected layers. The input user vectors for this network are vectors of relevant tags generated by Word2Vec. 

\medskip\noindent\textbf{RNN based methods.}
\citeSubject{pei2017interacting} devised a `RNN+Attention' based approach (IARN), which is similar as RRN (\citeSub{wu2016joint}). It uses two RNNs to capture both user and item dynamics. The estimated rating of user $u$ on item $i$ is the inner product of the corresponding hidden representations, which are transformations of the final hidden states in the two RNNs. Furthermore, it learns the attention scores of user and item history in an interactive way to capture the dependencies between the user and item dynamics. 
Feature encoder is used to fuse a set of categories $K$ that item $i$ belongs to, where a category $k\in K$ is modeled as a transformation function $\mathbf{M}_i^k$ that projects the item embedding $\mathbf{e}_i$ into a new space. Therefore, the influence of $K$ flat categories is denoted by the sum of their respective impacts, that is, $\sum_{k=1}^K\mathbf{M}^k_i\cdot\mathbf{e}_i$. 

\begin{table}[t]
\footnotesize
\addtolength{\tabcolsep}{-0.8mm}
\renewcommand{\arraystretch}{0.7}
\centering
\caption{Classifications of DLMs+FFs, where `AE' denotes auto-encoder; `Pre-f, Conc, Proj' represent 
\textit{Pre-filter, Concatenate, Projection}, repsectively.}\label{tab:dl+ff}
\vspace{-0.1in}
\begin{tabular}{l|cccc|ccc|l}
\specialrule{.15em}{.05em}{.05em}
\multirow{2}{*}{\textbf{Method}} &\multicolumn{4}{c|}{\textbf{Model Type}}&\multicolumn{3}{c|}{\textbf{FF Usage Type}}&\multirow{2}{*}{\textbf{Reference}} \\\cline{2-8}
&AE &MLP& CNN &RNN &Pre-f. &Conc. &Proj. & \\
\specialrule{.05em}{.05em}{.05em}
\specialrule{.05em}{.05em}{.05em}
ENR &$$\checkmark$$& &&&$$\checkmark$$&&&\citeSub{okura2017embedding}\\
HDS &$$\checkmark$$&&&&&&$$\checkmark$$&\citeSub{dong2017hybrid}\\
Wide\&Deep &&$$\checkmark$$&&&&$$\checkmark$$&&\citeSub{cheng2016wide}\\
DNN &&$$\checkmark$$&&&&$$\checkmark$$&&\citeSub{covington2016deep} \\
NPR &&$$\checkmark$$&&&&&$$\checkmark$$&\citeSub{niu2018neural}\\
CDL-Image &&&$$\checkmark$$&&&$$\checkmark$$&&\citeSub{lei2016comparative}\\
IARN &&&&$$\checkmark$$&&&$$\checkmark$$&\citeSub{pei2017interacting}\\
\specialrule{.15em}{.05em}{.05em}
\end{tabular}
\end{table}
\medskip\noindent\underline{Summary of DLMs+FFs.}
The flat features are generally incorporated into various DLMs in three different ways, as summarized in Table \ref{tab:dl+ff}: (1) \textit{pre-filtering} is the simplest way. For instance, ENR (\citeSub{okura2017embedding}) adopts item categories to pre-select the positive (within the same or similar categories) and negative (across different categories) articles; (2) \textit{concatenation} is the most straightforward way. For instance,  Wide\&Deep (\citeSub{cheng2016wide}), DNN (\citeSub{covington2016deep}), and CDL-Image (\citeSub{lei2016comparative}) directly concatenate all feature vectors together and feed them into the proposed network architecture. 
This is quite similar to what SVDFeature does in LFMs which directly adds feature representations to corresponding user and item representations. 
Despite their success on lifting accuracy, the flat features are mostly exploited in a coarse fashion by mere concatenation; and (3) \textit{projection} is the most fine-grained way in comparison with the above two ways. For example, HDS (\citeSub{dong2017hybrid}) and NPR (\citeSub{niu2018neural}) employ neural networks to learn deep user or item representations under different features, that is, the user and item contextual representations. By doing so, the special information from different flat features is taken into consideration. 
In other words, more elaborate design on feature fusion, together with the natural superiority of DLMs, may bring extra performance increments on recommendation.

\subsection{Deep learning models with network features (DLMs+NFs)}

In the proposed CNN based pair-wise image recommendation model (CDL-Image) (\citeSub{lei2016comparative}), social networks are utilized to help exclude negative images for each user. In particular, they assign the images to be negative, which do not have tags indicating the interests of the user and her friends. The social network is merely used to do pre-filter. 
\citeSubject{zhang2017next} proposed a MLP-based method, named NEXT for next POI recommendation. Similar to SVDFeature (\citeSub{chen2012svdfeature}), the embeddings of user and item are influenced by those of corresponding meta-data (i.e., social friends, item descriptions). Based on this, the user and item embeddings are further modeled by one layer of feed-forward neural network supercharged by ReLU (\citeSub{nair2010rectified}) for generating recommendations. The exploitation of social networks in this approach is exactly the same as that in LFMs. 

Later, \citeSubject{ding2017baydnn} designed a CNN-based method (BayDNN) for friend recommendation in social networks. It first exploits CNN to extract latent deep structural feature representations by regarding the input network data as an image and then adopts the Bayesian ranking to make friend recommendations.
This is fairly the first work that proposes an elegant and unified deep learning approach with social networks. 
After that, \citeSubject{fan2018deep} proposed a rating prediction model (DeepSoR), which first uses Node2Vec (\citeSub{grover2016node2vec}) to learn the embeddings of all users in the social network, and then takes the averaged embeddings associated with the $k$ most similar neighbours for each user, as the input of a MLP architecture, so as to learn non-linear features of each user from the social relations. Finally, the learned user features are integrated into the probabilistic matrix factorization for rating prediction.
Base on this, they further proposed a novel attention neural network (GraphRec) (\citeSub{fan2019graph}) to jointly model the user-item interactions and user social relations. 
First, the user embedding is learned via aggregating both (1) the user's opinions towards interacted items with different attention scores and (2) the influence of her social friends with different attention scores. 
Analogously, the item embedding is learned via attentively aggregating the opinions of users who have interacted with the item. 
All attention scores are automatically learned through a two-layer attention neural network. 
Finally, the user and item embeddings are concatenated and fed into a MLP for rating prediction. 

\medskip\noindent\underline{Summary of DLMs+NFs.}
First, similar as DLMs+FFs, the usage of NFs in DLMs evolves from simple pre-filtering, to moderate summation via SVDFeature, and finally to deep projection. This further helps support our point that the exploitation of side information in LFMs could provide better clues and guidance for DLMs.  
Second, extensive experimental results have demonstrated that DLMs+NFs consistently defeat LFMs+NFs in terms of their recommendation accuracy. 
However, we also notice that the studies on fusing NFs into DLMs are quite fewer than those on LFMs+NFs. On the other hand, there is still a great demand on social friend recommendations with the rapid development of social network platforms, such as Facebook, Twitter and so forth, which calls for more in-depth investigation on DLMs+NFs. Third, as the NFs can be considered as a graph, other graph related DLMs can also be adopted, including \textit{graph convolutional networks} (GCN)
(\citeSub{duvenaud2015convolutional}; 
\citeSub{niepert2016learning}) and \textit{graph neural network} (GNN) (\citeSub{scarselli2008graph}). They have attracted considerable and increasing attention due to their superior learning capability on the graph-structured data.
Last, the idea of leveraging distrust information and trust propagation in LFMs+NFs is still worthy of exploration in DLMs+NFs.  

\subsection{Deep learning models with feature hierarchies (DLMs+FHs)} 
\citeSubject{tuan20173d} proposed a 3D-CNN framework that combines session clicks and content features (i.e., item descriptions and category hierarchies) into a 3-dimensional CNN with character-level encoding of all the input data. To utilize the information encoded in the hierarchy, they concatenated the current category with all its ancestors up to the root and use the resulting sequence of characters as the category feature, for example, ``apple/iphone/iphone7/accessories".
However, this method cannot distinguish the different impacts of categories at different layers of the hierarchy. 
In IARN (\citeSub{pei2017interacting}), the feature encoder is used to fuse a set of categories $K$ that item $i$ belongs to, where a category $k$ is modeled as a transformation function $\mathbf{M}_i^k$ that projects the item embedding $\mathbf{e}_i$ into a new space. In terms of the hierarchical categories, it considers the recursive parent-children relations between categories from the root to leaf layer, that is, $\prod_{k=1}^{L}\mathbf{M}_i^k \cdot \mathbf{e}_i$.
With the recursive projection, the item can be gradually mapped into a more general feature space. 

\medskip\noindent\underline{Summary of DLMs+FHs.}
DLMs+FHs is also inadequately studied compared with
either LFMs+FHs or DLMs with other side information (e.g., FFs, KGs, TFs, IFs), although their effectiveness for high-quality recommendations has been empirically verified (\citeSub{tuan20173d}; \citeSub{pei2017interacting}). 
We argue that the advantages of investigation on DLMs+FHs lie in three aspects: (1) FHs are more easily and cost-effectively obtained in real-world applications (e.g., Amazon, Tmall) compared with other complex side information, such as knowledge graphs, textual reviews and visual images; (2) FHs provide human- and machine-readable descriptions about a set of features and their relations (e.g., affinity, alternation, and complement). The rich information encoded in FHs could enable more accurate and diverse recommendations; (3) the volumes of FHs are generally much smaller than other complicated side information, such as KGs, text reviews and visual images. Thus, the integration of FHs will deliver a flexible DLM with less computational cost and high efficiency.
In addition, the categories at different layers of the hierarchy have different impacts on characterizing user-item interactions, as already proven by LFMs+FHs. 
To achieve a better exploitation on FHs for more effective recommendation results, such different impacts may be learned by deep learning based advances (e.g., attention mechanism) in an automatic fashion. 

\subsection{Deep learning models with knowledge graphs (DLMs+KGs)}
According to how KGs are exploited, three types of approaches under DLMs+KGs are included: graph embedding based methods, path embedding based methods and propagation based methods.

\medskip\noindent\textbf{Graph embedding based methods.} 
Many KG-aware recommendation approaches directly make use of conventional graph embedding methods, such as TransE (\citeSub{bordes2013translating}),
TransR (\citeSub{lin2015learning}), TransH (\citeSub{yang2014embedding}) and TransD (\citeSub{ji2015knowledge}), to learn the embeddings for the KG. The learned embeddings are then incorporated into recommendation models.   
For instance, \citeSubject{zhang2016collaborative} proposed a \textit{collaborative knowledge graph embedding} method (CKE) that leverages TransR to learn better item representations.
This is jointly trained with the item visual and textual representation learning models in a unified Bayesian framework. 
\citeSubject{wang2018dkn} proposed DKN for news recommendation. It exploits CNN to generate embedding for a news $i$ based on its title $t_i$, where a sub-KG related to $t_i$ is extracted and learned based on conventional graph embedding methods like TransD. Then the learned embeddings of entities and corresponding context in the sub-KG, as well as the embeddings of words in $t_i$, are treated as different channels and stacked together as the input of a CNN to learn the news embeddings. 

Recently, \citeSubject{cao2019unifying} proposed KTUP to jointly learn the recommendation model and knowledge graph completion. 
Inspired by TransH, they came up with a new translation-based recommendation model (TUP) that automatically induces a preference for a user-item pair, and learns the embeddings of preference $\mathbf{p}$, user $\mathbf{u}$ and item $\mathbf{i}$, satisfying $\mathbf{u} + \mathbf{p} \approx \mathbf{i}$. 
KTUP further extends TUP to jointly optimize TUP (for the item recommendation) and TransH (for the KG completion) to enhance the item and preference modeling by transferring knowledge of entities and relations from the KG, respectively.
Similarly, \citeSubject{xin2019relational} proposed RCF to generate recommendations based on both the user-item interaction history and the item relational data in the KG. They developed a two-level hierarchical attention mechanism to model user preference: the first-level attention discriminates which types of relations are more important; and the second-level attention considers the specific relation values to estimate the contribution of an interacted item. Finally, they jointly modeled the user preference via the hierarchical attention mechanism and item relations via the KG embedding method (i.e., DistMult) (\citeSub{yang2014embedding}). 

Other methods adopt deep learning advances to learn the embeddings for the KG. For example, \citeSubject{wang2019multi} developed a multi-task learning approach, MKR, which utilizes the KG embedding task as the explicit constraint term to provide regularization for the recommendation task.
The recommendation module takes a user and an item as input, and uses MLP and cross\&compress units to output the predicted scores; the KG embedding module also uses MLP to extract features from the head $h$ and relation $r$ of a knowledge triple $\langle h, r, t \rangle$, and outputs the representation of the predicted tail $t$; and the two modules are bridged by the cross\&compress units, which automatically share the latent features and learn the high-order interactions between the items in recommender systems and the entities in the KG.

\medskip\noindent\textbf{Path embedding based methods.}
Typically, path embedding based methods extract connected paths with different semantics between the user-item pairs, and then encode these paths via DLMs.
For instance, \citeSubject{sun2018rkge} proposed a \textit{recurrent knowledge graph embedding} method (RKGE). It first extracts connected paths between a user-item pair in the KG, representing various semantic relations between the user and the item. Then these extracted paths are encoded via a batch of RNNs to learn the different path influences on characterizing the user-item interactions. After that, a pooling layer is incorporated to help distinguish the different saliency of these paths on modeling the user-item interaction. Finally, a fully-connected layer is employed to estimate the user's preference for each item.
Following the same idea, \citeSubject{wang2018explainable} later designed another KG-aware recommender, KPRN, by further taking into account different entity types and relations when encoding the extracted paths via RNNs. 

\medskip\noindent\textbf{Propagation based methods.}
These KG-aware recommendation approaches take advantage of both graph embedding based methods and path based methods. Instead of directly extracting paths in a KG, \textit{propagation} is adopted to discover high-order interactions between items in recommender systems and entities in the KG, which is equivalent to the automatic path mining. 

\citeauthor{wang2018ripplenet} (\citeyear{wang2018ripplenet,wang2019exploring,wang2019knowledgegraph,wang2019knowledge}) designed a series of KG-aware recommendation methods based on the idea of \textit{propagation}. 
Specifically, \citeauthor{wang2018ripplenet} (\citeyear{wang2018ripplenet,wang2019exploring}) developed RippleNet which naturally incorporates graph embedding based methods (i.e., three-way tensor factorization) by preference propagation. In particular, it treats user $u$'s interacted items as a set of seeds in the KG, and extends iteratively along the KG links to discover her $l$-order interests ($1\leqslant l \leqslant 3$). Based on this, it learns user $u$'s $l$-order preference with respect to item $v$, respectively, which are then accumulated together as user $u$'s hierarchical preference to item $v$.
Later, they further proposed KGCN (\citeSub{wang2019knowledgegraph}) based on \textit{graph convolutional networks} (GCN) (\citeSub{niepert2016learning}; \citeSub{duvenaud2015convolutional}), which outperforms RippleNet.  
It computes the user-specific item embeddings by first applying a trainable function that identifies important KG relations for a given user and then transforming the KG into a user-specific weighted graph. It further applies GCN to compute the embedding of an item by propagating and aggregating neighborhood information in the KG. 
After that, to provide better inductive bias, they upgraded KGCN to KGCN-LS (\citeSub{wang2019knowledge}) by using \textit{label smoothness} (LS), which provides regularization over edge weights and has proven to be equivalent to label propagation scheme on a graph. 

Recently, \citeSubject{wang2019kgat} designed KGAT, which recursively and attentively propagates the embeddings from a node to its neighbors in the KG to refine the node's embedding. Specifically, it first uses TransR (\citeSub{lin2015learning}) to learn the KG embeddings, and then employs the \textit{graph convolution network} (GCN) (\citeSub{duvenaud2015convolutional}; \citeSub{niepert2016learning}) to recursively propagate the embeddings along the high-order connectivity, and meanwhile generates attentive weights to reveal the saliency of such connectivity via graph attention network (\citeSub{velivckovic2017graph}).

\begin{table}[t]
\footnotesize
\addtolength{\tabcolsep}{-1.2mm}
\renewcommand{\arraystretch}{0.7}
\centering
\caption{Classifications of DLMs+KGs, where `AE' denotes auto-encoder; `Attn' means attention; `KGE' refers to knowledge graph embedding; and `Prop.' indicates propagation. }\label{tab:dl+kg}
\vspace{-0.1in}
\begin{tabular}{l|ccccc|ccc|l}
\specialrule{.15em}{.05em}{.05em}
\multirow{2}{*}{Method} &\multicolumn{5}{c|}{\textbf{Model Type}}&\multicolumn{3}{c|}{\textbf{KG Usage Type}}&\multirow{2}{*}{Reference} \\\cline{2-9}
&AE &MLP& CNN &RNN &Attn &KGE &Path &Prop. & \\
\specialrule{.05em}{.05em}{.05em}
\specialrule{.05em}{.05em}{.05em}
CKE &$$\checkmark$$&&&&&$$\checkmark$$&&&\citeSub{zhang2016collaborative}\\
KTUP &&&&&&$$\checkmark$$&&&\citeSub{cao2019unifying}\\
RCF &&$$\checkmark$$&&&$$\checkmark$$&$$\checkmark$$&&&\citeSub{xin2019relational}\\
DKN &&&$$\checkmark$$&&$$\checkmark$$&$$\checkmark$$&&&\citeSub{covington2016deep} \\
MKR &&$$\checkmark$$&&&&$$\checkmark$$&&&\citeSub{wang2019multi}\\
RKGE &&&&$$\checkmark$$&&&$$\checkmark$$&&\citeSub{sun2018rkge}\\
KPRN &&&&$$\checkmark$$&&&$$\checkmark$$&&\citeSub{wang2018explainable}\\
KGAT &&&$$\checkmark$$&&$$\checkmark$$&&&$$\checkmark$$&\citeSub{wang2019kgat}\\
RippleNet &&&&&$$\checkmark$$&&&$$\checkmark$$&\citeSub{wang2018ripplenet}\\
KGCN &&&$$\checkmark$$&&&&&$$\checkmark$$&
\citeauthor{wang2019knowledgegraph}
(\citeyear{wang2019knowledgegraph,wang2019knowledge})\\
\specialrule{.15em}{.05em}{.05em}
\end{tabular}
\end{table}

\medskip\noindent\underline{Summary of DLMs+KGs.}
Table \ref{tab:dl+kg} summarizes the classifications of DLMs+KGs in terms of fundamental model types and KG usage types. We further summarize the studies from the following perspectives.
First, DLMs+KGs show a great superiority compared to LFMs+KGs in terms of recommendation accuracy. However, the high computational cost limits the scalability of DLMs+KGs on large-scale datasets.
Hence, a promising direction for boosting DLMs+KGs approaches should focus on improving their scalability, and thus reduce time complexity. 
Second, empirical studies have demonstrated the strength of propagation (i.e., hybrid) based methods against those either exploiting graph embedding or path embedding only. 
Third, regardless of the KG usage types, most of these methods rely on conventional KG embedding methods such as TransE/R/H/D to incorporate KG for better recommendations. In particular, they learn embeddings for the KG based on the triple $\langle h, r, t\rangle$, where $h, t$ separately represent head and tail entities, and $r$ denotes entity relations. In contrast, several methods attempt to employ deep learning advances, for example, \textit{graph convolution network} (GCN) is developed to further boost the quality of recommendations. Thus, more further efforts should be devoted to this topic. 
Lastly, a better exploitation of the heterogeneity of KGs will facilitate more accurate recommendation results. For instance, by additionally distinguishing entity types and relation types, KPRN (\citeSub{wang2018explainable}) performs better than RKGE (\citeSub{sun2018rkge}), and by attentively identifying the saliency of different relation types, KGCN (\citeSub{wang2019knowledge}) outperforms RippleNet (\citeSub{wang2018ripplenet}).

\subsection{Deep learning models with text features (DLMs+TFs)}
There are a number of studies that incorporate textual features (e.g., reviews, tips, and item descriptions) into DLMs for better recommendations.

\medskip\noindent\textbf{Auto-Encoder based methods.}
\citeSubject{wang2015collaborative} proposed CDL based on the \textit{stacked denoising autoencoder} (SDAE)
(\citeSub{vincent2010stacked}) to learn the item representation with item content (i.e., paper abstract and movie plots) as input.
With the learned item representations by SDAE as bridge, CDL simultaneously minimizes the rating estimation error via matrix factorization and SDAE reconstruction error.
In CKE (\citeSub{zhang2016collaborative}), SDAE is used to extract item textual representations from textual knowledge (i.e., movie and book summaries). This is jointly trained with the matrix factorization and visual representation learning models.  In addition, 
ENR (\citeSub{okura2017embedding}) adopts a variant of a denoising autoencoder (\citeSub{vincent2008extracting}) to learn the distributed representations of articles for news recommendation.
\citeSubject{yin2017spatial} proposed SH-CDL which uses \textit{deep belief network} (DBN), an auto-encoder model (\citeSub{hinton2006fast}), to learn the item hidden representations $\mathbf{f}_i$ by feeding the textual content (i.e., categories, descriptions and comments). 
It unifies matrix factorization and 
DBN by linking the item latent vector $\mathbf{q}_i$ and its hidden representation $\mathbf{f}_i$ under the assumption that $\mathbf{q}_i$ follows a normal distribution with the mean $\mathbf{f}_i$. 

\medskip\noindent\textbf{MLP based methods.}
In the video recommendation method (DNN) (\citeSub{davidson2010youtube}), 
a user's watch/impression history, that is, a set of IDs associated with the videos that a user has visited before, and side information (e.g., the user's tokenized queries) are concatenated into a wide layer, followed by several layers of fully connected ReLU (\citeSub{nair2010rectified}) to rank the recommendation candidates.
In NEXT (\citeSub{zhang2017next}), the embedding of an item is influenced by that of its description (split into multiple words), and modeled via one fully-connected feed-forward layer supercharged by ReLU (\citeSub{nair2010rectified}).
\citeSubject{zhang2017joint} developed a joint representation learning approach (JRL), which jointly learns user and item representations from three types of sources: (1) textual reviews via PV-DBOW (\citeSub{le2014distributed}); (2) visual images via CNN; and (3) numerical ratings with a two-layer fully-connected neural network. 
The learned user and item representations from different sources are respectively concatenated together for pair-wise item ranking. 

\medskip\noindent\textbf{CNN based methods.}
The 3D-CNN framework (\citeSub{tuan20173d}) combines session clicks and content features (i.e., item descriptions and category hierarchies) into a 3-dimensional CNN with character-level encoding of all input data.
\citeSubject{zheng2017joint} designed DeepCoNN to learn both user and item representations in a joint manner using textual reviews. Parallel and identical user and item networks are involved: in the first layer, all reviews written by a user (or written for an item) are represented as matrices of word embeddings to capture the semantic information. The next layer employs CNN to extract the textual features for learning user and item representations. Finally, a shared layer is introduced on the top to couple these two networks together, which enables the learned user and item representations to interact with each other in a similar manner as factorization machine 
(\citeauthor{rendle2010factorization} \citeyear{rendle2010factorization,rendle2012factorization}). 

\citeSubject{catherine2017transnets} further extended DeepCoNN by proposing TransNets to help address the issue that a pair-wise review for the target user to the target item may not be available in testing procedure. 
It thus uses an additional transform layer to transform the latent representations of user and item into those of their pair-wise review. In the training, this layer is regularized to be similar to the real latent representation of the pair-wise review learned by the target network.  
Therefore, in the testing, an approximate representation of the pair-wise review can be generated and used for making predictions. 

\citeSubject{seo2017interpretable} developed D-Attn to jointly learn better user and item representations using CNN with local (L-Attn) and global (G-Attn) attentions. 
It uses the embeddings of words in the review as input, and adopts both L-Attn and G-Attn to learn the saliency of words with respect to a local window and the entire input texts.  These are then fed into two CNNs to learn the respective L-Attn and G-Attn representations of a user (an item), followed by a concatenate layer to get the final user (item) representation. Finally, the inner product of user and item representations is used to estimate a user's preference to an item.

\medskip\noindent\textbf{RNN based methods.}
\citeSubject{yao2017serm} proposed SERM for next POI recommendation, which jointly learns the embeddings of multiple features (user, location, time, keyword extracted from text messages) and the transition parameters of the RNN, to capture the spatial-temporal regularities, the activity semantics, as well as the user preferences in a unified way.
The embedding layer transforms all features into low-dimensional dense representations, and then concatenates them into a unified representation as the input of RNN module.
\citeSubject{wu2016joint} presented a joint \textit{review-rating recurrent recommender network} (RRN-Text). The rating model uses two LSTMs to capture the temporal dynamics of the user and movie representations, which are further combined with stationary user and item representations. Review texts are modeled by a character-level LSTM, and the input character embeddings are fused with both dynamic and stationary user and item representations in the rating model by the bottleneck layer.

\citeSubject{lu2018like} designed a multi-task learning model, MT,  for explainable recommendation. It extends \textit{matrix factorization} (MF) model by using the textual features extracted from reviews to serve as the regularizers for the user and item representations. In particular, the embedding of each word in the relevant review is sequentially fed into a bidirectional GRU (\citeSub{cho2014properties}) to learn the textual features for the user (item).   
Similarly, \citeSubject{li2017neural} introduced NRT that simultaneously predicts ratings and generates abstractive tips. It consists of two modules: (1) the neural rating regression module takes the user and item representations as input, and utilizes MLP to predict ratings; and (2) the tips generation module adopts GRU to generate concise tips. Its  hidden state is initialized by the user and item representations, and the vectorization of the predicted ratings, as well as the hidden variables from the review text.

\medskip\noindent\textbf{Attention based methods.}
D-Attn (\citeSub{seo2017interpretable}) uses two attention mechanisms to learn the importance of words with respect to a local window and the entire input text for better recommendation performance. Different from the existing DLMs with text features, which either adopt traditional network structures (e.g., TransNets (\citeSub{catherine2017transnets}) and DeepCoNN (\citeSub{zheng2017joint})), or utilize vanilla attention mechanisms to boost these structures (e.g., D-Attn (\citeSub{seo2017interpretable})), 
\citeSubject{tay2018multi} proposed a new network architecture, MPCN, to dynamically distinguish the importance of different reviews, instead of treating them equally.
In this method, each user (or item) is represented as a sequence of reviews, and each review is constructed from a sequence of words. It first leverages the review-level co-attention to select the most informative review pairs from the review bank of each user and item. Then, it adopts word-level co-attention to model the selected review pairs at word-level. Finally, the learned representations for each review pair at different levels are separately concatenated and passed into the factorization machine for rating prediction.

\begin{table}[t]
\centering
\footnotesize
\addtolength{\tabcolsep}{-0.5mm}
\renewcommand{\arraystretch}{0.9}
\caption{Classification of state-of-the-art methods in DLMs+TFs, where `AE' denotes auto-encoder; `Attn' means attention; the methods with early fusion are marked by `*', and the rest are late fusion methods. 
}\label{tab:dl+tf}
\vspace{-0.1in}
\begin{tabular}{l|l}
\specialrule{.15em}{.05em}{.05em}
\textbf{Type} &\textbf{Representative Method} \\
\specialrule{.05em}{.05em}{.05em}
\specialrule{.05em}{.05em}{.05em}
\multirow{2}{*}{\textbf{AE}} & (1) CDL (\citeSub{wang2015collaborative}) (2) CKE (\citeSub{zhang2016collaborative})  \\
& (3) ENR (\citeSub{okura2017embedding})* (4)  SH-CDL (\citeSub{yin2017spatial})\\
\specialrule{.05em}{.05em}{.05em}
\multirow{2}{*}{\textbf{MLP}} & (1) DNN (\citeSub{davidson2010youtube})* (2)  NEXT (\citeSub{zhang2017next})* \\
& (3) JRL (\citeSub{zhang2017joint}) \\
\specialrule{.05em}{.05em}{.05em}
\multirow{2}{*}{\textbf{CNN}} & (1) 3D-CNN (\citeSub{tuan20173d})* (2) DeepCoNN (\citeSub{zheng2017joint})*  \\
& (3) TransNets (\citeSub{catherine2017transnets})* (4) D-Attn (\citeSub{seo2017interpretable})* \\
\specialrule{.05em}{.05em}{.05em}
\multirow{2}{*}{\textbf{RNN}} & (1) SERM (\citeSub{yao2017serm})* (2) RRN-Text (\citeSub{wu2016joint}) \\ & (3) MT (\citeSub{lu2018like}) (4) NRT (\citeSub{li2017neural})  \\
\specialrule{.05em}{.05em}{.05em}
\textbf{Attn} & (1) D-Attn (\citeSub{seo2017interpretable})*; (2) MPCN (\citeSub{tay2018multi})*\\
\specialrule{.15em}{.05em}{.05em}
\end{tabular}
\end{table}

\medskip\noindent\underline{Summary of DLMs+TFs.}
Table \ref{tab:dl+tf} summarizes all the DLMs+TFs methods. First, compared with LFMs+TFs, which heavily depend on external toolkits to extract knowledge from TFs, DLMs based recommendation approaches seamlessly fuses TFs via deep learning advances such as CNN and RNN.
The homogeneity of underlying methodologies facilitates unified and elegant approaches with excellent recommendation accuracy. 
Second, the exploitation of TFs by DLMs+TFs is in a deeper and more fine-grained fashion. Recall that most LFMs+TFs leverage TFs in word-, aspect- and sentiment-level. For instance, they either simply apply averaged words embeddings to represent the text, or utilize the results of aspect extraction and sentiment analysis on the text to help infer user preferences. The topic-level methods, though, model latent topic distribution of text with fine granularity via conventional topic modelling models (e.g., LDA (\citeSub{blei2003latent})), cannot capture the complex relations (e.g., non-linearity) encoded in the text better. In contrast, DLMs+TFs take the text embedding at word-level as input, and feed it into deep learning advances (e.g., SDAE, MLP, CNN, RNN), and extract features of the text via multiple non-linear transformations. Meanwhile, neural attention mechanisms can be adopted to distinguish the saliency of each word for the text, and saliency of each review for users and items, so as to support more accurate recommendations. 
Third, in DLMs+TFs, text features are generally utilized to learn user (item) contextual representations. Based on the stage that text features are integrated, they can be classified into two types: (1) \textit{early fusion} and (2) late fusion. With early fusion, for instance, many methods concatenate the text embedding in word-level with the user (item) embedding, which are together fed into the network. This occurs with DNN (\citeSub{covington2016deep}), NEXT (\citeSub{zhang2017next}), 3D-CNN (\citeSub{tuan20173d}) and SERM (\citeSub{yao2017serm}). Other methods directly use relevant text embeddings of user (item) as input to learn user- and item-textual representations without concatenating user and item embeddings, like ENR (\citeSub{okura2017embedding}), DeepCoNN (\citeSub{zheng2017joint}), TransNets (\citeSub{catherine2017transnets}), and D-Attn (\citeSub{seo2017interpretable}), MPCN (\citeSub{tay2018multi}). Late fusion methods, in contrast, are often composed of two parallel modules, namely text module and rating module. The text module employs deep learning advances (e.g., SDAE, MLP, CNN) to help learn user (item) textual representations with the relevant text embedding as input; and the rating module takes another deep neural architecture to help learn the plain user and item representations with the user-item interaction data as input. The text module is responsible for regularizing the rating module to assist in learning better user and item representations, thus achieving outstanding recommendation results. These two modules are jointly trained and mutually enhanced. Typical approaches include CDL (\citeSub{wang2015collaborative}), CKE (\citeSub{zhang2016collaborative}), SH-CDL (\citeSub{yin2017spatial}) and JRL (\citeSub{zhang2017joint}).

\subsection{Deep learning models with image features (DLMs+IFs)}
In CKE (\citeSub{zhang2016collaborative}), 
the stacked convolutional auto-encoder (SCAE) (\citeSub{masci2011stacked}) is utilized to extract item visual representations from images (i.e., movie poster and book cover). This is jointly trained with the matrix factorization and textural representation learning models to achieve high-quality recommendations.   
In CDL-Image (\citeSub{lei2016comparative}), CNN is adopted to extract the high-level features and learn representations of images. The learned image representation together with the user representation learned via four fully-connected layers are fed into a distance calculation net to estimate the user's preference for each image.  
Another image recommender, NPR (\citeSub{niu2018neural}) utilizes the image visual features learned via CNN for better recommendations. After dimension reduction, the image visual representations are fed into a fully-connected layer to learn the representation of each user's contextual preference.
In JRL (\citeSub{zhang2017joint}), it jointly learns the user and item representations from three types of sources, namely reviews, images and ratings. It also utilizes a fully-connected layer to learn the image representation which is guided by the raw image features obtained via CNN.   

\citeSubject{alashkar2017examples} proposed a deep neural network for makeup recommendation with homogeneous style (Exp-Rul). The facial traits are classified automatically and coded as feature vectors, which are then fed into MLP to generate recommendations for each makeup element. The network is trained by examples and guided by rules. In particular, for an automatic analysis on the facial traits, $83$ facial landmarks are detected on $900$ facial images using the face++ framework (\url{www.faceplusplus.com}), and different regions of interest are extracted for different facial attributes. 
Also, \citeSubject{chen2017attentive} introduced ACF, a neural network consisting of two attention modules: (1) the component-level attention module learns the user's preference to the selected informative components inside each item, that is, the regions of image and frame of video; and (2) the item-level attention module helps learn the user's preference for the entire item by incorporating the learned component-level attentions with weighted combination.
They make use of the widely-used architecture, ResNet-152 (\citeSub{he2016deep}), to extract visual features from both the regions of images and frames of videos. 
The idea is quite similar to D-Attn (\citeSub{seo2017interpretable}) that learns user and item representations from the perspectives of local and global attentions towards the relevant reviews.

\medskip\noindent\underline{Summary of DLMs+IFs.}
Image features play a crucial role for recommendation tasks in domains such as fashion, restaurants and hotels, as well as image related platforms, such as Flickr and Instagram. They can be used to improve the attractiveness of the recommended items in addition to accuracy. Due to the superior capability of capturing local features, DLMs+IFs is mainly dominated by CNN structures through the use of CNNs to extract visual features from images to generate user (item) visual representations. These visual features are then fed into recommendation frameworks, so as to help regularize and learn high-quality user (item) representations. The major difference between LFMs+IFs and DLMs+IFs is the recommendation framework. That is, LFMs+IFs simply feeds the extracted visual features into linear latent factor models, whereas DLMs+IFs designs proper deep neural architectures with multiple non-linear hidden layers to better accommodate the visual features, and thus achieve more effective recommendation results.

\subsection{Deep learning models with video features (DLMs+VFs)}
The research studies on fusing videos into DLMs for recommendation tasks are far fewer than research about other types of side information. This may be mainly due to two reasons: (1) the video features are much difficult and time-consuming to be obtained and managed compared with other side information; and (2) the volume of video features is generally quite large, requiring a huge computational cost. 
One representative approach is the ACF (\citeSub{chen2017attentive}) mentioned above.
In particular, it adopts ResNet-152 (\citeSub{he2016deep}) to extract visual features for the frames of videos.
To further simplify the process, it uses the output of pool5 layer in ResNet-152, which is actually the mean pooling of the feature maps, as the feature vector for each frame of videos.

\subsection{Discussion of DLMs with side information}
The extensive and insightful analysis on DLMs with side information in the previous subsections leads to the following conclusions: (1) with deep architectures and non-linear transformations, DLMs have been empirically proven to be extraordinary effective in capturing the highly complex user-item interactions compared with the conventional models including MMs, LFMs and RLMs.
On the other hand, it also should be acknowledged that the performance improvements by deep learning advances are often accompanied with heavy computational cost and much longer training time due to the complexity of deep learning models. To meet this end, expensive computation devices such as powerful graph cards are necessary for effective training and inferring. In other words, conventional models are far more efficient than DLMs regarding time complexity in most cases; 
and (2) due to their high flexibility, DLMs can be easily extended to incorporate various side information. They have shown overwhelming superiority in coping with complex structural data like knowledge graphs and 
non-structural data including text features and image features. Moreover, conventional models normally need to first do feature engineering to fuse the side information and train the model, while DLMs seamlessly combine these two phases in an end-to-end manner. 

We now offer further details on DLMs with side information and provide a summary from the following perspectives.
(1) Regarding {how to integrate} various side information into DLMs, there are mainly three ways: 
(a) \textit{pre-filtering} is the simplest way to leverage side information to do data pre-processing; 
(b) \textit{concatenation} is the most straightforward approach because it directly concatenates all side information together; and (c) \textit{projection} is a more fine-grained way by mapping users (items) into low-dimensional space regarding the side information to learn the related contextual representations. 
(2) In terms of {when to integrate} the side information into DLMs, they can be broadly grouped into two types: (a) {early fusion} combines all available side information with user (item) in the input layer, and then feeds them into the network architecture to extract more high-level and complex features; and (b) {late fusion} uses two parallel modules, with the feature module that aims to learn user (item) contextual representations with respect to the side information, and the rating module that aims to learn plain user and item representations via user-item historical interaction data. The two modules are jointly trained and mutually benefit where the feature module regularizes the rating module, while the rating module in turn guides the feature module.
(3) Different types of side information improve recommendation performance from different aspects and are incorporated in different ways. For the first issue, in addition to accuracy, FHs and KGs can help with diversity, while text features facilitate explainable recommendations and image features may assist in attractiveness of recommendation. 
For the latter issue, simple side information like FFs is generally fused by {concatenation} in the {early fusion}, while the graph-structural data (e.g., NFs and KGs) can be accommodated with graph related deep learning advances (e.g., CNN, GCN, GNN). Similar to image features, text features (i.e., word embedding matrices) can be treated as images also, both thus can be fed into CNNs for feature extraction.

%% file: section/future.tex
\section{Future Directions}
In this Research Commentary, we surveyed recent developments in recommendation with side information. Despite all the progress, there are many challenges to be addressed and plenty of room for improvement. In this section, we identify key challenges and opportunities which we believe can shape future research on this topic. We mainly discuss future directions by considering the following research questions:
\begin{itemize}[leftmargin=0.4cm]
    \item How to further improve deep learning based recommendation with side information in complex structures?
    \item How to obtain high-quality side information to improve recommendation?
    \item For which recommendation techniques can side information play an important role, thus should be taken into account?
    \item In which recommendation scenarios can side information be most valuable?
\end{itemize}

To answer these questions, we discuss the following challenges and future research directions: deep learning with structured information, leveraging crowdsourcing as means to solicit side information for recommendation; side information for specific recommendation techniques such as reinforcement learning and adversarial recommendation; and side information as an important data source for improving recommendation in specific recommendation scenarios such as cross-domain and package recommendation.

\medskip\noindent\textbf{Deep recommenders with structured side information.}
Integrating side information into deep learning based recommendation is currently an active research topic. Existing approaches, as we have discussed, are highly limited in exploiting the full potential of structured side information for recommendation. The challenges mainly arise from two complications: the intrinsic complexity of structured side information and the difficulty in adapting deep learning models for incorporating structured information. 

With knowledge graphs as an example, the current deep learning based recommendation approaches are limited to the usage of the most basic information in a knowledge graph, for example, paths (\citeSub{sun2018rkge}) or meta-paths (\citeSub{hu2018leveraging}). There is room for improvement by considering higher-level information such as meta-graphs, that is, a collection of linked meta-paths; and hyper-graphs, that is, an abstract view to knowledge graphs that considers entities of the same type as a hyper-node and the connections between hyper-nodes as hyper-edges. Recent work can be found on the hyper-graph based approach for recommendation in  \cite{dingqi2019revisiting}, though the authors only  considered random walk in the hyper-graph for location recommendation in LBSNs. More research is needed to take advantage of meta-graphs or hyper-graphs for deep learning based recommendation.

We observe that the majority of existing methods process structured information into a data format that is consumable by common neural network architectures (e.g., convolutional or recurrent). An alternative approach to using structured side information in deep learning based recommendation is adapting deep neural networks such that they can directly model structured information. Seminal work has been carried out by \citeauthor{wang2019knowledge} (\citeyear{wang2019knowledgegraph,wang2019knowledge}), in which graph convolutional networks (\citeSub{kipf2016semi}) have been used to incorporate knowledge graphs for recommendation. We note that graph convolutional networks are a special class of graph networks (\citeSub{battaglia2018relational}), which are designed to process structured information as an intrinsic capability. The investigation of graph networks is an active ongoing research topic in the machine learning community and we expect that developments on this topic can nurture future advances in deep learning based recommendation with structured side information. 

\medskip\noindent\textbf{Crowdsourcing side information for recommendation.}
Crowdsourcing provides an efficient and cost-effective mean for data collection and augmentation. User feedback that is used as the main input to various recommendation methods, can be viewed as the result of crowdsourced feedback collection where the crowd is the large amount of users in the recommender system. From this perspective, the main forms of user feedback that have been considered in recommender systems are rather restricted: we either consider explicit feedback such as ratings or implicit feedback such as clicks, views, or check-ins. The development of recommendation techniques that incorporate various side information opens up new research directions to leverage crowdsourcing for collecting much more types of data as side information for improving recommendation. In this sense, crowdsourcing has the potential to become an integral component of recommender systems enhanced by side information. 

Existing work on the intersection of recommendation and crowdsourcing mainly studies recommendation within crowdsourcing platforms or leverages existing crowdsourced data on the Web for recommendation. For example, \citeSubject{leal2019scalable} studied the problem of recommending wiki pages -- a popular example of crowdsourced knowledge repository -- with different publisher profiling strategies.  \cite{veloso2019online} proposed a stream recommendation engine that leverages crowdsourced information for hotel recommendation. They specifically considered crowdsourced data streams contributed by tourists in tourism information sharing platforms such as TripAdvisor, Expedia or Booking.com. In contrast, much less work has been carried out to actively involve users to provide the data that is most beneficial for boosting recommendation. We note that this topic is related to research on interactive, controllable recommender systems (\citeSub{zhao2013interactive}), where users are explicitly asked to express their preferences. Research on this topic, so far, has been mostly theoretical and limited to small scale user studies. It remains key research questions which type of side information is most suitable to be crowdsourced and how to best leverage crowdsourcing techniques to improve recommendation. To bring forward research in this direction, existing literature in crowdsourcing can be inspirational, for example, (deep) active learning from crowds (\citeSub{yang2018leveraging}; \citeSub{ostapuk2019activelink}) and gamification
(\citeSub{morschheuser2016gamification}).

\medskip\noindent\textbf{Side information for reinforcement \& adversarial recommendation.}
Reinforcement learning is an effective approach to quickly identify items for recomendation in a dynamic Web-based environment where new items are continuously generated
(\citeSub{li2010contextual}; 
\citeSub{li2011unbiased}), for example, news in a news recommender system. The basic idea is balancing between exploration and exploitation, that is, recommending most relevant items to users to maximize user satisfaction, while collecting user feedback on less relevant items so as to improve recommendation performance in the long run. A major challenge in reinforcement learning based recommendation is the large space of possible actions to choose from, namely, which items to present to the users. Being able to leverage similarities between items for knowledge transfer is, therefore, of key importance to reduce the action space. In this respect, side information can play a critical role for enhancing reinforcement learning based recommenders. We note that the importance of content features of items have been widely recognized in this context. Existing research, however, has not tapped into the rich structure of side information. 

Adversarial recommendation is a more recent recommendation technique, where the goal is to improve recommendation performance by leveraging adversarial examples, either by directly sampling from the item pool (\citeSub{wang2017irgan}) or by performing perturbations on the embedding parameters of items (\citeSub{he2018adversarial}). In this context, side information has the potential to help better sample the adversarial examples or choose items for parameter perturbations. Research on this topic is still at the infant stage and there are plenty of gaps to be filled.

\medskip\noindent\textbf{Side information for cross-domain \& package recommendation.}
Here, we discuss two recommendation scenarios where side information can play an important role to enhance recommendation performance, namely, cross-domain recommendation and package recommendation. Cross-domain recommendation addresses the problem of leveraging data in different domains to generate recommendations (\citeSub{berkovsky2007cross}; \citeSub{fernandez2012cross}). Assuming that there is certain overlap of information between a source and a target domain, the main idea underlying this class of recommendation methods is to transfer knowledge from the source domain to the target domain, thus addressing data sparsity or cold start problems in the target domain. In this context, side information that describes the common type information of items in the two domains can be highly valuable for knowledge transfer, for example, a topic taxonomy for both books and movies. Besides, side information about users that is indicative of user preferences in different domains can also bridge the source and the target domains, thus useful to improve cross-domain recommendation. 

Package recommendation is relevant for scenarios where a package of items is necessary to be recommended (\citeSub{adomavicius2011multi}; \citeSub{wibowo2017matrix}), for example, a list of POIs for tourism or a basket of products. For this kind of recommendation scenarios, it is important to take into account the relationships between items in the recommended package, for example, the geographical proximity of POIs in the recommended POI list or the complementary relationships among products in the recommended basket. Existing research on individual item recommendation has shown that structured side information can be highly beneficial for identifying relationships among items (\citeSub{yang2016learning}; \citeSub{sun2017exploiting}), in improving recommendation performance and in providing explanations. Uncovering item relationships encoded in structured side information specifically for package recommendation is, however, an underdeveloped topic, which calls for more attention from the research community.

%% file: section/conclusion.tex
\section{Conclusion}
This Research Commentary surveyed a considerable amount of state-of-the-art recommendation algorithms with the incorporation of side information from two orthogonal angles: (1) different fundamental methodologies of recommendation, including memory-based methods, latent factor, representation learning and deep learning models; and (2) different representations of side information including structural data (flat features, network features, hierarchical features and knowledge graphs) and non-structural data (text features, image features and video features).
In addition, we further discussed the challenges and provided new potential directions in recommendation with side information. By doing so, a comprehensive and systematic survey was delivered to benefit both researchers and practitioners in the area of recommender systems.

%% file: main.bbl

\begin{thebibliography}{242}


\ifx \showCODEN    \undefined \def \showCODEN     #1{\unskip}     \fi
\ifx \showDOI      \undefined \def \showDOI       #1{#1}\fi
\ifx \showISBNx    \undefined \def \showISBNx     #1{\unskip}     \fi
\ifx \showISBNxiii \undefined \def \showISBNxiii  #1{\unskip}     \fi
\ifx \showISSN     \undefined \def \showISSN      #1{\unskip}     \fi
\ifx \showLCCN     \undefined \def \showLCCN      #1{\unskip}     \fi
\ifx \shownote     \undefined \def \shownote      #1{#1}          \fi
\ifx \showarticletitle \undefined \def \showarticletitle #1{#1}   \fi
\ifx \showURL      \undefined \def \showURL       {\relax}        \fi
\providecommand\bibfield[2]{#2}
\providecommand\bibinfo[2]{#2}
\providecommand\natexlab[1]{#1}
\providecommand\showeprint[2][]{arXiv:#2}

\bibitem[\protect\citeauthoryear{Adomavicius, Manouselis, and Kwon}{Adomavicius
  et~al\mbox{.}}{2011}]%
        {adomavicius2011multi}
\bibfield{author}{\bibinfo{person}{Gediminas Adomavicius},
  \bibinfo{person}{Nikos Manouselis}, {and} \bibinfo{person}{YoungOk Kwon}.}
  \bibinfo{year}{2011}\natexlab{}.
\newblock \showarticletitle{Multi-criteria recommender systems}.
\newblock In \bibinfo{booktitle}{\emph{Recommender Systems Handbook}}.
  \bibinfo{publisher}{Boston, MA: Springer}, \bibinfo{pages}{769--803}.
\newblock


\bibitem[\protect\citeauthoryear{Adomavicius and Tuzhilin}{Adomavicius and
  Tuzhilin}{2005}]%
        {adomavicius2005toward}
\bibfield{author}{\bibinfo{person}{Gediminas Adomavicius} {and}
  \bibinfo{person}{Alexander Tuzhilin}.} \bibinfo{year}{2005}\natexlab{}.
\newblock \showarticletitle{Toward the next generation of recommender systems:
  A survey of the state-of-the-art and possible extensions}.
\newblock \bibinfo{journal}{\emph{IEEE Transactions on Knowledge and Data
  Engineering}}  \bibinfo{volume}{6} (\bibinfo{year}{2005}),
  \bibinfo{pages}{734--749}.
\newblock


\bibitem[\protect\citeauthoryear{Alashkar, Jiang, Wang, and Fu}{Alashkar
  et~al\mbox{.}}{2017}]%
        {alashkar2017examples}
\bibfield{author}{\bibinfo{person}{Taleb Alashkar}, \bibinfo{person}{Songyao
  Jiang}, \bibinfo{person}{Shuyang Wang}, {and} \bibinfo{person}{Yun Fu}.}
  \bibinfo{year}{2017}\natexlab{}.
\newblock \showarticletitle{Examples-Rules Guided Deep Neural Network for
  Makeup Recommendation}. In \bibinfo{booktitle}{\emph{Proceedings of the 31st
  AAAI Conference on Artificial Intelligence}}. Menlo Park, CA: AAAI Press,
  \bibinfo{pages}{941--947}.
\newblock


\bibitem[\protect\citeauthoryear{Bao, Zheng, and Mokbel}{Bao
  et~al\mbox{.}}{2012}]%
        {bao2012location}
\bibfield{author}{\bibinfo{person}{Jie Bao}, \bibinfo{person}{Yu Zheng}, {and}
  \bibinfo{person}{Mohamed~F Mokbel}.} \bibinfo{year}{2012}\natexlab{}.
\newblock \showarticletitle{Location-based and preference-aware recommendation
  using sparse geo-social networking data}. In
  \bibinfo{booktitle}{\emph{Proceedings of the 20th International Conference on
  Advances in Geographic Information Systems}}. New York: ACM Press,
  \bibinfo{pages}{199--208}.
\newblock


\bibitem[\protect\citeauthoryear{Bao, Fang, and Zhang}{Bao
  et~al\mbox{.}}{2014}]%
        {bao2014topicmf}
\bibfield{author}{\bibinfo{person}{Yang Bao}, \bibinfo{person}{Hui Fang}, {and}
  \bibinfo{person}{Jie Zhang}.} \bibinfo{year}{2014}\natexlab{}.
\newblock \showarticletitle{Topicmf: Simultaneously exploiting ratings and
  reviews for recommendation}. In \bibinfo{booktitle}{\emph{Proceedings of the
  28th AAAI Conference on Artificial Intelligence}}. Menlo Park, CA: AAAI
  Press, \bibinfo{pages}{2--8}.
\newblock


\bibitem[\protect\citeauthoryear{Barkan and Koenigstein}{Barkan and
  Koenigstein}{2016}]%
        {barkan2016item2vec}
\bibfield{author}{\bibinfo{person}{Oren Barkan} {and} \bibinfo{person}{Noam
  Koenigstein}.} \bibinfo{year}{2016}\natexlab{}.
\newblock \showarticletitle{Item2vec: neural item embedding for collaborative
  filtering}. In \bibinfo{booktitle}{\emph{2016 IEEE 26th International
  Workshop on Machine Learning for Signal Processing (MLSP)}}. Washington, DC:
  IEEE Computing Society Press, \bibinfo{pages}{1--6}.
\newblock


\bibitem[\protect\citeauthoryear{Battaglia, Hamrick, Bapst, Sanchez-Gonzalez,
  Zambaldi, Malinowski, Tacchetti, Raposo, Santoro, Faulkner, Gulcehre, Song,
  Ballard, Gilmer, Dahl, Vaswani, Allen, Nash, Langston, Dyer, Heess, Wierstra,
  Kohli, Botvinick, Vinyals, Li, and Pascanu}{Battaglia et~al\mbox{.}}{2018}]%
        {battaglia2018relational}
\bibfield{author}{\bibinfo{person}{Peter~W Battaglia},
  \bibinfo{person}{Jessica~B Hamrick}, \bibinfo{person}{Victor Bapst},
  \bibinfo{person}{Alvaro Sanchez-Gonzalez}, \bibinfo{person}{Vinicius
  Zambaldi}, \bibinfo{person}{Mateusz Malinowski}, \bibinfo{person}{Andrea
  Tacchetti}, \bibinfo{person}{David Raposo}, \bibinfo{person}{Adam Santoro},
  \bibinfo{person}{Ryan Faulkner}, \bibinfo{person}{Caglar Gulcehre},
  \bibinfo{person}{Francis Song}, \bibinfo{person}{Andrew Ballard},
  \bibinfo{person}{Justin Gilmer}, \bibinfo{person}{George Dahl},
  \bibinfo{person}{Ashish Vaswani}, \bibinfo{person}{Kelsey Allen},
  \bibinfo{person}{Charles Nash}, \bibinfo{person}{Victoria Langston},
  \bibinfo{person}{Chris Dyer}, \bibinfo{person}{Nicolas Heess},
  \bibinfo{person}{Daan Wierstra}, \bibinfo{person}{Pushmeet Kohli},
  \bibinfo{person}{Matt Botvinick}, \bibinfo{person}{Oriol Vinyals},
  \bibinfo{person}{Yujia Li}, {and} \bibinfo{person}{Razvan Pascanu}.}
  \bibinfo{year}{2018}\natexlab{}.
\newblock \bibinfo{title}{Relational inductive biases, deep learning, and graph
  networks}.
\newblock \bibinfo{howpublished}{arXiv preprint arXiv:1806.01261}.
\newblock


\bibitem[\protect\citeauthoryear{Bellog{\'\i}n, Cantador, D{\'\i}ez, Castells,
  and Chavarriaga}{Bellog{\'\i}n et~al\mbox{.}}{2013}]%
        {bellogin2013empirical}
\bibfield{author}{\bibinfo{person}{Alejandro Bellog{\'\i}n},
  \bibinfo{person}{Iv{\'a}n Cantador}, \bibinfo{person}{Fernando D{\'\i}ez},
  \bibinfo{person}{Pablo Castells}, {and} \bibinfo{person}{Enrique
  Chavarriaga}.} \bibinfo{year}{2013}\natexlab{}.
\newblock \showarticletitle{An empirical comparison of social, collaborative
  filtering, and hybrid recommenders}.
\newblock \bibinfo{journal}{\emph{ACM Transactions on Intelligent Systems and
  Technology (TIST)}} \bibinfo{volume}{4}, \bibinfo{number}{1}
  (\bibinfo{year}{2013}), \bibinfo{pages}{14}.
\newblock


\bibitem[\protect\citeauthoryear{Bengio, Ducharme, Vincent, and Jauvin}{Bengio
  et~al\mbox{.}}{2003}]%
        {bengio2003neural}
\bibfield{author}{\bibinfo{person}{Yoshua Bengio}, \bibinfo{person}{R{\'e}jean
  Ducharme}, \bibinfo{person}{Pascal Vincent}, {and} \bibinfo{person}{Christian
  Jauvin}.} \bibinfo{year}{2003}\natexlab{}.
\newblock \showarticletitle{A neural probabilistic language model}.
\newblock \bibinfo{journal}{\emph{Journal of Machine Learning Research}}
  \bibinfo{volume}{3}, \bibinfo{number}{Feb} (\bibinfo{year}{2003}),
  \bibinfo{pages}{1137--1155}.
\newblock


\bibitem[\protect\citeauthoryear{Berkovsky, Kuflik, and Ricci}{Berkovsky
  et~al\mbox{.}}{2007}]%
        {berkovsky2007cross}
\bibfield{author}{\bibinfo{person}{Shlomo Berkovsky}, \bibinfo{person}{Tsvi
  Kuflik}, {and} \bibinfo{person}{Francesco Ricci}.}
  \bibinfo{year}{2007}\natexlab{}.
\newblock \showarticletitle{Cross-domain mediation in collaborative filtering}.
  In \bibinfo{booktitle}{\emph{International Conference on User Modeling}}.
  Berlin-Heidelberg, Germany: Springer, \bibinfo{pages}{355--359}.
\newblock


\bibitem[\protect\citeauthoryear{Bhargava, Phan, Zhou, and Lee}{Bhargava
  et~al\mbox{.}}{2015}]%
        {bhargava2015and}
\bibfield{author}{\bibinfo{person}{Preeti Bhargava}, \bibinfo{person}{Thomas
  Phan}, \bibinfo{person}{Jiayu Zhou}, {and} \bibinfo{person}{Juhan Lee}.}
  \bibinfo{year}{2015}\natexlab{}.
\newblock \showarticletitle{Who, what, when, and where: Multi-dimensional
  collaborative recommendations using tensor factorization on sparse
  user-generated data}. In \bibinfo{booktitle}{\emph{Proceedings of the 24th
  International Conference on World Wide Web}}. New York: ACM Press,
  \bibinfo{pages}{130--140}.
\newblock


\bibitem[\protect\citeauthoryear{Blei, Ng, and Jordan}{Blei
  et~al\mbox{.}}{2003}]%
        {blei2003latent}
\bibfield{author}{\bibinfo{person}{David~M Blei}, \bibinfo{person}{Andrew~Y
  Ng}, {and} \bibinfo{person}{Michael~I Jordan}.}
  \bibinfo{year}{2003}\natexlab{}.
\newblock \showarticletitle{Latent {Dirichlet} allocation}.
\newblock \bibinfo{journal}{\emph{Journal of Machine Learning Research}}
  \bibinfo{volume}{3}, \bibinfo{number}{Jan} (\bibinfo{year}{2003}),
  \bibinfo{pages}{993--1022}.
\newblock


\bibitem[\protect\citeauthoryear{Bobadilla, Ortega, Hernando, and
  Guti{\'e}rrez}{Bobadilla et~al\mbox{.}}{2013}]%
        {bobadilla2013recommender}
\bibfield{author}{\bibinfo{person}{Jes{\'u}s Bobadilla},
  \bibinfo{person}{Fernando Ortega}, \bibinfo{person}{Antonio Hernando}, {and}
  \bibinfo{person}{Abraham Guti{\'e}rrez}.} \bibinfo{year}{2013}\natexlab{}.
\newblock \showarticletitle{Recommender systems survey}.
\newblock \bibinfo{journal}{\emph{Knowledge-Based System}}
  \bibinfo{volume}{46} (\bibinfo{year}{2013}), \bibinfo{pages}{109--132}.
\newblock


\bibitem[\protect\citeauthoryear{Bordes, Usunier, Garcia-Duran, Weston, and
  Yakhnenko}{Bordes et~al\mbox{.}}{2013}]%
        {bordes2013translating}
\bibfield{author}{\bibinfo{person}{Antoine Bordes}, \bibinfo{person}{Nicolas
  Usunier}, \bibinfo{person}{Alberto Garcia-Duran}, \bibinfo{person}{Jason
  Weston}, {and} \bibinfo{person}{Oksana Yakhnenko}.}
  \bibinfo{year}{2013}\natexlab{}.
\newblock \showarticletitle{Translating embeddings for modeling
  multi-relational data}. In \bibinfo{booktitle}{\emph{Advances in Neural
  Information Processing Systems}}. South Lake Tahoe, CA, December,
  \bibinfo{pages}{2787--2795}.
\newblock


\bibitem[\protect\citeauthoryear{Breese, Heckerman, and Kadie}{Breese
  et~al\mbox{.}}{1998}]%
        {breese1998empirical}
\bibfield{author}{\bibinfo{person}{John~S Breese}, \bibinfo{person}{David
  Heckerman}, {and} \bibinfo{person}{Carl Kadie}.}
  \bibinfo{year}{1998}\natexlab{}.
\newblock \showarticletitle{Empirical analysis of predictive algorithms for
  collaborative filtering}. In \bibinfo{booktitle}{\emph{Proceedings of the
  14th Conference on Uncertainty in Artificial Intelligence}}. San Francisco:
  CA, Morgan Kaufmann Publishers Inc., \bibinfo{pages}{43--52}.
\newblock


\bibitem[\protect\citeauthoryear{Bruno, Fatima, Benedita, and Juan}{Bruno
  et~al\mbox{.}}{2019}]%
        {veloso2019online}
\bibfield{author}{\bibinfo{person}{Veloso Bruno}, \bibinfo{person}{Leal
  Fatima}, \bibinfo{person}{Malheiro Benedita}, {and}
  \bibinfo{person}{Carlos~Burguillo Juan}.} \bibinfo{year}{2019}\natexlab{}.
\newblock \showarticletitle{On-line Guest Profiling and Hotel Recommendation}.
\newblock \bibinfo{journal}{\emph{Electronic Commerce Research and
  Applications}}  \bibinfo{volume}{34} (\bibinfo{year}{2019}),
  \bibinfo{pages}{100832}.
\newblock


\bibitem[\protect\citeauthoryear{Burke}{Burke}{2002}]%
        {burke2002hybrid}
\bibfield{author}{\bibinfo{person}{Robin Burke}.}
  \bibinfo{year}{2002}\natexlab{}.
\newblock \showarticletitle{Hybrid recommender systems: Survey and
  experiments}.
\newblock \bibinfo{journal}{\emph{User Modeling and User-Adapted Interaction}}
  \bibinfo{volume}{12}, \bibinfo{number}{4} (\bibinfo{year}{2002}),
  \bibinfo{pages}{331--370}.
\newblock


\bibitem[\protect\citeauthoryear{Cai, Zheng, and Chang}{Cai
  et~al\mbox{.}}{2018}]%
        {cai2018comprehensive}
\bibfield{author}{\bibinfo{person}{Hongyun Cai}, \bibinfo{person}{Vincent~W
  Zheng}, {and} \bibinfo{person}{Kevin Chen-Chuan Chang}.}
  \bibinfo{year}{2018}\natexlab{}.
\newblock \showarticletitle{A comprehensive survey of graph embedding:
  Problems, techniques, and applications}.
\newblock \bibinfo{journal}{\emph{IEEE Transactions on Knowledge and Data
  Engineering}} \bibinfo{volume}{30}, \bibinfo{number}{9}
  (\bibinfo{year}{2018}), \bibinfo{pages}{1616--1637}.
\newblock


\bibitem[\protect\citeauthoryear{Cao, Wang, He, Hu, and Chua}{Cao
  et~al\mbox{.}}{2019}]%
        {cao2019unifying}
\bibfield{author}{\bibinfo{person}{Yixin Cao}, \bibinfo{person}{Xiang Wang},
  \bibinfo{person}{Xiangnan He}, \bibinfo{person}{Zikun Hu}, {and}
  \bibinfo{person}{Tat-Seng Chua}.} \bibinfo{year}{2019}\natexlab{}.
\newblock \bibinfo{title}{Unifying Knowledge Graph Learning and Recommendation:
  Towards a Better Understanding of User Preferences}.
\newblock \bibinfo{howpublished}{arXiv preprint arXiv:1902.06236}.
\newblock


\bibitem[\protect\citeauthoryear{Catherine and Cohen}{Catherine and
  Cohen}{2016}]%
        {catherine2016personalized}
\bibfield{author}{\bibinfo{person}{Rose Catherine} {and}
  \bibinfo{person}{William Cohen}.} \bibinfo{year}{2016}\natexlab{}.
\newblock \showarticletitle{Personalized recommendations using knowledge
  graphs: A probabilistic logic programming approach}. In
  \bibinfo{booktitle}{\emph{Proceedings of the 11th ACM Conference on
  Recommender Systems}}. New York: ACM Press, \bibinfo{pages}{325--332}.
\newblock


\bibitem[\protect\citeauthoryear{Catherine and Cohen}{Catherine and
  Cohen}{2017}]%
        {catherine2017transnets}
\bibfield{author}{\bibinfo{person}{Rose Catherine} {and}
  \bibinfo{person}{William Cohen}.} \bibinfo{year}{2017}\natexlab{}.
\newblock \showarticletitle{Transnets: Learning to transform for
  recommendation}. In \bibinfo{booktitle}{\emph{Proceedings of the 12th ACM
  Conference on Recommender Systems}}. New York: ACM Press,
  \bibinfo{pages}{288--296}.
\newblock


\bibitem[\protect\citeauthoryear{Chen, Zhang, He, Nie, Liu, and Chua}{Chen
  et~al\mbox{.}}{2017}]%
        {chen2017attentive}
\bibfield{author}{\bibinfo{person}{Jingyuan Chen}, \bibinfo{person}{Hanwang
  Zhang}, \bibinfo{person}{Xiangnan He}, \bibinfo{person}{Liqiang Nie},
  \bibinfo{person}{Wei Liu}, {and} \bibinfo{person}{Tat-Seng Chua}.}
  \bibinfo{year}{2017}\natexlab{}.
\newblock \showarticletitle{Attentive collaborative filtering: Multimedia
  recommendation with item-and component-level attention}. In
  \bibinfo{booktitle}{\emph{Proceedings of the 40th International ACM SIGIR
  conference on Research and Development in Information Retrieval}}. New York:
  ACM Press, \bibinfo{pages}{335--344}.
\newblock


\bibitem[\protect\citeauthoryear{Chen, Zhang, Lu, Chen, Zheng, and Yu}{Chen
  et~al\mbox{.}}{2012}]%
        {chen2012svdfeature}
\bibfield{author}{\bibinfo{person}{Tianqi Chen}, \bibinfo{person}{Weinan
  Zhang}, \bibinfo{person}{Qiuxia Lu}, \bibinfo{person}{Kailong Chen},
  \bibinfo{person}{Zhao Zheng}, {and} \bibinfo{person}{Yong Yu}.}
  \bibinfo{year}{2012}\natexlab{}.
\newblock \showarticletitle{SVDFeature: a toolkit for feature-based
  collaborative filtering}.
\newblock \bibinfo{journal}{\emph{Journal of Machine Learning Research}}
  \bibinfo{volume}{13}, \bibinfo{number}{Dec} (\bibinfo{year}{2012}),
  \bibinfo{pages}{3619--3622}.
\newblock


\bibitem[\protect\citeauthoryear{Cheng, Koc, Harmsen, Shaked, Chandra, Aradhye,
  Anderson, Corrado, Chai, Ispir, Anil, Haque, Hong, Jain, Liu, and Shah}{Cheng
  et~al\mbox{.}}{2016}]%
        {cheng2016wide}
\bibfield{author}{\bibinfo{person}{Heng-Tze Cheng}, \bibinfo{person}{Levent
  Koc}, \bibinfo{person}{Jeremiah Harmsen}, \bibinfo{person}{Tal Shaked},
  \bibinfo{person}{Tushar Chandra}, \bibinfo{person}{Hrishi Aradhye},
  \bibinfo{person}{Glen Anderson}, \bibinfo{person}{Greg Corrado},
  \bibinfo{person}{Wei Chai}, \bibinfo{person}{Mustafa Ispir},
  \bibinfo{person}{Rohan Anil}, \bibinfo{person}{Zakaria Haque},
  \bibinfo{person}{Lichan Hong}, \bibinfo{person}{Vihan Jain},
  \bibinfo{person}{Xiaobing Liu}, {and} \bibinfo{person}{Hemal Shah}.}
  \bibinfo{year}{2016}\natexlab{}.
\newblock \showarticletitle{Wide \& deep learning for recommender systems}. In
  \bibinfo{booktitle}{\emph{Proceedings of the 11th ACM Conference on
  Recommender Systems}}. New York: ACM Press, \bibinfo{pages}{7--10}.
\newblock


\bibitem[\protect\citeauthoryear{Cho, Van-Merrienboer, Bahdanau, and
  Bengio}{Cho et~al\mbox{.}}{2014a}]%
        {cho2014properties}
\bibfield{author}{\bibinfo{person}{Kyunghyun Cho}, \bibinfo{person}{Bart
  Van-Merrienboer}, \bibinfo{person}{Dzmitry Bahdanau}, {and}
  \bibinfo{person}{Yoshua Bengio}.} \bibinfo{year}{2014}\natexlab{a}.
\newblock \bibinfo{title}{On the Properties of Neural Machine Translation:
  Encoder-Decoder Approaches}.
\newblock \bibinfo{howpublished}{arXiv preprint arXiv:1409.1259}.
\newblock


\bibitem[\protect\citeauthoryear{Cho, Van~Merri{\"e}nboer, Gulcehre, Bahdanau,
  Bougares, Schwenk, and Bengio}{Cho et~al\mbox{.}}{2014b}]%
        {cho2014learning}
\bibfield{author}{\bibinfo{person}{Kyunghyun Cho}, \bibinfo{person}{Bart
  Van~Merri{\"e}nboer}, \bibinfo{person}{Caglar Gulcehre},
  \bibinfo{person}{Dzmitry Bahdanau}, \bibinfo{person}{Fethi Bougares},
  \bibinfo{person}{Holger Schwenk}, {and} \bibinfo{person}{Yoshua Bengio}.}
  \bibinfo{year}{2014}\natexlab{b}.
\newblock \bibinfo{title}{Learning phrase representations using RNN
  encoder-decoder for statistical machine translation}.
\newblock \bibinfo{howpublished}{arXiv preprint arXiv:1406.1078}.
\newblock


\bibitem[\protect\citeauthoryear{Chu and Tsai}{Chu and Tsai}{2017}]%
        {chu2017hybrid}
\bibfield{author}{\bibinfo{person}{Wei-Ta Chu} {and} \bibinfo{person}{Ya-Lun
  Tsai}.} \bibinfo{year}{2017}\natexlab{}.
\newblock \showarticletitle{A hybrid recommendation system considering visual
  information for predicting favorite restaurants}.
\newblock \bibinfo{journal}{\emph{World Wide Web}} \bibinfo{volume}{20},
  \bibinfo{number}{6} (\bibinfo{year}{2017}), \bibinfo{pages}{1313--1331}.
\newblock


\bibitem[\protect\citeauthoryear{Collobert, Weston, Bottou, Karlen,
  Kavukcuoglu, and Kuksa}{Collobert et~al\mbox{.}}{2011}]%
        {collobert2011natural}
\bibfield{author}{\bibinfo{person}{Ronan Collobert}, \bibinfo{person}{Jason
  Weston}, \bibinfo{person}{L{\'e}on Bottou}, \bibinfo{person}{Michael Karlen},
  \bibinfo{person}{Koray Kavukcuoglu}, {and} \bibinfo{person}{Pavel Kuksa}.}
  \bibinfo{year}{2011}\natexlab{}.
\newblock \showarticletitle{Natural language processing (almost) from scratch}.
\newblock \bibinfo{journal}{\emph{Journal of Machine Learning Research}}
  \bibinfo{volume}{12}, \bibinfo{number}{Aug} (\bibinfo{year}{2011}),
  \bibinfo{pages}{2493--2537}.
\newblock


\bibitem[\protect\citeauthoryear{Covington, Adams, and Sargin}{Covington
  et~al\mbox{.}}{2016}]%
        {covington2016deep}
\bibfield{author}{\bibinfo{person}{Paul Covington}, \bibinfo{person}{Jay
  Adams}, {and} \bibinfo{person}{Emre Sargin}.}
  \bibinfo{year}{2016}\natexlab{}.
\newblock \showarticletitle{Deep neural networks for youtube recommendations}.
  In \bibinfo{booktitle}{\emph{Proceedings of the 11th ACM Conference on
  Recommender Systems}}. New York: ACM Press, \bibinfo{pages}{191--198}.
\newblock


\bibitem[\protect\citeauthoryear{Davidson, Liebald, Liu, Nandy, Van~Vleet,
  Gargi, Gupta, He, Lambert, Livingston, et~al\mbox{.}}{Davidson
  et~al\mbox{.}}{2010}]%
        {davidson2010youtube}
\bibfield{author}{\bibinfo{person}{James Davidson}, \bibinfo{person}{Benjamin
  Liebald}, \bibinfo{person}{Junning Liu}, \bibinfo{person}{Palash Nandy},
  \bibinfo{person}{Taylor Van~Vleet}, \bibinfo{person}{Ullas Gargi},
  \bibinfo{person}{Sujoy Gupta}, \bibinfo{person}{Yu He}, \bibinfo{person}{Mike
  Lambert}, \bibinfo{person}{Blake Livingston}, {et~al\mbox{.}}}
  \bibinfo{year}{2010}\natexlab{}.
\newblock \showarticletitle{The YouTube video recommendation system}. In
  \bibinfo{booktitle}{\emph{Proceedings of the 5th ACM Conference on
  Recommender Systems}}. New York: ACM Press, \bibinfo{pages}{293--296}.
\newblock


\bibitem[\protect\citeauthoryear{Desrosiers and Karypis}{Desrosiers and
  Karypis}{2011}]%
        {desrosiers2011comprehensive}
\bibfield{author}{\bibinfo{person}{Christian Desrosiers} {and}
  \bibinfo{person}{George Karypis}.} \bibinfo{year}{2011}\natexlab{}.
\newblock \showarticletitle{A comprehensive survey of neighborhood-based
  recommendation methods}.
\newblock In \bibinfo{booktitle}{\emph{Recommender systems handbook}}.
  \bibinfo{publisher}{Boston MA: Springer}, \bibinfo{pages}{107--144}.
\newblock


\bibitem[\protect\citeauthoryear{Ding, Zhang, Li, Tang, Chen, and Zhou}{Ding
  et~al\mbox{.}}{2017}]%
        {ding2017baydnn}
\bibfield{author}{\bibinfo{person}{Daizong Ding}, \bibinfo{person}{Mi Zhang},
  \bibinfo{person}{Shao-Yuan Li}, \bibinfo{person}{Jie Tang},
  \bibinfo{person}{Xiaotie Chen}, {and} \bibinfo{person}{Zhi-Hua Zhou}.}
  \bibinfo{year}{2017}\natexlab{}.
\newblock \showarticletitle{BayDNN: Friend Recommendation with Bayesian
  Personalized Ranking Deep Neural Network}. In
  \bibinfo{booktitle}{\emph{Proceedings of the 26th ACM International
  Conference on Information and Knowledge Management}}. New York: ACM Press,
  \bibinfo{pages}{1479--1488}.
\newblock


\bibitem[\protect\citeauthoryear{Dingqi, Qu, Jie, and Cudr{\'e}-Mauroux}{Dingqi
  et~al\mbox{.}}{2019}]%
        {dingqi2019revisiting}
\bibfield{author}{\bibinfo{person}{Yang Dingqi}, \bibinfo{person}{Bingqing Qu},
  \bibinfo{person}{Yang Jie}, {and} \bibinfo{person}{Philippe
  Cudr{\'e}-Mauroux}.} \bibinfo{year}{2019}\natexlab{}.
\newblock \showarticletitle{Revisiting user mobility and social relationships
  in lbsns: a hypergraph embedding approach}. In
  \bibinfo{booktitle}{\emph{Proceedings of the 2019 World Wide Web Conference
  on World Wide Web}}. New York: ACM Press, \bibinfo{pages}{2147--2157}.
\newblock


\bibitem[\protect\citeauthoryear{Dong, Yu, Wu, Sun, Yuan, and Zhang}{Dong
  et~al\mbox{.}}{2017}]%
        {dong2017hybrid}
\bibfield{author}{\bibinfo{person}{Xin Dong}, \bibinfo{person}{Lei Yu},
  \bibinfo{person}{Zhonghuo Wu}, \bibinfo{person}{Yuxia Sun},
  \bibinfo{person}{Lingfeng Yuan}, {and} \bibinfo{person}{Fangxi Zhang}.}
  \bibinfo{year}{2017}\natexlab{}.
\newblock \showarticletitle{A hybrid collaborative filtering model with deep
  structure for recommender systems}. In \bibinfo{booktitle}{\emph{Proceedings
  of the 31st AAAI Conference on Artificial Intelligence}}. Menlo Park, CA:
  AAAI Press, \bibinfo{pages}{1309--1315}.
\newblock


\bibitem[\protect\citeauthoryear{Duvenaud, Maclaurin, Iparraguirre, Bombarell,
  Hirzel, Aspuru-Guzik, and Adams}{Duvenaud et~al\mbox{.}}{2015}]%
        {duvenaud2015convolutional}
\bibfield{author}{\bibinfo{person}{David~K Duvenaud}, \bibinfo{person}{Dougal
  Maclaurin}, \bibinfo{person}{Jorge Iparraguirre}, \bibinfo{person}{Rafael
  Bombarell}, \bibinfo{person}{Timothy Hirzel}, \bibinfo{person}{Al{\'a}n
  Aspuru-Guzik}, {and} \bibinfo{person}{Ryan~P Adams}.}
  \bibinfo{year}{2015}\natexlab{}.
\newblock \showarticletitle{Convolutional networks on graphs for learning
  molecular fingerprints}. In \bibinfo{booktitle}{\emph{Advances in Neural
  Information Processing Systems}}. Montreal, Canada,
  \bibinfo{pages}{2224--2232}.
\newblock


\bibitem[\protect\citeauthoryear{Ekstrand, Riedl, Konstan,
  et~al\mbox{.}}{Ekstrand et~al\mbox{.}}{2011}]%
        {ekstrand2011collaborative}
\bibfield{author}{\bibinfo{person}{Michael~D Ekstrand}, \bibinfo{person}{John~T
  Riedl}, \bibinfo{person}{Joseph~A Konstan}, {et~al\mbox{.}}}
  \bibinfo{year}{2011}\natexlab{}.
\newblock \showarticletitle{Collaborative filtering recommender systems}.
\newblock \bibinfo{journal}{\emph{Foundations and Trends{\textregistered} in
  Human Computer Interaction}} \bibinfo{volume}{4}, \bibinfo{number}{2}
  (\bibinfo{year}{2011}), \bibinfo{pages}{81--173}.
\newblock


\bibitem[\protect\citeauthoryear{Fan, Li, and Cheng}{Fan et~al\mbox{.}}{2018}]%
        {fan2018deep}
\bibfield{author}{\bibinfo{person}{Wenqi Fan}, \bibinfo{person}{Qing Li}, {and}
  \bibinfo{person}{Min Cheng}.} \bibinfo{year}{2018}\natexlab{}.
\newblock \showarticletitle{Deep Modeling of Social Relations for
  Recommendation}. In \bibinfo{booktitle}{\emph{Proceedings of the 32nd AAAI
  Conference on Artificial Intelligence}}. Menlo Park, CA: AAAI Press,
  \bibinfo{pages}{8075--8076}.
\newblock


\bibitem[\protect\citeauthoryear{Fan, Ma, Li, He, Zhao, Tang, and Yin}{Fan
  et~al\mbox{.}}{2019}]%
        {fan2019graph}
\bibfield{author}{\bibinfo{person}{Wenqi Fan}, \bibinfo{person}{Yao Ma},
  \bibinfo{person}{Qing Li}, \bibinfo{person}{Yuan He}, \bibinfo{person}{Eric
  Zhao}, \bibinfo{person}{Jiliang Tang}, {and} \bibinfo{person}{Dawei Yin}.}
  \bibinfo{year}{2019}\natexlab{}.
\newblock \bibinfo{title}{Graph Neural Networks for Social Recommendation}.
\newblock \bibinfo{howpublished}{arXiv preprint arXiv:1902.07243}.
\newblock


\bibitem[\protect\citeauthoryear{Fang, Bao, and Zhang}{Fang
  et~al\mbox{.}}{2014}]%
        {bao2014leveraging}
\bibfield{author}{\bibinfo{person}{Hui Fang}, \bibinfo{person}{Yang Bao}, {and}
  \bibinfo{person}{Jie Zhang}.} \bibinfo{year}{2014}\natexlab{}.
\newblock \showarticletitle{Leveraging decomposed trust in probabilistic matrix
  factorization for effective recommendation}. In
  \bibinfo{booktitle}{\emph{Proceedings of the 28st AAAI Conference on
  Artificial Intelligence}}, Vol.~\bibinfo{volume}{350}. Menlo Park, CA: AAAI
  Press, \bibinfo{pages}{30--36}.
\newblock


\bibitem[\protect\citeauthoryear{Fang, Guo, and Zhang}{Fang
  et~al\mbox{.}}{2015}]%
        {fang2015multi}
\bibfield{author}{\bibinfo{person}{Hui Fang}, \bibinfo{person}{Guibing Guo},
  {and} \bibinfo{person}{Jie Zhang}.} \bibinfo{year}{2015}\natexlab{}.
\newblock \showarticletitle{Multi-faceted trust and distrust prediction for
  recommender systems}.
\newblock \bibinfo{journal}{\emph{Decision Support Systems}}
  \bibinfo{volume}{71} (\bibinfo{year}{2015}), \bibinfo{pages}{37--47}.
\newblock


\bibitem[\protect\citeauthoryear{Feng, Li, Zhang, Sun, Meng, Guo, and Jin}{Feng
  et~al\mbox{.}}{2018}]%
        {feng2018deepmove}
\bibfield{author}{\bibinfo{person}{Jie Feng}, \bibinfo{person}{Yong Li},
  \bibinfo{person}{Chao Zhang}, \bibinfo{person}{Funing Sun},
  \bibinfo{person}{Fanchao Meng}, \bibinfo{person}{Ang Guo}, {and}
  \bibinfo{person}{Depeng Jin}.} \bibinfo{year}{2018}\natexlab{}.
\newblock \showarticletitle{DeepMove: Predicting Human Mobility with
  Attentional Recurrent Networks}. In \bibinfo{booktitle}{\emph{Proceedings of
  the 2018 World Wide Web Conference on World Wide Web}}. New York: ACM Press,
  \bibinfo{pages}{1459--1468}.
\newblock


\bibitem[\protect\citeauthoryear{Feng, Cong, An, and Chee}{Feng
  et~al\mbox{.}}{2017}]%
        {feng2017poi2vec}
\bibfield{author}{\bibinfo{person}{Shanshan Feng}, \bibinfo{person}{Gao Cong},
  \bibinfo{person}{Bo An}, {and} \bibinfo{person}{Yeow~Meng Chee}.}
  \bibinfo{year}{2017}\natexlab{}.
\newblock \showarticletitle{POI2Vec: Geographical Latent Representation for
  Predicting Future Visitors}. In \bibinfo{booktitle}{\emph{Proceedings of the
  31st AAAI Conference on Artificial Intelligence}}. Menlo Park, CA: AAAI
  Press, \bibinfo{pages}{102--108}.
\newblock


\bibitem[\protect\citeauthoryear{Fern{\'a}ndez-Tob{\'\i}as, Cantador,
  Kaminskas, and Ricci}{Fern{\'a}ndez-Tob{\'\i}as et~al\mbox{.}}{2012}]%
        {fernandez2012cross}
\bibfield{author}{\bibinfo{person}{Ignacio Fern{\'a}ndez-Tob{\'\i}as},
  \bibinfo{person}{Iv{\'a}n Cantador}, \bibinfo{person}{Marius Kaminskas},
  {and} \bibinfo{person}{Francesco Ricci}.} \bibinfo{year}{2012}\natexlab{}.
\newblock \showarticletitle{Cross-domain recommender systems: A survey of the
  state of the art}. In \bibinfo{booktitle}{\emph{Spanish Conference on
  Information Retrieval}}. \bibinfo{pages}{1--12}.
\newblock


\bibitem[\protect\citeauthoryear{Forsati, Mahdavi, Shamsfard, and
  Sarwat}{Forsati et~al\mbox{.}}{2014}]%
        {forsati2014matrix}
\bibfield{author}{\bibinfo{person}{Rana Forsati}, \bibinfo{person}{Mehrdad
  Mahdavi}, \bibinfo{person}{Mehrnoush Shamsfard}, {and}
  \bibinfo{person}{Mohamed Sarwat}.} \bibinfo{year}{2014}\natexlab{}.
\newblock \showarticletitle{Matrix factorization with explicit trust and
  distrust side information for improved social recommendation}.
\newblock \bibinfo{journal}{\emph{ACM Transactions on Information Systems
  (TOIS)}} \bibinfo{volume}{32}, \bibinfo{number}{4} (\bibinfo{year}{2014}),
  \bibinfo{pages}{17}.
\newblock


\bibitem[\protect\citeauthoryear{Gao, Tang, Hu, and Liu}{Gao
  et~al\mbox{.}}{2015}]%
        {gao2015content}
\bibfield{author}{\bibinfo{person}{Huiji Gao}, \bibinfo{person}{Jiliang Tang},
  \bibinfo{person}{Xia Hu}, {and} \bibinfo{person}{Huan Liu}.}
  \bibinfo{year}{2015}\natexlab{}.
\newblock \showarticletitle{Content-Aware Point of Interest Recommendation on
  Location-Based Social Networks}. In \bibinfo{booktitle}{\emph{Proceedings of
  the 29th AAAI Conference on Artificial Intelligence}}. Menlo Park, CA: AAAI
  Press, \bibinfo{pages}{1721--1727}.
\newblock


\bibitem[\protect\citeauthoryear{Gomez-Uribe and Hunt}{Gomez-Uribe and
  Hunt}{2016}]%
        {gomez2016netflix}
\bibfield{author}{\bibinfo{person}{Carlos~A Gomez-Uribe} {and}
  \bibinfo{person}{Neil Hunt}.} \bibinfo{year}{2016}\natexlab{}.
\newblock \showarticletitle{The netflix recommender system: Algorithms,
  business value, and innovation}.
\newblock \bibinfo{journal}{\emph{ACM Transactions on Management Information
  Systems (TMIS)}} \bibinfo{volume}{6}, \bibinfo{number}{4}
  (\bibinfo{year}{2016}), \bibinfo{pages}{13}.
\newblock


\bibitem[\protect\citeauthoryear{Grad-Gyenge, Filzmoser, and
  Werthner}{Grad-Gyenge et~al\mbox{.}}{2015}]%
        {grad2015recommendations}
\bibfield{author}{\bibinfo{person}{L{\'a}szl{\'o} Grad-Gyenge},
  \bibinfo{person}{Peter Filzmoser}, {and} \bibinfo{person}{Hannes Werthner}.}
  \bibinfo{year}{2015}\natexlab{}.
\newblock \showarticletitle{Recommendations on a knowledge graph}. In
  \bibinfo{booktitle}{\emph{1st International Workshop on Machine Learning
  Methods for Recommender Systems (MLRec)}}. Vancouver, Canada, April 30-May 2,
  \bibinfo{pages}{13--20}.
\newblock


\bibitem[\protect\citeauthoryear{Grbovic, Radosavljevic, Djuric, Bhamidipati,
  Savla, Bhagwan, and Sharp}{Grbovic et~al\mbox{.}}{2015}]%
        {grbovic2015commerce}
\bibfield{author}{\bibinfo{person}{Mihajlo Grbovic}, \bibinfo{person}{Vladan
  Radosavljevic}, \bibinfo{person}{Nemanja Djuric}, \bibinfo{person}{Narayan
  Bhamidipati}, \bibinfo{person}{Jaikit Savla}, \bibinfo{person}{Varun
  Bhagwan}, {and} \bibinfo{person}{Doug Sharp}.}
  \bibinfo{year}{2015}\natexlab{}.
\newblock \showarticletitle{E-commerce in your inbox: Product recommendations
  at scale}. In \bibinfo{booktitle}{\emph{Proceedings of the 21st ACM SIGKDD
  International Conference on Knowledge Discovery and Data Mining}}. New York:
  ACM Press, \bibinfo{pages}{1809--1818}.
\newblock


\bibitem[\protect\citeauthoryear{Grover and Leskovec}{Grover and
  Leskovec}{2016}]%
        {grover2016node2vec}
\bibfield{author}{\bibinfo{person}{Aditya Grover} {and} \bibinfo{person}{Jure
  Leskovec}.} \bibinfo{year}{2016}\natexlab{}.
\newblock \showarticletitle{Node2vec: Scalable feature learning for networks}.
  In \bibinfo{booktitle}{\emph{Proceedings of the 22nd ACM SIGKDD International
  Conference on Knowledge Discovery and Data Mining}}. New York: ACM Press,
  \bibinfo{pages}{855--864}.
\newblock


\bibitem[\protect\citeauthoryear{Guo}{Guo}{2012}]%
        {guo2012resolving}
\bibfield{author}{\bibinfo{person}{Guibing Guo}.}
  \bibinfo{year}{2012}\natexlab{}.
\newblock \showarticletitle{Resolving data sparsity and cold start in
  recommender systems}. In \bibinfo{booktitle}{\emph{International Conference
  on User Modeling, Adaptation, and Personalization}}. Berlin-Heidelberg,
  Germany: Springer, \bibinfo{pages}{361--364}.
\newblock


\bibitem[\protect\citeauthoryear{Guo}{Guo}{2013}]%
        {guo2013integrating}
\bibfield{author}{\bibinfo{person}{Guibing Guo}.}
  \bibinfo{year}{2013}\natexlab{}.
\newblock \showarticletitle{Integrating trust and similarity to ameliorate the
  data sparsity and cold start for recommender systems}. In
  \bibinfo{booktitle}{\emph{Proceedings of the 7th ACM Conference on
  Recommender Systems}}. New York: ACM Press, \bibinfo{pages}{451--454}.
\newblock


\bibitem[\protect\citeauthoryear{Guo, Zhang, and Thalmann}{Guo
  et~al\mbox{.}}{2012}]%
        {guo2012simple}
\bibfield{author}{\bibinfo{person}{Guibing Guo}, \bibinfo{person}{Jie Zhang},
  {and} \bibinfo{person}{Daniel Thalmann}.} \bibinfo{year}{2012}\natexlab{}.
\newblock \showarticletitle{A simple but effective method to incorporate
  trusted neighbors in recommender systems}. In
  \bibinfo{booktitle}{\emph{International Conference on User Modeling,
  Adaptation, and Personalization}}. Berlin-Heidelberg, Germany: Springer,
  \bibinfo{pages}{114--125}.
\newblock


\bibitem[\protect\citeauthoryear{Guo, Zhang, and Thalmann}{Guo
  et~al\mbox{.}}{2014}]%
        {guo2014merging}
\bibfield{author}{\bibinfo{person}{Guibing Guo}, \bibinfo{person}{Jie Zhang},
  {and} \bibinfo{person}{Daniel Thalmann}.} \bibinfo{year}{2014}\natexlab{}.
\newblock \showarticletitle{Merging trust in collaborative filtering to
  alleviate data sparsity and cold start}.
\newblock \bibinfo{journal}{\emph{Knowledge-Based Systems}}
  \bibinfo{volume}{57} (\bibinfo{year}{2014}), \bibinfo{pages}{57--68}.
\newblock


\bibitem[\protect\citeauthoryear{Guo, Zhang, and Yorke-Smith}{Guo
  et~al\mbox{.}}{2015a}]%
        {guo2015leveraging}
\bibfield{author}{\bibinfo{person}{Guibing Guo}, \bibinfo{person}{Jie Zhang},
  {and} \bibinfo{person}{Neil Yorke-Smith}.} \bibinfo{year}{2015}\natexlab{a}.
\newblock \showarticletitle{Leveraging multiviews of trust and similarity to
  enhance clustering-based recommender systems}.
\newblock \bibinfo{journal}{\emph{Knowledge-Based Systems}}
  \bibinfo{volume}{74} (\bibinfo{year}{2015}), \bibinfo{pages}{14--27}.
\newblock


\bibitem[\protect\citeauthoryear{Guo, Zhang, and Yorke-Smith}{Guo
  et~al\mbox{.}}{2015b}]%
        {guo2015trustsvd}
\bibfield{author}{\bibinfo{person}{Guibing Guo}, \bibinfo{person}{Jie Zhang},
  {and} \bibinfo{person}{Neil Yorke-Smith}.} \bibinfo{year}{2015}\natexlab{b}.
\newblock \showarticletitle{TrustSVD: Collaborative Filtering with Both the
  Explicit and Implicit Influence of User Trust and of Item Ratings}. In
  \bibinfo{booktitle}{\emph{International Joint Conference on Artificial
  Intelligence}}, Vol.~\bibinfo{volume}{15}. Menlo Park: AAAI Press,
  \bibinfo{pages}{123--125}.
\newblock


\bibitem[\protect\citeauthoryear{Guo, Sun, and Theng}{Guo
  et~al\mbox{.}}{2019}]%
        {guo2019exploiting}
\bibfield{author}{\bibinfo{person}{Qing Guo}, \bibinfo{person}{Zhu Sun}, {and}
  \bibinfo{person}{Yin-Leng Theng}.} \bibinfo{year}{2019}\natexlab{}.
\newblock \showarticletitle{Exploiting side information for recommendation}. In
  \bibinfo{booktitle}{\emph{International Conference on Web Engineering}}.
  Berlin-Heidelberg, Germany: Springer, \bibinfo{pages}{569--573}.
\newblock


\bibitem[\protect\citeauthoryear{Guo, Sun, Zhang, Chen, and Theng}{Guo
  et~al\mbox{.}}{2017}]%
        {guo2017aspect}
\bibfield{author}{\bibinfo{person}{Qing Guo}, \bibinfo{person}{Zhu Sun},
  \bibinfo{person}{Jie Zhang}, \bibinfo{person}{Qi Chen}, {and}
  \bibinfo{person}{Yin-Leng Theng}.} \bibinfo{year}{2017}\natexlab{}.
\newblock \showarticletitle{Aspect-aware point-of-interest recommendation with
  geo-social influence}. In \bibinfo{booktitle}{\emph{International Conference
  on User Modeling, Adaptation, and Personalization}}. New York: ACM Press,
  \bibinfo{pages}{17--22}.
\newblock


\bibitem[\protect\citeauthoryear{He, Zhang, Ren, and Sun}{He
  et~al\mbox{.}}{2016c}]%
        {he2016deep}
\bibfield{author}{\bibinfo{person}{Kaiming He}, \bibinfo{person}{Xiangyu
  Zhang}, \bibinfo{person}{Shaoqing Ren}, {and} \bibinfo{person}{Jian Sun}.}
  \bibinfo{year}{2016}\natexlab{c}.
\newblock \showarticletitle{Deep residual learning for image recognition}. In
  \bibinfo{booktitle}{\emph{Proceedings of the IEEE Conference on Computer
  Vision and Pattern Recognition}}. Washington, DC: IEEE Computer Society
  Press, \bibinfo{pages}{770--778}.
\newblock


\bibitem[\protect\citeauthoryear{He, Fang, Wang, and McAuley}{He
  et~al\mbox{.}}{2016a}]%
        {he2016vista}
\bibfield{author}{\bibinfo{person}{Ruining He}, \bibinfo{person}{Chen Fang},
  \bibinfo{person}{Zhaowen Wang}, {and} \bibinfo{person}{Julian McAuley}.}
  \bibinfo{year}{2016}\natexlab{a}.
\newblock \showarticletitle{Vista: A visually, socially, and temporally-aware
  model for artistic recommendation}. In \bibinfo{booktitle}{\emph{Proceedings
  of the 11th ACM Conference on Recommender Systems}}. New York: ACM Press,
  \bibinfo{pages}{309--316}.
\newblock


\bibitem[\protect\citeauthoryear{He, Lin, Wang, and McAuley}{He
  et~al\mbox{.}}{2016b}]%
        {he2016sherlock}
\bibfield{author}{\bibinfo{person}{Ruining He}, \bibinfo{person}{Chunbin Lin},
  \bibinfo{person}{Jianguo Wang}, {and} \bibinfo{person}{Julian McAuley}.}
  \bibinfo{year}{2016}\natexlab{b}.
\newblock \showarticletitle{Sherlock: sparse hierarchical embeddings for
  visually-aware one-class collaborative filtering}. In
  \bibinfo{booktitle}{\emph{International Joint Conference on Artificial
  Intelligence}}. Menlo Park: AAAI Press, \bibinfo{pages}{3740--3746}.
\newblock


\bibitem[\protect\citeauthoryear{He and McAuley}{He and McAuley}{2016a}]%
        {he2016ups}
\bibfield{author}{\bibinfo{person}{Ruining He} {and} \bibinfo{person}{Julian
  McAuley}.} \bibinfo{year}{2016}\natexlab{a}.
\newblock \showarticletitle{Ups and downs: Modeling the visual evolution of
  fashion trends with one-class collaborative filtering}. In
  \bibinfo{booktitle}{\emph{Proceedings of the 2016 World Wide Web Conference
  on World Wide Web}}. New York: ACM Press, \bibinfo{pages}{507--517}.
\newblock


\bibitem[\protect\citeauthoryear{He and McAuley}{He and McAuley}{2016b}]%
        {he2016vbpr}
\bibfield{author}{\bibinfo{person}{Ruining He} {and} \bibinfo{person}{Julian
  McAuley}.} \bibinfo{year}{2016}\natexlab{b}.
\newblock \showarticletitle{VBPR: Visual {Bayesian} Personalized Ranking from
  Implicit Feedback}. In \bibinfo{booktitle}{\emph{Proceedings of the 30th AAAI
  Conference on Artificial Intelligence}}. Menlo Park, CA: AAAI Press,
  \bibinfo{pages}{144--150}.
\newblock


\bibitem[\protect\citeauthoryear{He, Chen, Kan, and Chen}{He
  et~al\mbox{.}}{2015}]%
        {he2015trirank}
\bibfield{author}{\bibinfo{person}{Xiangnan He}, \bibinfo{person}{Tao Chen},
  \bibinfo{person}{Min-Yen Kan}, {and} \bibinfo{person}{Xiao Chen}.}
  \bibinfo{year}{2015}\natexlab{}.
\newblock \showarticletitle{Trirank: Review-aware explainable recommendation by
  modeling aspects}. In \bibinfo{booktitle}{\emph{Proceedings of the 24th ACM
  International on Conference on Information and Knowledge Management}}. New
  York: ACM Press, \bibinfo{pages}{1661--1670}.
\newblock


\bibitem[\protect\citeauthoryear{He, Du, Wang, Tian, Tang, and Chua}{He
  et~al\mbox{.}}{2018a}]%
        {he2018outer}
\bibfield{author}{\bibinfo{person}{Xiangnan He}, \bibinfo{person}{Xiaoyu Du},
  \bibinfo{person}{Xiang Wang}, \bibinfo{person}{Feng Tian},
  \bibinfo{person}{Jinhui Tang}, {and} \bibinfo{person}{Tat-Seng Chua}.}
  \bibinfo{year}{2018}\natexlab{a}.
\newblock \bibinfo{title}{Outer product-based neural collaborative filtering}.
\newblock \bibinfo{howpublished}{arXiv preprint arXiv:1808.03912}.
\newblock


\bibitem[\protect\citeauthoryear{He, He, Du, and Chua}{He
  et~al\mbox{.}}{2018b}]%
        {he2018adversarial}
\bibfield{author}{\bibinfo{person}{Xiangnan He}, \bibinfo{person}{Zhankui He},
  \bibinfo{person}{Xiaoyu Du}, {and} \bibinfo{person}{Tat-Seng Chua}.}
  \bibinfo{year}{2018}\natexlab{b}.
\newblock \showarticletitle{Adversarial personalized ranking for
  recommendation}. In \bibinfo{booktitle}{\emph{Proceedings of the 41st
  International ACM SIGIR conference on Research and Development in Information
  Retrieval}}. New York: ACM Press, \bibinfo{pages}{355--364}.
\newblock


\bibitem[\protect\citeauthoryear{He, Liao, Zhang, Nie, Hu, and Chua}{He
  et~al\mbox{.}}{2017}]%
        {he2017neural}
\bibfield{author}{\bibinfo{person}{Xiangnan He}, \bibinfo{person}{Lizi Liao},
  \bibinfo{person}{Hanwang Zhang}, \bibinfo{person}{Liqiang Nie},
  \bibinfo{person}{Xia Hu}, {and} \bibinfo{person}{Tat-Seng Chua}.}
  \bibinfo{year}{2017}\natexlab{}.
\newblock \showarticletitle{Neural collaborative filtering}. In
  \bibinfo{booktitle}{\emph{Proceedings of the 2017 World Wide Web Conference
  on World Wide Web}}. New York: ACM Press, \bibinfo{pages}{173--182}.
\newblock


\bibitem[\protect\citeauthoryear{Hershey and Olsen}{Hershey and Olsen}{2007}]%
        {hershey2007approximating}
\bibfield{author}{\bibinfo{person}{John~R Hershey} {and}
  \bibinfo{person}{Peder~A Olsen}.} \bibinfo{year}{2007}\natexlab{}.
\newblock \showarticletitle{Approximating the Kullback Leibler divergence
  between Gaussian mixture models}. In \bibinfo{booktitle}{\emph{2007 IEEE
  International Conference on Acoustics, Speech and Signal Processing}},
  Vol.~\bibinfo{volume}{4}. Los Alamitos, CA: IEEE Computer Society Press,
  \bibinfo{pages}{IV--317}.
\newblock


\bibitem[\protect\citeauthoryear{Hidasi, Karatzoglou, Baltrunas, and
  Tikk}{Hidasi et~al\mbox{.}}{2015}]%
        {hidasi2015session}
\bibfield{author}{\bibinfo{person}{Bal{\'a}zs Hidasi},
  \bibinfo{person}{Alexandros Karatzoglou}, \bibinfo{person}{Linas Baltrunas},
  {and} \bibinfo{person}{Domonkos Tikk}.} \bibinfo{year}{2015}\natexlab{}.
\newblock \bibinfo{title}{Session-based recommendations with recurrent neural
  networks}.
\newblock \bibinfo{howpublished}{arXiv preprint arXiv:1511.06939}.
\newblock


\bibitem[\protect\citeauthoryear{Hinton, Osindero, and Teh}{Hinton
  et~al\mbox{.}}{2006}]%
        {hinton2006fast}
\bibfield{author}{\bibinfo{person}{Geoffrey~E Hinton}, \bibinfo{person}{Simon
  Osindero}, {and} \bibinfo{person}{Yee-Whye Teh}.}
  \bibinfo{year}{2006}\natexlab{}.
\newblock \showarticletitle{A fast learning algorithm for deep belief nets}.
\newblock \bibinfo{journal}{\emph{Neural Computation}} \bibinfo{volume}{18},
  \bibinfo{number}{7} (\bibinfo{year}{2006}), \bibinfo{pages}{1527--1554}.
\newblock


\bibitem[\protect\citeauthoryear{Hochreiter and Schmidhuber}{Hochreiter and
  Schmidhuber}{1997}]%
        {hochreiter1997long}
\bibfield{author}{\bibinfo{person}{Sepp Hochreiter} {and}
  \bibinfo{person}{J{\"u}rgen Schmidhuber}.} \bibinfo{year}{1997}\natexlab{}.
\newblock \showarticletitle{Long short-term memory}.
\newblock \bibinfo{journal}{\emph{Neural Computation}} \bibinfo{volume}{9},
  \bibinfo{number}{8} (\bibinfo{year}{1997}), \bibinfo{pages}{1735--1780}.
\newblock


\bibitem[\protect\citeauthoryear{Hosseini, Khodadadi, Alizadeh, Arabzadeh,
  Farajtabar, Zha, and Rabiee}{Hosseini et~al\mbox{.}}{2018}]%
        {hosseini2018recurrent}
\bibfield{author}{\bibinfo{person}{Seyedabbas Hosseini}, \bibinfo{person}{Ali
  Khodadadi}, \bibinfo{person}{Keivan Alizadeh}, \bibinfo{person}{Ali
  Arabzadeh}, \bibinfo{person}{Mehrdad Farajtabar}, \bibinfo{person}{Hongyuan
  Zha}, {and} \bibinfo{person}{Hamid~RR Rabiee}.}
  \bibinfo{year}{2018}\natexlab{}.
\newblock \bibinfo{title}{Recurrent poisson factorization for temporal
  recommendation}.
\newblock \bibinfo{howpublished}{IEEE Transactions on Knowledge and Data
  Engineering}.
\newblock
\newblock
\shownote{in press.}


\bibitem[\protect\citeauthoryear{Hu, Shi, Zhao, and Yu}{Hu
  et~al\mbox{.}}{2018}]%
        {hu2018leveraging}
\bibfield{author}{\bibinfo{person}{Binbin Hu}, \bibinfo{person}{Chuan Shi},
  \bibinfo{person}{Wayne~Xin Zhao}, {and} \bibinfo{person}{Philip~S Yu}.}
  \bibinfo{year}{2018}\natexlab{}.
\newblock \showarticletitle{Leveraging meta-path based context for top-n
  recommendation with a neural co-attention model}. In
  \bibinfo{booktitle}{\emph{Proceedings of the 24th ACM SIGKDD International
  Conference on Knowledge Discovery and Data Mining}}. New York: ACM Press,
  \bibinfo{pages}{1531--1540}.
\newblock


\bibitem[\protect\citeauthoryear{Hu, Sun, and Liu}{Hu et~al\mbox{.}}{2014}]%
        {hu2014your}
\bibfield{author}{\bibinfo{person}{Longke Hu}, \bibinfo{person}{Aixin Sun},
  {and} \bibinfo{person}{Yong Liu}.} \bibinfo{year}{2014}\natexlab{}.
\newblock \showarticletitle{Your neighbors affect your ratings: on geographical
  neighborhood influence to rating prediction}. In
  \bibinfo{booktitle}{\emph{Proceedings of the 37th International ACM SIGIR
  Conference on Research \& Development in Information Retrieval}}. New York:
  ACM Press, \bibinfo{pages}{345--354}.
\newblock


\bibitem[\protect\citeauthoryear{Hu, Huang, Deng, Gao, and Xing}{Hu
  et~al\mbox{.}}{2015}]%
        {hu2015entity}
\bibfield{author}{\bibinfo{person}{Zhiting Hu}, \bibinfo{person}{Poyao Huang},
  \bibinfo{person}{Yuntian Deng}, \bibinfo{person}{Yingkai Gao}, {and}
  \bibinfo{person}{Eric Xing}.} \bibinfo{year}{2015}\natexlab{}.
\newblock \showarticletitle{Entity hierarchy embedding}. In
  \bibinfo{booktitle}{\emph{Proceedings of the 53rd Annual Meeting of the
  Association for Computational Linguistics and the 7th International Joint
  Conference on Natural Language Processing (1, Long Papers)}},
  Vol.~\bibinfo{volume}{1}. Menlo Park: AAAI Press,
  \bibinfo{pages}{1292--1300}.
\newblock


\bibitem[\protect\citeauthoryear{Hwang, Lee, Kim, and Lee}{Hwang
  et~al\mbox{.}}{2012}]%
        {hwang2012using}
\bibfield{author}{\bibinfo{person}{Won-Seok Hwang}, \bibinfo{person}{Ho-Jong
  Lee}, \bibinfo{person}{Sang-Wook Kim}, {and} \bibinfo{person}{Minsoo Lee}.}
  \bibinfo{year}{2012}\natexlab{}.
\newblock \showarticletitle{On using category experts for improving the
  performance and accuracy in recommender systems}. In
  \bibinfo{booktitle}{\emph{Proceedings of the 21st ACM International
  Conference on Information and Knowledge Management}}. New York: ACM Press,
  \bibinfo{pages}{2355--2358}.
\newblock


\bibitem[\protect\citeauthoryear{Jamali and Ester}{Jamali and Ester}{2010}]%
        {jamali2010matrix}
\bibfield{author}{\bibinfo{person}{Mohsen Jamali} {and} \bibinfo{person}{Martin
  Ester}.} \bibinfo{year}{2010}\natexlab{}.
\newblock \showarticletitle{A matrix factorization technique with trust
  propagation for recommendation in social networks}. In
  \bibinfo{booktitle}{\emph{Proceedings of the 5th ACM Conference on
  Recommender Systems}}. New York: ACM Press, \bibinfo{pages}{135--142}.
\newblock


\bibitem[\protect\citeauthoryear{Jenatton, Mairal, Obozinski, and
  Bach}{Jenatton et~al\mbox{.}}{2010}]%
        {jenatton2010proximal}
\bibfield{author}{\bibinfo{person}{Rodolphe Jenatton}, \bibinfo{person}{Julien
  Mairal}, \bibinfo{person}{Guillaume Obozinski}, {and}
  \bibinfo{person}{Francis~R Bach}.} \bibinfo{year}{2010}\natexlab{}.
\newblock \showarticletitle{Proximal Methods for Sparse Hierarchical Dictionary
  Learning}. In \bibinfo{booktitle}{\emph{International Conference on Machine
  Learning}}. Haifa, Israel: JMLR.org, \bibinfo{pages}{487--494}.
\newblock


\bibitem[\protect\citeauthoryear{Ji, He, Xu, Liu, and Zhao}{Ji
  et~al\mbox{.}}{2015}]%
        {ji2015knowledge}
\bibfield{author}{\bibinfo{person}{Guoliang Ji}, \bibinfo{person}{Shizhu He},
  \bibinfo{person}{Liheng Xu}, \bibinfo{person}{Kang Liu}, {and}
  \bibinfo{person}{Jun Zhao}.} \bibinfo{year}{2015}\natexlab{}.
\newblock \showarticletitle{Knowledge graph embedding via dynamic mapping
  matrix}. In \bibinfo{booktitle}{\emph{Proceedings of the 53rd Annual Meeting
  of the Association for Computational Linguistics and the 7th International
  Joint Conference on Natural Language Processing (1, Long Papers)}},
  Vol.~\bibinfo{volume}{1}. \bibinfo{pages}{687--696}.
\newblock


\bibitem[\protect\citeauthoryear{Ji, Shen, Tian, Wu, and Wu}{Ji
  et~al\mbox{.}}{2014}]%
        {ji2014two}
\bibfield{author}{\bibinfo{person}{Ke Ji}, \bibinfo{person}{Hong Shen},
  \bibinfo{person}{Hui Tian}, \bibinfo{person}{Yanbo Wu}, {and}
  \bibinfo{person}{Jun Wu}.} \bibinfo{year}{2014}\natexlab{}.
\newblock \showarticletitle{Two-phase layered learning recommendation via
  category structure}. In \bibinfo{booktitle}{\emph{Pacific-Asia Conference on
  Knowledge Discovery and Data Mining}}. Berlin-Heidelberg, Germany: Springer,
  \bibinfo{pages}{13--24}.
\newblock


\bibitem[\protect\citeauthoryear{Jia, Shelhamer, Donahue, Karayev, Long,
  Girshick, Guadarrama, and Darrell}{Jia et~al\mbox{.}}{2014}]%
        {jia2013caffe}
\bibfield{author}{\bibinfo{person}{Yangqing Jia}, \bibinfo{person}{Evan
  Shelhamer}, \bibinfo{person}{Jeff Donahue}, \bibinfo{person}{Sergey Karayev},
  \bibinfo{person}{Jonathan Long}, \bibinfo{person}{Ross Girshick},
  \bibinfo{person}{Sergio Guadarrama}, {and} \bibinfo{person}{Trevor Darrell}.}
  \bibinfo{year}{2014}\natexlab{}.
\newblock \showarticletitle{Caffe: Convolutional architecture for fast feature
  embedding}. In \bibinfo{booktitle}{\emph{Proceedings of the 22nd ACM
  International Conference on Multimedia}}. New York: ACM Press,
  \bibinfo{pages}{675--678}.
\newblock


\bibitem[\protect\citeauthoryear{Jing and Smola}{Jing and Smola}{2017}]%
        {jing2017neural}
\bibfield{author}{\bibinfo{person}{How Jing} {and} \bibinfo{person}{Alexander~J
  Smola}.} \bibinfo{year}{2017}\natexlab{}.
\newblock \showarticletitle{Neural survival recommender}. In
  \bibinfo{booktitle}{\emph{Proceedings of the 11th ACM International
  Conference on Web Search and Data Mining}}. New York: ACM Press,
  \bibinfo{pages}{515--524}.
\newblock


\bibitem[\protect\citeauthoryear{Kanagal, Ahmed, Pandey, Josifovski, Yuan, and
  Garcia-Pueyo}{Kanagal et~al\mbox{.}}{2012}]%
        {kanagal2012supercharging}
\bibfield{author}{\bibinfo{person}{Bhargav Kanagal}, \bibinfo{person}{Amr
  Ahmed}, \bibinfo{person}{Sandeep Pandey}, \bibinfo{person}{Vanja Josifovski},
  \bibinfo{person}{Jeff Yuan}, {and} \bibinfo{person}{Lluis Garcia-Pueyo}.}
  \bibinfo{year}{2012}\natexlab{}.
\newblock \showarticletitle{Supercharging recommender systems using taxonomies
  for learning user purchase behavior}.
\newblock \bibinfo{journal}{\emph{Proceedings of the VLDB Endowment}}
  \bibinfo{volume}{5}, \bibinfo{number}{10} (\bibinfo{year}{2012}),
  \bibinfo{pages}{956--967}.
\newblock


\bibitem[\protect\citeauthoryear{Kang and McAuley}{Kang and McAuley}{2018}]%
        {kang2018self}
\bibfield{author}{\bibinfo{person}{Wang-Cheng Kang} {and}
  \bibinfo{person}{Julian McAuley}.} \bibinfo{year}{2018}\natexlab{}.
\newblock \showarticletitle{Self-Attentive Sequential Recommendation}. In
  \bibinfo{booktitle}{\emph{2018 IEEE International Conference on Data
  Mining}}. Washington, DC: IEEE Computer Society Press,
  \bibinfo{pages}{197--206}.
\newblock


\bibitem[\protect\citeauthoryear{Karatzoglou, Amatriain, Baltrunas, and
  Oliver}{Karatzoglou et~al\mbox{.}}{2010}]%
        {karatzoglou2010multiverse}
\bibfield{author}{\bibinfo{person}{Alexandros Karatzoglou},
  \bibinfo{person}{Xavier Amatriain}, \bibinfo{person}{Linas Baltrunas}, {and}
  \bibinfo{person}{Nuria Oliver}.} \bibinfo{year}{2010}\natexlab{}.
\newblock \showarticletitle{Multiverse recommendation: n-dimensional tensor
  factorization for context-aware collaborative filtering}. In
  \bibinfo{booktitle}{\emph{Proceedings of the 4th ACM conference on
  Recommender systems}}. New York: ACM Press, \bibinfo{pages}{79--86}.
\newblock


\bibitem[\protect\citeauthoryear{Kim and Kim}{Kim and Kim}{2003}]%
        {kim2003recommendation}
\bibfield{author}{\bibinfo{person}{Choonho Kim} {and} \bibinfo{person}{Juntae
  Kim}.} \bibinfo{year}{2003}\natexlab{}.
\newblock \showarticletitle{A recommendation algorithm using multi-level
  association rules}. In \bibinfo{booktitle}{\emph{Proceedings IEEE/WIC
  International Conference on Web Intelligence}}. Washington, DC: IEEE Computer
  Society Press, \bibinfo{pages}{524--527}.
\newblock


\bibitem[\protect\citeauthoryear{Kim and Xing}{Kim and Xing}{2010}]%
        {kim2010tree}
\bibfield{author}{\bibinfo{person}{Seyoung Kim} {and} \bibinfo{person}{Eric~P
  Xing}.} \bibinfo{year}{2010}\natexlab{}.
\newblock \showarticletitle{Tree-guided group lasso for multi-task regression
  with structured sparsity}. In \bibinfo{booktitle}{\emph{International
  Conference on Machine Learning}}. Haifa, Israel: JMLR.org,
  \bibinfo{pages}{543--550}.
\newblock


\bibitem[\protect\citeauthoryear{Kipf and Welling}{Kipf and Welling}{2016}]%
        {kipf2016semi}
\bibfield{author}{\bibinfo{person}{Thomas~N Kipf} {and} \bibinfo{person}{Max
  Welling}.} \bibinfo{year}{2016}\natexlab{}.
\newblock \bibinfo{title}{Semi-supervised classification with graph
  convolutional networks}.
\newblock \bibinfo{howpublished}{arXiv preprint arXiv:1609.02907}.
\newblock


\bibitem[\protect\citeauthoryear{Koenigstein, Dror, and Koren}{Koenigstein
  et~al\mbox{.}}{2011}]%
        {koenigstein2011yahoo}
\bibfield{author}{\bibinfo{person}{Noam Koenigstein}, \bibinfo{person}{Gideon
  Dror}, {and} \bibinfo{person}{Yehuda Koren}.}
  \bibinfo{year}{2011}\natexlab{}.
\newblock \showarticletitle{Yahoo! music recommendations: modeling music
  ratings with temporal dynamics and item taxonomy}. In
  \bibinfo{booktitle}{\emph{Proceedings of the 5th ACM Conference on
  Recommender Systems}}. New York: ACM Press, \bibinfo{pages}{165--172}.
\newblock


\bibitem[\protect\citeauthoryear{Koren}{Koren}{2008}]%
        {koren2008factorization}
\bibfield{author}{\bibinfo{person}{Yehuda Koren}.}
  \bibinfo{year}{2008}\natexlab{}.
\newblock \showarticletitle{Factorization meets the neighborhood: a
  multifaceted collaborative filtering model}. In
  \bibinfo{booktitle}{\emph{Proceedings of the 14th ACM SIGKDD International
  Conference on Knowledge Discovery and Data Mining}}. New York: ACM Press,
  \bibinfo{pages}{426--434}.
\newblock


\bibitem[\protect\citeauthoryear{Koren}{Koren}{2009}]%
        {koren2009collaborative}
\bibfield{author}{\bibinfo{person}{Yehuda Koren}.}
  \bibinfo{year}{2009}\natexlab{}.
\newblock \showarticletitle{Collaborative filtering with temporal dynamics}. In
  \bibinfo{booktitle}{\emph{Proceedings of the 15th ACM SIGKDD International
  Conference on Knowledge Discovery and Data Mining}}. New York: ACM Press,
  \bibinfo{pages}{447--456}.
\newblock


\bibitem[\protect\citeauthoryear{Koren, Bell, and Volinsky}{Koren
  et~al\mbox{.}}{2009}]%
        {koren2009matrix}
\bibfield{author}{\bibinfo{person}{Yehuda Koren}, \bibinfo{person}{Robert
  Bell}, {and} \bibinfo{person}{Chris Volinsky}.}
  \bibinfo{year}{2009}\natexlab{}.
\newblock \showarticletitle{Matrix factorization techniques for recommender
  systems}.
\newblock \bibinfo{journal}{\emph{Computer}}  \bibinfo{volume}{8}
  (\bibinfo{year}{2009}), \bibinfo{pages}{30--37}.
\newblock


\bibitem[\protect\citeauthoryear{Krizhevsky, Sutskever, and Hinton}{Krizhevsky
  et~al\mbox{.}}{2012}]%
        {krizhevsky2012imagenet}
\bibfield{author}{\bibinfo{person}{Alex Krizhevsky}, \bibinfo{person}{Ilya
  Sutskever}, {and} \bibinfo{person}{Geoffrey~E Hinton}.}
  \bibinfo{year}{2012}\natexlab{}.
\newblock \showarticletitle{Imagenet classification with deep convolutional
  neural networks}. In \bibinfo{booktitle}{\emph{Advances in Neural Information
  Processing Systems}}. Harrahs and Harveys, Lake Tahoe,
  \bibinfo{pages}{1097--1105}.
\newblock


\bibitem[\protect\citeauthoryear{Le and Mikolov}{Le and Mikolov}{2014}]%
        {le2014distributed}
\bibfield{author}{\bibinfo{person}{Quoc Le} {and} \bibinfo{person}{Tomas
  Mikolov}.} \bibinfo{year}{2014}\natexlab{}.
\newblock \showarticletitle{Distributed representations of sentences and
  documents}. In \bibinfo{booktitle}{\emph{International Conference on Machine
  Learning}}. Beijing China: JMLR.org, \bibinfo{pages}{1188--1196}.
\newblock


\bibitem[\protect\citeauthoryear{Leal, Veloso, Malheiro,
  Gonz{\'a}lez-V{\'e}lez, and Burguillo}{Leal et~al\mbox{.}}{2019}]%
        {leal2019scalable}
\bibfield{author}{\bibinfo{person}{F{\'a}tima Leal}, \bibinfo{person}{Bruno~M
  Veloso}, \bibinfo{person}{Benedita Malheiro}, \bibinfo{person}{Horacio
  Gonz{\'a}lez-V{\'e}lez}, {and} \bibinfo{person}{Juan~Carlos Burguillo}.}
  \bibinfo{year}{2019}\natexlab{}.
\newblock \showarticletitle{Scalable modelling and recommendation using
  wiki-based crowdsourced repositories}.
\newblock \bibinfo{journal}{\emph{Electronic Commerce Research and
  Applications}}  \bibinfo{volume}{33} (\bibinfo{year}{2019}),
  \bibinfo{pages}{100817}.
\newblock


\bibitem[\protect\citeauthoryear{LeCun and Bengio}{LeCun and Bengio}{1995}]%
        {lecun1995convolutional}
\bibfield{author}{\bibinfo{person}{Yann LeCun} {and} \bibinfo{person}{Yoshua
  Bengio}.} \bibinfo{year}{1995}\natexlab{}.
\newblock \bibinfo{title}{Convolutional networks for images, speech, and time
  series}.
\newblock \bibinfo{howpublished}{Cambridge, MA: MIT Press}.
\newblock


\bibitem[\protect\citeauthoryear{Lei, Liu, Li, Zha, and Li}{Lei
  et~al\mbox{.}}{2016}]%
        {lei2016comparative}
\bibfield{author}{\bibinfo{person}{Chenyi Lei}, \bibinfo{person}{Dong Liu},
  \bibinfo{person}{Weiping Li}, \bibinfo{person}{Zheng-Jun Zha}, {and}
  \bibinfo{person}{Houqiang Li}.} \bibinfo{year}{2016}\natexlab{}.
\newblock \showarticletitle{Comparative deep learning of hybrid representations
  for image recommendations}. In \bibinfo{booktitle}{\emph{Proceedings of the
  IEEE Conference on Computer Vision and Pattern Recognition}}. Washington, DC:
  IEEE Computer Society Press, \bibinfo{pages}{2545--2553}.
\newblock


\bibitem[\protect\citeauthoryear{Li, Ge, Hong, and Zhu}{Li
  et~al\mbox{.}}{2016}]%
        {li2016point}
\bibfield{author}{\bibinfo{person}{Huayu Li}, \bibinfo{person}{Yong Ge},
  \bibinfo{person}{Richang Hong}, {and} \bibinfo{person}{Hengshu Zhu}.}
  \bibinfo{year}{2016}\natexlab{}.
\newblock \showarticletitle{Point-of-interest recommendations: Learning
  potential check-ins from friends}. In \bibinfo{booktitle}{\emph{Proceedings
  of the 22th ACM SIGKDD International Conference on Knowledge Discovery and
  Data Mining}}. New York: ACM Press, \bibinfo{pages}{975--984}.
\newblock


\bibitem[\protect\citeauthoryear{Li, Chu, Langford, and Schapire}{Li
  et~al\mbox{.}}{2010}]%
        {li2010contextual}
\bibfield{author}{\bibinfo{person}{Lihong Li}, \bibinfo{person}{Wei Chu},
  \bibinfo{person}{John Langford}, {and} \bibinfo{person}{Robert~E Schapire}.}
  \bibinfo{year}{2010}\natexlab{}.
\newblock \showarticletitle{A contextual-bandit approach to personalized news
  article recommendation}. In \bibinfo{booktitle}{\emph{Proceedings of the 2010
  World Wide Web Conference on World Wide Web}}. New York: ACM Press,
  \bibinfo{pages}{661--670}.
\newblock


\bibitem[\protect\citeauthoryear{Li, Chu, Langford, and Wang}{Li
  et~al\mbox{.}}{2011}]%
        {li2011unbiased}
\bibfield{author}{\bibinfo{person}{Lihong Li}, \bibinfo{person}{Wei Chu},
  \bibinfo{person}{John Langford}, {and} \bibinfo{person}{Xuanhui Wang}.}
  \bibinfo{year}{2011}\natexlab{}.
\newblock \showarticletitle{Unbiased offline evaluation of
  contextual-bandit-based news article recommendation algorithms}. In
  \bibinfo{booktitle}{\emph{Proceedings of the 4th ACM International Conference
  on Web Search and Data Mining}}. New York: ACM Press,
  \bibinfo{pages}{297--306}.
\newblock


\bibitem[\protect\citeauthoryear{Li, Wang, Ren, Bing, and Lam}{Li
  et~al\mbox{.}}{2017}]%
        {li2017neural}
\bibfield{author}{\bibinfo{person}{Piji Li}, \bibinfo{person}{Zihao Wang},
  \bibinfo{person}{Zhaochun Ren}, \bibinfo{person}{Lidong Bing}, {and}
  \bibinfo{person}{Wai Lam}.} \bibinfo{year}{2017}\natexlab{}.
\newblock \showarticletitle{Neural rating regression with abstractive tips
  generation for recommendation}. In \bibinfo{booktitle}{\emph{Proceedings of
  the 40th International ACM SIGIR Conference on Research and Development in
  Information Retrieval}}. New York: ACM Press, \bibinfo{pages}{345--354}.
\newblock


\bibitem[\protect\citeauthoryear{Lian, Zhao, Xie, Sun, Chen, and Rui}{Lian
  et~al\mbox{.}}{2014}]%
        {lian2014geomf}
\bibfield{author}{\bibinfo{person}{Defu Lian}, \bibinfo{person}{Cong Zhao},
  \bibinfo{person}{Xing Xie}, \bibinfo{person}{Guangzhong Sun},
  \bibinfo{person}{Enhong Chen}, {and} \bibinfo{person}{Yong Rui}.}
  \bibinfo{year}{2014}\natexlab{}.
\newblock \showarticletitle{GeoMF: joint geographical modeling and matrix
  factorization for point-of-interest recommendation}. In
  \bibinfo{booktitle}{\emph{Proceedings of the 20th ACM SIGKDD International
  Conference on Knowledge Discovery and Data Mining}}. New York: ACM Press,
  \bibinfo{pages}{831--840}.
\newblock


\bibitem[\protect\citeauthoryear{Liang, Altosaar, Charlin, and Blei}{Liang
  et~al\mbox{.}}{2016}]%
        {liang2016factorization}
\bibfield{author}{\bibinfo{person}{Dawen Liang}, \bibinfo{person}{Jaan
  Altosaar}, \bibinfo{person}{Laurent Charlin}, {and} \bibinfo{person}{David~M
  Blei}.} \bibinfo{year}{2016}\natexlab{}.
\newblock \showarticletitle{Factorization meets the item embedding:
  Regularizing matrix factorization with item co-occurrence}. In
  \bibinfo{booktitle}{\emph{Proceedings of the 11th ACM Conference on
  Recommender Systems}}. New York: ACM Press, \bibinfo{pages}{59--66}.
\newblock


\bibitem[\protect\citeauthoryear{Lin, Liu, Sun, Liu, and Zhu}{Lin
  et~al\mbox{.}}{2015}]%
        {lin2015learning}
\bibfield{author}{\bibinfo{person}{Yankai Lin}, \bibinfo{person}{Zhiyuan Liu},
  \bibinfo{person}{Maosong Sun}, \bibinfo{person}{Yang Liu}, {and}
  \bibinfo{person}{Xuan Zhu}.} \bibinfo{year}{2015}\natexlab{}.
\newblock \showarticletitle{Learning entity and relation embeddings for
  knowledge graph completion}. In \bibinfo{booktitle}{\emph{Proceedings of the
  29th AAAI Conference on Artificial Intelligence}}, Vol.~\bibinfo{volume}{15}.
  Menlo Park, CA: AAAI Press, \bibinfo{pages}{2181--2187}.
\newblock


\bibitem[\protect\citeauthoryear{Linden, Smith, and York}{Linden
  et~al\mbox{.}}{2003}]%
        {linden2003amazon}
\bibfield{author}{\bibinfo{person}{Greg Linden}, \bibinfo{person}{Brent Smith},
  {and} \bibinfo{person}{Jeremy York}.} \bibinfo{year}{2003}\natexlab{}.
\newblock \showarticletitle{Amazon.com recommendations: Item-to-item
  collaborative filtering}.
\newblock \bibinfo{journal}{\emph{IEEE Internet Computing}}
  \bibinfo{volume}{1}, \bibinfo{number}{1} (\bibinfo{year}{2003}),
  \bibinfo{pages}{76--80}.
\newblock


\bibitem[\protect\citeauthoryear{Lippert, Weber, Huang, Tresp, Schubert, and
  Kriegel}{Lippert et~al\mbox{.}}{2008}]%
        {lippert2008relation}
\bibfield{author}{\bibinfo{person}{Christoph Lippert},
  \bibinfo{person}{Stefan~Hagen Weber}, \bibinfo{person}{Yi Huang},
  \bibinfo{person}{Volker Tresp}, \bibinfo{person}{Matthias Schubert}, {and}
  \bibinfo{person}{Hans-Peter Kriegel}.} \bibinfo{year}{2008}\natexlab{}.
\newblock \showarticletitle{Relation prediction in multi-relational domains
  using matrix factorization}. In \bibinfo{booktitle}{\emph{Proceedings of the
  NIPS 2008 Workshop: Structured Input-Structured Output, Vancouver, Canada}}.
  \bibinfo{pages}{1--4}.
\newblock


\bibitem[\protect\citeauthoryear{Liu, Wu, and Wang}{Liu et~al\mbox{.}}{2017}]%
        {liu2017deepstyle}
\bibfield{author}{\bibinfo{person}{Qiang Liu}, \bibinfo{person}{Shu Wu}, {and}
  \bibinfo{person}{Liang Wang}.} \bibinfo{year}{2017}\natexlab{}.
\newblock \showarticletitle{DeepStyle: Learning user preferences for visual
  recommendation}. In \bibinfo{booktitle}{\emph{Proceedings of the 40th
  International ACM SIGIR conference on Research and Development in Information
  Retrieval}}. New York: ACM Press, \bibinfo{pages}{841--844}.
\newblock


\bibitem[\protect\citeauthoryear{Liu, Wu, Wang, and Tan}{Liu
  et~al\mbox{.}}{2016b}]%
        {liu2016predicting}
\bibfield{author}{\bibinfo{person}{Qiang Liu}, \bibinfo{person}{Shu Wu},
  \bibinfo{person}{Liang Wang}, {and} \bibinfo{person}{Tieniu Tan}.}
  \bibinfo{year}{2016}\natexlab{b}.
\newblock \showarticletitle{Predicting the Next Location: A Recurrent Model
  with Spatial and Temporal Contexts}. In \bibinfo{booktitle}{\emph{Proceedings
  of the 30th AAAI Conference on Artificial Intelligence}}. Menlo Park, CA:
  AAAI Press, \bibinfo{pages}{194--200}.
\newblock


\bibitem[\protect\citeauthoryear{Liu, Liu, Aberer, and Miao}{Liu
  et~al\mbox{.}}{2013}]%
        {liu2013personalized}
\bibfield{author}{\bibinfo{person}{Xin Liu}, \bibinfo{person}{Yong Liu},
  \bibinfo{person}{Karl Aberer}, {and} \bibinfo{person}{Chunyan Miao}.}
  \bibinfo{year}{2013}\natexlab{}.
\newblock \showarticletitle{Personalized point-of-interest recommendation by
  mining users' preference transition}. In
  \bibinfo{booktitle}{\emph{Proceedings of the 22nd ACM International
  Conference on Information \& Knowledge Management}}. New York: ACM Press,
  \bibinfo{pages}{733--738}.
\newblock


\bibitem[\protect\citeauthoryear{Liu, Liu, and Li}{Liu et~al\mbox{.}}{2016a}]%
        {liu2016exploring}
\bibfield{author}{\bibinfo{person}{Xin Liu}, \bibinfo{person}{Yong Liu}, {and}
  \bibinfo{person}{Xiaoli Li}.} \bibinfo{year}{2016}\natexlab{a}.
\newblock \showarticletitle{Exploring the Context of Locations for Personalized
  Location Recommendations}. In \bibinfo{booktitle}{\emph{International Joint
  Conference on Artificial Intelligence}}. Menlo Park: AAAI Press,
  \bibinfo{pages}{1188--1194}.
\newblock


\bibitem[\protect\citeauthoryear{Lops, De~Gemmis, and Semeraro}{Lops
  et~al\mbox{.}}{2011}]%
        {lops2011content}
\bibfield{author}{\bibinfo{person}{Pasquale Lops}, \bibinfo{person}{Marco
  De~Gemmis}, {and} \bibinfo{person}{Giovanni Semeraro}.}
  \bibinfo{year}{2011}\natexlab{}.
\newblock \showarticletitle{Content-based recommender systems: State of the art
  and trends}.
\newblock In \bibinfo{booktitle}{\emph{Recommender Systems Handbook}}.
  \bibinfo{publisher}{Boston MA: Springer}, \bibinfo{pages}{73--105}.
\newblock


\bibitem[\protect\citeauthoryear{Lu, Dong, and Smyth}{Lu et~al\mbox{.}}{2018}]%
        {lu2018like}
\bibfield{author}{\bibinfo{person}{Yichao Lu}, \bibinfo{person}{Ruihai Dong},
  {and} \bibinfo{person}{Barry Smyth}.} \bibinfo{year}{2018}\natexlab{}.
\newblock \showarticletitle{Why I like it: multi-task learning for
  recommendation and explanation}. In \bibinfo{booktitle}{\emph{Proceedings of
  the 13th ACM Conference on Recommender Systems}}. New York: ACM Press,
  \bibinfo{pages}{4--12}.
\newblock


\bibitem[\protect\citeauthoryear{Luo, Pang, Wang, and Lin}{Luo
  et~al\mbox{.}}{2014}]%
        {luo2014hete}
\bibfield{author}{\bibinfo{person}{Chen Luo}, \bibinfo{person}{Wei Pang},
  \bibinfo{person}{Zhe Wang}, {and} \bibinfo{person}{Chenghua Lin}.}
  \bibinfo{year}{2014}\natexlab{}.
\newblock \showarticletitle{Hete-cf: Social-based collaborative filtering
  recommendation using heterogeneous relations}. In
  \bibinfo{booktitle}{\emph{2014 IEEE International Conference on Data
  Mining}}. Washington, DC: IEEE Computer Society Press,
  \bibinfo{pages}{917--922}.
\newblock


\bibitem[\protect\citeauthoryear{Luong, Pham, and Manning}{Luong
  et~al\mbox{.}}{2015}]%
        {luong2015effective}
\bibfield{author}{\bibinfo{person}{Minh-Thang Luong}, \bibinfo{person}{Hieu
  Pham}, {and} \bibinfo{person}{Christopher~D Manning}.}
  \bibinfo{year}{2015}\natexlab{}.
\newblock \bibinfo{title}{Effective approaches to attention-based neural
  machine translation}.
\newblock \bibinfo{howpublished}{arXiv preprint arXiv:1508.04025}.
\newblock


\bibitem[\protect\citeauthoryear{Ma}{Ma}{2013}]%
        {ma2013experimental}
\bibfield{author}{\bibinfo{person}{Hao Ma}.} \bibinfo{year}{2013}\natexlab{}.
\newblock \showarticletitle{An experimental study on implicit social
  recommendation}. In \bibinfo{booktitle}{\emph{Proceedings of the 36th
  International ACM SIGIR conference on Research and Development in Information
  Retrieval}}. New York: ACM Press, \bibinfo{pages}{73--82}.
\newblock


\bibitem[\protect\citeauthoryear{Ma, King, and Lyu}{Ma et~al\mbox{.}}{2009a}]%
        {ma2009learning}
\bibfield{author}{\bibinfo{person}{Hao Ma}, \bibinfo{person}{Irwin King}, {and}
  \bibinfo{person}{Michael~R Lyu}.} \bibinfo{year}{2009}\natexlab{a}.
\newblock \showarticletitle{Learning to recommend with social trust ensemble}.
  In \bibinfo{booktitle}{\emph{Proceedings of the 32nd International ACM SIGIR
  conference on Research and Development in Information Retrieval}}. New York:
  ACM Press, \bibinfo{pages}{203--210}.
\newblock


\bibitem[\protect\citeauthoryear{Ma, King, and Lyu}{Ma et~al\mbox{.}}{2011a}]%
        {ma2011learning}
\bibfield{author}{\bibinfo{person}{Hao Ma}, \bibinfo{person}{Irwin King}, {and}
  \bibinfo{person}{Michael~R Lyu}.} \bibinfo{year}{2011}\natexlab{a}.
\newblock \showarticletitle{Learning to recommend with explicit and implicit
  social relations}.
\newblock \bibinfo{journal}{\emph{ACM Transactions on Intelligent Systems and
  Technology (TIST)}} \bibinfo{volume}{2}, \bibinfo{number}{3}
  (\bibinfo{year}{2011}), \bibinfo{pages}{29}.
\newblock


\bibitem[\protect\citeauthoryear{Ma, Lyu, and King}{Ma et~al\mbox{.}}{2009b}]%
        {ma2009distrust}
\bibfield{author}{\bibinfo{person}{Hao Ma}, \bibinfo{person}{Michael~R Lyu},
  {and} \bibinfo{person}{Irwin King}.} \bibinfo{year}{2009}\natexlab{b}.
\newblock \showarticletitle{Learning to recommend with trust and distrust
  relationships}. In \bibinfo{booktitle}{\emph{Proceedings of the 4th ACM
  Conference on Recommender Systems}}. New York: ACM Press,
  \bibinfo{pages}{189--196}.
\newblock


\bibitem[\protect\citeauthoryear{Ma, Yang, Lyu, and King}{Ma
  et~al\mbox{.}}{2008}]%
        {ma2008sorec}
\bibfield{author}{\bibinfo{person}{Hao Ma}, \bibinfo{person}{Haixuan Yang},
  \bibinfo{person}{Michael~R Lyu}, {and} \bibinfo{person}{Irwin King}.}
  \bibinfo{year}{2008}\natexlab{}.
\newblock \showarticletitle{Sorec: social recommendation using probabilistic
  matrix factorization}. In \bibinfo{booktitle}{\emph{Proceedings of the 17th
  ACM International Conference on Information and Knowledge Management}}. New
  York: ACM Press, \bibinfo{pages}{931--940}.
\newblock


\bibitem[\protect\citeauthoryear{Ma, Zhou, Liu, Lyu, and King}{Ma
  et~al\mbox{.}}{2011b}]%
        {ma2011recommender}
\bibfield{author}{\bibinfo{person}{Hao Ma}, \bibinfo{person}{Dengyong Zhou},
  \bibinfo{person}{Chao Liu}, \bibinfo{person}{Michael~R Lyu}, {and}
  \bibinfo{person}{Irwin King}.} \bibinfo{year}{2011}\natexlab{b}.
\newblock \showarticletitle{Recommender systems with social regularization}. In
  \bibinfo{booktitle}{\emph{Proceedings of the 4th ACM International Conference
  on Web Search and Data Mining}}. New York: ACM Press,
  \bibinfo{pages}{287--296}.
\newblock


\bibitem[\protect\citeauthoryear{Masci, Meier, Cire{\c{s}}an, and
  Schmidhuber}{Masci et~al\mbox{.}}{2011}]%
        {masci2011stacked}
\bibfield{author}{\bibinfo{person}{Jonathan Masci}, \bibinfo{person}{Ueli
  Meier}, \bibinfo{person}{Dan Cire{\c{s}}an}, {and}
  \bibinfo{person}{J{\"u}rgen Schmidhuber}.} \bibinfo{year}{2011}\natexlab{}.
\newblock \showarticletitle{Stacked convolutional auto-encoders for
  hierarchical feature extraction}. In \bibinfo{booktitle}{\emph{International
  Conference on Artificial Neural Networks}}. Berlin-Heidelberg, Germany:
  Springer, \bibinfo{pages}{52--59}.
\newblock


\bibitem[\protect\citeauthoryear{McAuley and Leskovec}{McAuley and
  Leskovec}{2013}]%
        {mcauley2013hidden}
\bibfield{author}{\bibinfo{person}{Julian McAuley} {and} \bibinfo{person}{Jure
  Leskovec}.} \bibinfo{year}{2013}\natexlab{}.
\newblock \showarticletitle{Hidden factors and hidden topics: understanding
  rating dimensions with review text}. In \bibinfo{booktitle}{\emph{Proceedings
  of the 8th ACM Conference on Recommender Systems}}. New York: ACM Press,
  \bibinfo{pages}{165--172}.
\newblock


\bibitem[\protect\citeauthoryear{McAuley, Targett, Shi, and Van
  Den~Hengel}{McAuley et~al\mbox{.}}{2015}]%
        {mcauley2015image}
\bibfield{author}{\bibinfo{person}{Julian McAuley},
  \bibinfo{person}{Christopher Targett}, \bibinfo{person}{Qinfeng Shi}, {and}
  \bibinfo{person}{Anton Van Den~Hengel}.} \bibinfo{year}{2015}\natexlab{}.
\newblock \showarticletitle{Image-based recommendations on styles and
  substitutes}. In \bibinfo{booktitle}{\emph{Proceedings of the 38th
  International ACM SIGIR conference on Research and Development in Information
  Retrieval}}. New York: ACM Press, \bibinfo{pages}{43--52}.
\newblock


\bibitem[\protect\citeauthoryear{Melville, Mooney, and Nagarajan}{Melville
  et~al\mbox{.}}{2002}]%
        {melville2002content}
\bibfield{author}{\bibinfo{person}{Prem Melville}, \bibinfo{person}{Raymond~J
  Mooney}, {and} \bibinfo{person}{Ramadass Nagarajan}.}
  \bibinfo{year}{2002}\natexlab{}.
\newblock \showarticletitle{Content-boosted collaborative filtering for
  improved recommendations}.
\newblock \bibinfo{journal}{\emph{Proceedings of the 16th AAAI Conference on
  Artificial Intelligence}}  \bibinfo{volume}{23} (\bibinfo{year}{2002}),
  \bibinfo{pages}{187--192}.
\newblock


\bibitem[\protect\citeauthoryear{Menon, Chitrapura, Garg, Agarwal, and
  Kota}{Menon et~al\mbox{.}}{2011}]%
        {menon2011response}
\bibfield{author}{\bibinfo{person}{Aditya~Krishna Menon},
  \bibinfo{person}{Krishna-Prasad Chitrapura}, \bibinfo{person}{Sachin Garg},
  \bibinfo{person}{Deepak Agarwal}, {and} \bibinfo{person}{Nagaraj Kota}.}
  \bibinfo{year}{2011}\natexlab{}.
\newblock \showarticletitle{Response prediction using collaborative filtering
  with hierarchies and side-information}. In
  \bibinfo{booktitle}{\emph{Proceedings of the 17th ACM SIGKDD International
  Conference on Knowledge Discovery and Data Mining}}. New York: ACM Press,
  \bibinfo{pages}{141--149}.
\newblock


\bibitem[\protect\citeauthoryear{Mikolov, Chen, Corrado, and Dean}{Mikolov
  et~al\mbox{.}}{2013}]%
        {mikolov2013efficient}
\bibfield{author}{\bibinfo{person}{Tomas Mikolov}, \bibinfo{person}{Kai Chen},
  \bibinfo{person}{Greg Corrado}, {and} \bibinfo{person}{Jeffrey Dean}.}
  \bibinfo{year}{2013}\natexlab{}.
\newblock \bibinfo{title}{Efficient estimation of word representations in
  vector space}.
\newblock \bibinfo{howpublished}{arXiv preprint arXiv:1301.3781}.
\newblock


\bibitem[\protect\citeauthoryear{Mnih}{Mnih}{2011}]%
        {mnih2011taxonomy}
\bibfield{author}{\bibinfo{person}{Andriy Mnih}.}
  \bibinfo{year}{2011}\natexlab{}.
\newblock \showarticletitle{Taxonomy-informed latent factor models for implicit
  feedback}. In \bibinfo{booktitle}{\emph{Proceedings of the 2011 International
  Conference on KDD Cup 2011, 18}}. JMLR.org, \bibinfo{pages}{169--181}.
\newblock


\bibitem[\protect\citeauthoryear{Mnih and Salakhutdinov}{Mnih and
  Salakhutdinov}{2008}]%
        {mnih2008probabilistic}
\bibfield{author}{\bibinfo{person}{Andriy Mnih} {and} \bibinfo{person}{Ruslan~R
  Salakhutdinov}.} \bibinfo{year}{2008}\natexlab{}.
\newblock \showarticletitle{Probabilistic matrix factorization}. In
  \bibinfo{booktitle}{\emph{Advances in Neural Information Processing
  Systems}}. Vancouver, B.C., Canada, \bibinfo{pages}{1257--1264}.
\newblock


\bibitem[\protect\citeauthoryear{Mnih and Teh}{Mnih and Teh}{2012}]%
        {mnih2012learning}
\bibfield{author}{\bibinfo{person}{Andriy Mnih} {and} \bibinfo{person}{Yee~W
  Teh}.} \bibinfo{year}{2012}\natexlab{}.
\newblock \showarticletitle{Learning label trees for probabilistic modelling of
  implicit feedback}. In \bibinfo{booktitle}{\emph{Advances in Neural
  Information Processing Systems}}. Harrahs and Harveys, Lake Tahoe,
  \bibinfo{pages}{2816--2824}.
\newblock


\bibitem[\protect\citeauthoryear{Morschheuser, Hamari, and
  Koivisto}{Morschheuser et~al\mbox{.}}{2016}]%
        {morschheuser2016gamification}
\bibfield{author}{\bibinfo{person}{Benedikt Morschheuser},
  \bibinfo{person}{Juho Hamari}, {and} \bibinfo{person}{Jonna Koivisto}.}
  \bibinfo{year}{2016}\natexlab{}.
\newblock \showarticletitle{Gamification in crowdsourcing: a review}. In
  \bibinfo{booktitle}{\emph{The 49th Hawaii International Conference on System
  Sciences (HICSS)}}. Washington, DC: IEEE Computer Society Press,
  \bibinfo{pages}{4375--4384}.
\newblock


\bibitem[\protect\citeauthoryear{Nair and Hinton}{Nair and Hinton}{2010}]%
        {nair2010rectified}
\bibfield{author}{\bibinfo{person}{Vinod Nair} {and}
  \bibinfo{person}{Geoffrey~E Hinton}.} \bibinfo{year}{2010}\natexlab{}.
\newblock \showarticletitle{Rectified linear units improve restricted boltzmann
  machines}. In \bibinfo{booktitle}{\emph{International Conference on Machine
  Learning}}. Haifa, Israel: JMLR.org, \bibinfo{pages}{807--814}.
\newblock


\bibitem[\protect\citeauthoryear{Niepert, Ahmed, and Kutzkov}{Niepert
  et~al\mbox{.}}{2016}]%
        {niepert2016learning}
\bibfield{author}{\bibinfo{person}{Mathias Niepert}, \bibinfo{person}{Mohamed
  Ahmed}, {and} \bibinfo{person}{Konstantin Kutzkov}.}
  \bibinfo{year}{2016}\natexlab{}.
\newblock \showarticletitle{Learning convolutional neural networks for graphs}.
  In \bibinfo{booktitle}{\emph{International Conference on Machine Learning}}.
  New York City, NY: JMLR.org, \bibinfo{pages}{2014--2023}.
\newblock


\bibitem[\protect\citeauthoryear{Niu, Caverlee, and Lu}{Niu
  et~al\mbox{.}}{2018}]%
        {niu2018neural}
\bibfield{author}{\bibinfo{person}{Wei Niu}, \bibinfo{person}{James Caverlee},
  {and} \bibinfo{person}{Haokai Lu}.} \bibinfo{year}{2018}\natexlab{}.
\newblock \showarticletitle{Neural Personalized Ranking for Image
  Recommendation}. In \bibinfo{booktitle}{\emph{Proceedings of the 11th ACM
  International Conference on Web Search and Data Mining}}. New York: ACM
  Press, \bibinfo{pages}{423--431}.
\newblock


\bibitem[\protect\citeauthoryear{Okura, Tagami, Ono, and Tajima}{Okura
  et~al\mbox{.}}{2017}]%
        {okura2017embedding}
\bibfield{author}{\bibinfo{person}{Shumpei Okura}, \bibinfo{person}{Yukihiro
  Tagami}, \bibinfo{person}{Shingo Ono}, {and} \bibinfo{person}{Akira Tajima}.}
  \bibinfo{year}{2017}\natexlab{}.
\newblock \showarticletitle{Embedding-based news recommendation for millions of
  users}. In \bibinfo{booktitle}{\emph{Proceedings of the 23rd ACM SIGKDD
  International Conference on Knowledge Discovery and Data Mining}}. New York:
  ACM Press, \bibinfo{pages}{1933--1942}.
\newblock


\bibitem[\protect\citeauthoryear{Ostapuk, Yang, and Cudre-Mauroux}{Ostapuk
  et~al\mbox{.}}{2019}]%
        {ostapuk2019activelink}
\bibfield{author}{\bibinfo{person}{Natalia Ostapuk}, \bibinfo{person}{Jie
  Yang}, {and} \bibinfo{person}{Philippe Cudre-Mauroux}.}
  \bibinfo{year}{2019}\natexlab{}.
\newblock \showarticletitle{ActiveLink: Deep Active Learning for Link
  Prediction in Knowledge Graphs}. In \bibinfo{booktitle}{\emph{Proceedings of
  the 2019 World Wide Web Conference on World Wide Web}}. New York: ACM Press,
  \bibinfo{pages}{1398--1408}.
\newblock


\bibitem[\protect\citeauthoryear{Pappas and Popescu-Belis}{Pappas and
  Popescu-Belis}{2013}]%
        {pappas2013sentiment}
\bibfield{author}{\bibinfo{person}{Nikolaos Pappas} {and}
  \bibinfo{person}{Andrei Popescu-Belis}.} \bibinfo{year}{2013}\natexlab{}.
\newblock \showarticletitle{Sentiment analysis of user comments for one-class
  collaborative filtering over ted talks}. In
  \bibinfo{booktitle}{\emph{Proceedings of the 36th International ACM SIGIR
  conference on Research and Development in Information Retrieval}}. New York:
  ACM Press, \bibinfo{pages}{773--776}.
\newblock


\bibitem[\protect\citeauthoryear{Park, Pennock, Madani, Good, and DeCoste}{Park
  et~al\mbox{.}}{2006}]%
        {park2006naive}
\bibfield{author}{\bibinfo{person}{Seung-Taek Park}, \bibinfo{person}{David
  Pennock}, \bibinfo{person}{Omid Madani}, \bibinfo{person}{Nathan Good}, {and}
  \bibinfo{person}{Dennis DeCoste}.} \bibinfo{year}{2006}\natexlab{}.
\newblock \showarticletitle{Na{\"\i}ve filterbots for robust cold-start
  recommendations}. In \bibinfo{booktitle}{\emph{Proceedings of the 12th ACM
  SIGKDD International Conference on Knowledge Discovery and Data Mining}}. New
  York: ACM Press, \bibinfo{pages}{699--705}.
\newblock


\bibitem[\protect\citeauthoryear{Pasricha and McAuley}{Pasricha and
  McAuley}{2018}]%
        {pasricha2018translation}
\bibfield{author}{\bibinfo{person}{Rajiv Pasricha} {and}
  \bibinfo{person}{Julian McAuley}.} \bibinfo{year}{2018}\natexlab{}.
\newblock \showarticletitle{Translation-based factorization machines for
  sequential recommendation}. In \bibinfo{booktitle}{\emph{Proceedings of the
  13th ACM Conference on Recommender Systems}}. New York: ACM Press,
  \bibinfo{pages}{63--71}.
\newblock


\bibitem[\protect\citeauthoryear{Paterek}{Paterek}{2007}]%
        {paterek2007improving}
\bibfield{author}{\bibinfo{person}{Arkadiusz Paterek}.}
  \bibinfo{year}{2007}\natexlab{}.
\newblock \showarticletitle{Improving regularized singular value decomposition
  for collaborative filtering}. In \bibinfo{booktitle}{\emph{Proceedings of KDD
  Cup and Workshop}}, Vol.~\bibinfo{volume}{2007}. New York: ACM Press,
  \bibinfo{pages}{5--8}.
\newblock


\bibitem[\protect\citeauthoryear{Pei, Baltrusaitis, Tax, and Morency}{Pei
  et~al\mbox{.}}{2017a}]%
        {pei2017temporal}
\bibfield{author}{\bibinfo{person}{Wenjie Pei}, \bibinfo{person}{Tadas
  Baltrusaitis}, \bibinfo{person}{David~MJ Tax}, {and}
  \bibinfo{person}{Louis-Philippe Morency}.} \bibinfo{year}{2017}\natexlab{a}.
\newblock \showarticletitle{Temporal attention-gated model for robust sequence
  classification}. In \bibinfo{booktitle}{\emph{Proceedings of the IEEE
  Conference on Computer Vision and Pattern Recognition}}. Washington, DC: IEEE
  Computer Society Press, \bibinfo{pages}{6730--6739}.
\newblock


\bibitem[\protect\citeauthoryear{Pei, Yang, Sun, Zhang, Bozzon, and Tax}{Pei
  et~al\mbox{.}}{2017b}]%
        {pei2017interacting}
\bibfield{author}{\bibinfo{person}{Wenjie Pei}, \bibinfo{person}{Jie Yang},
  \bibinfo{person}{Zhu Sun}, \bibinfo{person}{Jie Zhang},
  \bibinfo{person}{Alessandro Bozzon}, {and} \bibinfo{person}{David~MJ Tax}.}
  \bibinfo{year}{2017}\natexlab{b}.
\newblock \showarticletitle{Interacting attention-gated recurrent networks for
  recommendation}. In \bibinfo{booktitle}{\emph{Proceedings of the 2017 ACM on
  Conference on Information and Knowledge Management}}. New York: ACM Press,
  \bibinfo{pages}{1459--1468}.
\newblock


\bibitem[\protect\citeauthoryear{Pero and Horv{\'a}th}{Pero and
  Horv{\'a}th}{2013}]%
        {pero2013opinion}
\bibfield{author}{\bibinfo{person}{{\v{S}}tefan Pero} {and}
  \bibinfo{person}{Tom{\'a}{\v{s}} Horv{\'a}th}.}
  \bibinfo{year}{2013}\natexlab{}.
\newblock \showarticletitle{Opinion-driven matrix factorization for rating
  prediction}. In \bibinfo{booktitle}{\emph{International Conference on User
  Modeling, Adaptation, and Personalization}}. Berlin-Heidelberg, Germany:
  Springer, \bibinfo{pages}{1--13}.
\newblock


\bibitem[\protect\citeauthoryear{Pourgholamali, Kahani, Bagheri, and
  Noorian}{Pourgholamali et~al\mbox{.}}{2017}]%
        {pourgholamali2017embedding}
\bibfield{author}{\bibinfo{person}{Fatemeh Pourgholamali},
  \bibinfo{person}{Mohsen Kahani}, \bibinfo{person}{Ebrahim Bagheri}, {and}
  \bibinfo{person}{Zeinab Noorian}.} \bibinfo{year}{2017}\natexlab{}.
\newblock \showarticletitle{Embedding unstructured side information in product
  recommendation}.
\newblock \bibinfo{journal}{\emph{Electronic Commerce Research and
  Applications}}  \bibinfo{volume}{25} (\bibinfo{year}{2017}),
  \bibinfo{pages}{70--85}.
\newblock


\bibitem[\protect\citeauthoryear{Ramos et~al\mbox{.}}{Ramos
  et~al\mbox{.}}{2003}]%
        {ramos2003using}
\bibfield{author}{\bibinfo{person}{Juan Ramos} {et~al\mbox{.}}}
  \bibinfo{year}{2003}\natexlab{}.
\newblock \showarticletitle{Using tf-idf to determine word relevance in
  document queries}. In \bibinfo{booktitle}{\emph{Proceedings of the 1st
  Instructional Conference on Machine Learning}}, Vol.~\bibinfo{volume}{242}.
  Piscataway, NJ, \bibinfo{pages}{133--142}.
\newblock


\bibitem[\protect\citeauthoryear{Rendle}{Rendle}{2010}]%
        {rendle2010factorization}
\bibfield{author}{\bibinfo{person}{Steffen Rendle}.}
  \bibinfo{year}{2010}\natexlab{}.
\newblock \showarticletitle{Factorization machines}. In
  \bibinfo{booktitle}{\emph{2010 IEEE International Conference on Data
  Mining}}. Washington, DC: IEEE Computer Society Press,
  \bibinfo{pages}{995--1000}.
\newblock


\bibitem[\protect\citeauthoryear{Rendle}{Rendle}{2012}]%
        {rendle2012factorization}
\bibfield{author}{\bibinfo{person}{Steffen Rendle}.}
  \bibinfo{year}{2012}\natexlab{}.
\newblock \showarticletitle{Factorization machines with libfm}.
\newblock \bibinfo{journal}{\emph{ACM Transactions on Intelligent Systems and
  Technology (TIST)}} \bibinfo{volume}{3}, \bibinfo{number}{3}
  (\bibinfo{year}{2012}), \bibinfo{pages}{57}.
\newblock


\bibitem[\protect\citeauthoryear{Rendle, Freudenthaler, Gantner, and
  Schmidt-Thieme}{Rendle et~al\mbox{.}}{2009}]%
        {rendle2009bpr}
\bibfield{author}{\bibinfo{person}{Steffen Rendle}, \bibinfo{person}{Christoph
  Freudenthaler}, \bibinfo{person}{Zeno Gantner}, {and} \bibinfo{person}{Lars
  Schmidt-Thieme}.} \bibinfo{year}{2009}\natexlab{}.
\newblock \showarticletitle{BPR: Bayesian personalized ranking from implicit
  feedback}. In \bibinfo{booktitle}{\emph{Proceedings of the 25th Conference on
  Uncertainty in Artificial Intelligence}}. Montreal, Canada: AUAI Press,
  \bibinfo{pages}{452--461}.
\newblock


\bibitem[\protect\citeauthoryear{Rendle, Freudenthaler, and
  Schmidt-Thieme}{Rendle et~al\mbox{.}}{2010}]%
        {rendle2010factorizing}
\bibfield{author}{\bibinfo{person}{Steffen Rendle}, \bibinfo{person}{Christoph
  Freudenthaler}, {and} \bibinfo{person}{Lars Schmidt-Thieme}.}
  \bibinfo{year}{2010}\natexlab{}.
\newblock \showarticletitle{Factorizing personalized markov chains for
  next-basket recommendation}. In \bibinfo{booktitle}{\emph{Proceedings of the
  2010 World Wide Web Conference on World Wide Web}}. New York: ACM Press,
  \bibinfo{pages}{811--820}.
\newblock


\bibitem[\protect\citeauthoryear{Ricci, Rokach, and Shapira}{Ricci
  et~al\mbox{.}}{2015}]%
        {ricci2015recommender}
\bibfield{author}{\bibinfo{person}{Francesco Ricci}, \bibinfo{person}{Lior
  Rokach}, {and} \bibinfo{person}{Bracha Shapira}.}
  \bibinfo{year}{2015}\natexlab{}.
\newblock \showarticletitle{Recommender systems: introduction and challenges}.
\newblock In \bibinfo{booktitle}{\emph{Recommender Systems Handbook}}.
  \bibinfo{publisher}{Boston MA: Springer}, \bibinfo{pages}{1--34}.
\newblock


\bibitem[\protect\citeauthoryear{Rumelhart, Hinton, and Williams}{Rumelhart
  et~al\mbox{.}}{1985}]%
        {rumelhart1985learning}
\bibfield{author}{\bibinfo{person}{David~E Rumelhart},
  \bibinfo{person}{Geoffrey~E Hinton}, {and} \bibinfo{person}{Ronald~J
  Williams}.} \bibinfo{year}{1985}\natexlab{}.
\newblock \bibinfo{booktitle}{\emph{Learning internal representations by error
  propagation}}.
\newblock \bibinfo{type}{{T}echnical {R}eport}. \bibinfo{institution}{La Jolla
  Institute for Cognitive Science, University of California, San Diego}.
\newblock


\bibitem[\protect\citeauthoryear{Sarwar, Karypis, Konstan, and Riedl}{Sarwar
  et~al\mbox{.}}{2001}]%
        {sarwar2001item}
\bibfield{author}{\bibinfo{person}{Badrul Sarwar}, \bibinfo{person}{George
  Karypis}, \bibinfo{person}{Joseph Konstan}, {and} \bibinfo{person}{John
  Riedl}.} \bibinfo{year}{2001}\natexlab{}.
\newblock \showarticletitle{Item-based collaborative filtering recommendation
  algorithms}. In \bibinfo{booktitle}{\emph{Proceedings of the 2001 World Wide
  Web Conference on World Wide Web}}. New York: ACM Press,
  \bibinfo{pages}{285--295}.
\newblock


\bibitem[\protect\citeauthoryear{Scarselli, Gori, Tsoi, Hagenbuchner, and
  Monfardini}{Scarselli et~al\mbox{.}}{2008}]%
        {scarselli2008graph}
\bibfield{author}{\bibinfo{person}{Franco Scarselli}, \bibinfo{person}{Marco
  Gori}, \bibinfo{person}{Ah~Chung Tsoi}, \bibinfo{person}{Markus
  Hagenbuchner}, {and} \bibinfo{person}{Gabriele Monfardini}.}
  \bibinfo{year}{2008}\natexlab{}.
\newblock \showarticletitle{The graph neural network model}.
\newblock \bibinfo{journal}{\emph{IEEE Transactions on Neural Networks}}
  \bibinfo{volume}{20}, \bibinfo{number}{1} (\bibinfo{year}{2008}),
  \bibinfo{pages}{61--80}.
\newblock


\bibitem[\protect\citeauthoryear{Schafer, Frankowski, Herlocker, and
  Sen}{Schafer et~al\mbox{.}}{2007}]%
        {schafer2007collaborative}
\bibfield{author}{\bibinfo{person}{J~Ben Schafer}, \bibinfo{person}{Dan
  Frankowski}, \bibinfo{person}{Jon Herlocker}, {and} \bibinfo{person}{Shilad
  Sen}.} \bibinfo{year}{2007}\natexlab{}.
\newblock \showarticletitle{Collaborative filtering recommender systems}.
\newblock In \bibinfo{booktitle}{\emph{The Adaptive Web}}.
  \bibinfo{publisher}{Berlin-Heidelberg, Germany: Springer},
  \bibinfo{pages}{291--324}.
\newblock


\bibitem[\protect\citeauthoryear{Schmidhuber}{Schmidhuber}{2015}]%
        {schmidhuber2015deep}
\bibfield{author}{\bibinfo{person}{J{\"u}rgen Schmidhuber}.}
  \bibinfo{year}{2015}\natexlab{}.
\newblock \showarticletitle{Deep learning in neural networks: An overview}.
\newblock \bibinfo{journal}{\emph{Neural Networks}}  \bibinfo{volume}{61}
  (\bibinfo{year}{2015}), \bibinfo{pages}{85--117}.
\newblock


\bibitem[\protect\citeauthoryear{Sedhain, Menon, Sanner, and Xie}{Sedhain
  et~al\mbox{.}}{2015}]%
        {sedhain2015autorec}
\bibfield{author}{\bibinfo{person}{Suvash Sedhain},
  \bibinfo{person}{Aditya~Krishna Menon}, \bibinfo{person}{Scott Sanner}, {and}
  \bibinfo{person}{Lexing Xie}.} \bibinfo{year}{2015}\natexlab{}.
\newblock \showarticletitle{Autorec: Autoencoders meet collaborative
  filtering}. In \bibinfo{booktitle}{\emph{Proceedings of the 2015 World Wide
  Web Conference on World Wide Web}}. New York: ACM Press,
  \bibinfo{pages}{111--112}.
\newblock


\bibitem[\protect\citeauthoryear{Seo, Huang, Yang, and Liu}{Seo
  et~al\mbox{.}}{2017}]%
        {seo2017interpretable}
\bibfield{author}{\bibinfo{person}{Sungyong Seo}, \bibinfo{person}{Jing Huang},
  \bibinfo{person}{Hao Yang}, {and} \bibinfo{person}{Yan Liu}.}
  \bibinfo{year}{2017}\natexlab{}.
\newblock \showarticletitle{Interpretable convolutional neural networks with
  dual local and global attention for review rating prediction}. In
  \bibinfo{booktitle}{\emph{Proceedings of the 12th ACM Conference on
  Recommender Systems}}. New York: ACM Press, \bibinfo{pages}{297--305}.
\newblock


\bibitem[\protect\citeauthoryear{Sharma, Reddy, Kiran, and Ragunathan}{Sharma
  et~al\mbox{.}}{2011}]%
        {sharma2011improving}
\bibfield{author}{\bibinfo{person}{Mohak Sharma}, \bibinfo{person}{P~Krishna
  Reddy}, \bibinfo{person}{R~Uday Kiran}, {and} \bibinfo{person}{Thirumalaisamy
  Ragunathan}.} \bibinfo{year}{2011}\natexlab{}.
\newblock \showarticletitle{Improving the performance of recommender system by
  exploiting the categories of products}. In
  \bibinfo{booktitle}{\emph{International Workshop on Databases in Networked
  Information Systems}}. Berlin-Heidelberg, Germany: Springer,
  \bibinfo{pages}{137--146}.
\newblock


\bibitem[\protect\citeauthoryear{Shi, Liu, Zhuang, Philip, and Wu}{Shi
  et~al\mbox{.}}{2016}]%
        {shi2016integrating}
\bibfield{author}{\bibinfo{person}{Chuan Shi}, \bibinfo{person}{Jian Liu},
  \bibinfo{person}{Fuzhen Zhuang}, \bibinfo{person}{S~Yu Philip}, {and}
  \bibinfo{person}{Bin Wu}.} \bibinfo{year}{2016}\natexlab{}.
\newblock \showarticletitle{Integrating heterogeneous information via flexible
  regularization framework for recommendation}.
\newblock \bibinfo{journal}{\emph{Knowledge and Information Systems}}
  \bibinfo{volume}{49}, \bibinfo{number}{3} (\bibinfo{year}{2016}),
  \bibinfo{pages}{835--859}.
\newblock


\bibitem[\protect\citeauthoryear{Shi, Zhang, Luo, Yu, Yue, and Wu}{Shi
  et~al\mbox{.}}{2015}]%
        {shi2015semantic}
\bibfield{author}{\bibinfo{person}{Chuan Shi}, \bibinfo{person}{Zhiqiang
  Zhang}, \bibinfo{person}{Ping Luo}, \bibinfo{person}{Philip~S Yu},
  \bibinfo{person}{Yading Yue}, {and} \bibinfo{person}{Bin Wu}.}
  \bibinfo{year}{2015}\natexlab{}.
\newblock \showarticletitle{Semantic path based personalized recommendation on
  weighted heterogeneous information networks}. In
  \bibinfo{booktitle}{\emph{Proceedings of the 24th ACM International
  Conference on Information and Knowledge Management}}. New York: ACM Press,
  \bibinfo{pages}{453--462}.
\newblock


\bibitem[\protect\citeauthoryear{Shi, Larson, and Hanjalic}{Shi
  et~al\mbox{.}}{2011}]%
        {shi2011tags}
\bibfield{author}{\bibinfo{person}{Yue Shi}, \bibinfo{person}{Martha Larson},
  {and} \bibinfo{person}{Alan Hanjalic}.} \bibinfo{year}{2011}\natexlab{}.
\newblock \showarticletitle{Tags as bridges between domains: Improving
  recommendation with tag-induced cross-domain collaborative filtering}. In
  \bibinfo{booktitle}{\emph{International Conference on User Modeling,
  Adaptation, and Personalization}}. Berlin-Heidelberg, Germany: Springer,
  \bibinfo{pages}{305--316}.
\newblock


\bibitem[\protect\citeauthoryear{Shi, Larson, and Hanjalic}{Shi
  et~al\mbox{.}}{2014}]%
        {shi2014collaborative}
\bibfield{author}{\bibinfo{person}{Yue Shi}, \bibinfo{person}{Martha Larson},
  {and} \bibinfo{person}{Alan Hanjalic}.} \bibinfo{year}{2014}\natexlab{}.
\newblock \showarticletitle{Collaborative filtering beyond the user-item
  matrix: A survey of the state of the art and future challenges}.
\newblock \bibinfo{journal}{\emph{ACM Computing Surveys (CSUR)}}
  \bibinfo{volume}{47}, \bibinfo{number}{1} (\bibinfo{year}{2014}),
  \bibinfo{pages}{3}.
\newblock


\bibitem[\protect\citeauthoryear{Singh and Gordon}{Singh and Gordon}{2008}]%
        {singh2008relational}
\bibfield{author}{\bibinfo{person}{Ajit~P Singh} {and}
  \bibinfo{person}{Geoffrey~J Gordon}.} \bibinfo{year}{2008}\natexlab{}.
\newblock \showarticletitle{Relational learning via collective matrix
  factorization}. In \bibinfo{booktitle}{\emph{Proceedings of the 14th ACM
  SIGKDD International Conference on Knowledge Discovery and Data Mining}}. New
  York: ACM Press, \bibinfo{pages}{650--658}.
\newblock


\bibitem[\protect\citeauthoryear{Smola and Kondor}{Smola and Kondor}{2003}]%
        {smola2003kernels}
\bibfield{author}{\bibinfo{person}{Alexander~J Smola} {and}
  \bibinfo{person}{Risi Kondor}.} \bibinfo{year}{2003}\natexlab{}.
\newblock \showarticletitle{Kernels and regularization on graphs}.
\newblock In \bibinfo{booktitle}{\emph{Learning Theory and Kernel Machines}}.
  \bibinfo{publisher}{Berlin-Heidelberg, Germany: Springer},
  \bibinfo{pages}{144--158}.
\newblock


\bibitem[\protect\citeauthoryear{Socher, Lin, Manning, and Ng}{Socher
  et~al\mbox{.}}{2011}]%
        {socher2011parsing}
\bibfield{author}{\bibinfo{person}{Richard Socher}, \bibinfo{person}{Cliff~C
  Lin}, \bibinfo{person}{Chris Manning}, {and} \bibinfo{person}{Andrew~Y Ng}.}
  \bibinfo{year}{2011}\natexlab{}.
\newblock \showarticletitle{Parsing natural scenes and natural language with
  recursive neural networks}. In \bibinfo{booktitle}{\emph{International
  Conference on Machine Learning}}. Bellevue, Washington: JMLR.org,
  \bibinfo{pages}{129--136}.
\newblock


\bibitem[\protect\citeauthoryear{Song, Dixon, and Pearce}{Song
  et~al\mbox{.}}{2012}]%
        {song2012survey}
\bibfield{author}{\bibinfo{person}{Yading Song}, \bibinfo{person}{Simon Dixon},
  {and} \bibinfo{person}{Marcus Pearce}.} \bibinfo{year}{2012}\natexlab{}.
\newblock \showarticletitle{A survey of music recommendation systems and future
  perspectives}. In \bibinfo{booktitle}{\emph{9th International Symposium on
  Computer Music Modeling and Retrieval}}, Vol.~\bibinfo{volume}{4}. London,
  UK, \bibinfo{pages}{395--410}.
\newblock


\bibitem[\protect\citeauthoryear{Su and Khoshgoftaar}{Su and
  Khoshgoftaar}{2009}]%
        {su2009survey}
\bibfield{author}{\bibinfo{person}{Xiaoyuan Su} {and} \bibinfo{person}{Taghi~M
  Khoshgoftaar}.} \bibinfo{year}{2009}\natexlab{}.
\newblock \showarticletitle{A survey of collaborative filtering techniques}.
\newblock \bibinfo{journal}{\emph{Advances in Artificial Intelligence}}
  \bibinfo{volume}{2009}, \bibinfo{number}{1} (\bibinfo{year}{2009}),
  \bibinfo{pages}{395--410}.
\newblock


\bibitem[\protect\citeauthoryear{Sun, Han, Yan, Yu, and Wu}{Sun
  et~al\mbox{.}}{2011}]%
        {sun2011pathsim}
\bibfield{author}{\bibinfo{person}{Yizhou Sun}, \bibinfo{person}{Jiawei Han},
  \bibinfo{person}{Xifeng Yan}, \bibinfo{person}{Philip~S Yu}, {and}
  \bibinfo{person}{Tianyi Wu}.} \bibinfo{year}{2011}\natexlab{}.
\newblock \showarticletitle{Pathsim: Meta path-based top-k similarity search in
  heterogeneous information networks}.
\newblock \bibinfo{journal}{\emph{Proceedings of the VLDB Endowment}}
  \bibinfo{volume}{4}, \bibinfo{number}{11} (\bibinfo{year}{2011}),
  \bibinfo{pages}{992--1003}.
\newblock


\bibitem[\protect\citeauthoryear{Sun}{Sun}{2015}]%
        {sun2015exploiting}
\bibfield{author}{\bibinfo{person}{Zhu Sun}.} \bibinfo{year}{2015}\natexlab{}.
\newblock \showarticletitle{Exploiting Item and User Relationships for
  Recommender Systems}. In \bibinfo{booktitle}{\emph{International Conference
  on User Modeling, Adaptation, and Personalization}}. Berlin-Heidelberg,
  Germany: Springer, \bibinfo{pages}{397--402}.
\newblock


\bibitem[\protect\citeauthoryear{Sun, Guo, and Zhang}{Sun
  et~al\mbox{.}}{2016}]%
        {sun2016effective}
\bibfield{author}{\bibinfo{person}{Zhu Sun}, \bibinfo{person}{Guibing Guo},
  {and} \bibinfo{person}{Jie Zhang}.} \bibinfo{year}{2016}\natexlab{}.
\newblock \showarticletitle{Effective recommendation with category hierarchy}.
  In \bibinfo{booktitle}{\emph{Proceedings of the 2016 Conference on User
  Modeling Adaptation and Personalization}}. New York: ACM Press,
  \bibinfo{pages}{299--300}.
\newblock


\bibitem[\protect\citeauthoryear{Sun, Guo, Zhang, and Xu}{Sun
  et~al\mbox{.}}{2017a}]%
        {sun2017unified}
\bibfield{author}{\bibinfo{person}{Zhu Sun}, \bibinfo{person}{Guibing Guo},
  \bibinfo{person}{Jie Zhang}, {and} \bibinfo{person}{Chi Xu}.}
  \bibinfo{year}{2017}\natexlab{a}.
\newblock \showarticletitle{A Unified Latent Factor Model for Effective
  Category-Aware Recommendation}. In \bibinfo{booktitle}{\emph{International
  Conference on User Modeling, Adaptation, and Personalization}}. New York: ACM
  Press, \bibinfo{pages}{389--390}.
\newblock


\bibitem[\protect\citeauthoryear{Sun, Yang, Zhang, and Bozzon}{Sun
  et~al\mbox{.}}{2017b}]%
        {sun2017exploiting}
\bibfield{author}{\bibinfo{person}{Zhu Sun}, \bibinfo{person}{Jie Yang},
  \bibinfo{person}{Jie Zhang}, {and} \bibinfo{person}{Alessandro Bozzon}.}
  \bibinfo{year}{2017}\natexlab{b}.
\newblock \showarticletitle{Exploiting both vertical and horizontal dimensions
  of feature hierarchy for effective recommendation}. In
  \bibinfo{booktitle}{\emph{Proceedings of the 31st AAAI Conference on
  Artificial Intelligence}}. Menlo Park, CA: AAAI Press,
  \bibinfo{pages}{189--195}.
\newblock


\bibitem[\protect\citeauthoryear{Sun, Yang, Zhang, Bozzon, Chen, and Xu}{Sun
  et~al\mbox{.}}{2017c}]%
        {sun2017mrlr}
\bibfield{author}{\bibinfo{person}{Zhu Sun}, \bibinfo{person}{Jie Yang},
  \bibinfo{person}{Jie Zhang}, \bibinfo{person}{Alessandro Bozzon},
  \bibinfo{person}{Yu Chen}, {and} \bibinfo{person}{Chi Xu}.}
  \bibinfo{year}{2017}\natexlab{c}.
\newblock \showarticletitle{MRLR: Multi-level Representation Learning for
  Personalized Ranking in Recommendation}. In
  \bibinfo{booktitle}{\emph{International Joint Conference on Artificial
  Intelligence}}. Menlo Park: AAAI Press, \bibinfo{pages}{2807--2813}.
\newblock


\bibitem[\protect\citeauthoryear{Sun, Yang, Zhang, Bozzon, Huang, and Xu}{Sun
  et~al\mbox{.}}{2018}]%
        {sun2018rkge}
\bibfield{author}{\bibinfo{person}{Zhu Sun}, \bibinfo{person}{Jie Yang},
  \bibinfo{person}{Jie Zhang}, \bibinfo{person}{Alessandro Bozzon},
  \bibinfo{person}{Long-Kai Huang}, {and} \bibinfo{person}{Chi Xu}.}
  \bibinfo{year}{2018}\natexlab{}.
\newblock \showarticletitle{Recurrent Knowledge Graph Embedding for Effective
  Recommendation}. In \bibinfo{booktitle}{\emph{Proceedings of the 13th ACM
  Conference on Recommender Systems}}. \bibinfo{pages}{297--305}.
\newblock


\bibitem[\protect\citeauthoryear{Symeonidis, Nanopoulos, and
  Manolopoulos}{Symeonidis et~al\mbox{.}}{2010}]%
        {symeonidis2010unified}
\bibfield{author}{\bibinfo{person}{Panagiotis Symeonidis},
  \bibinfo{person}{Alexandros Nanopoulos}, {and} \bibinfo{person}{Yannis
  Manolopoulos}.} \bibinfo{year}{2010}\natexlab{}.
\newblock \showarticletitle{A unified framework for providing recommendations
  in social tagging systems based on ternary semantic analysis}.
\newblock \bibinfo{journal}{\emph{IEEE Transactions on Knowledge and Data
  Engineering}} \bibinfo{volume}{22}, \bibinfo{number}{2}
  (\bibinfo{year}{2010}), \bibinfo{pages}{179--192}.
\newblock


\bibitem[\protect\citeauthoryear{Tang and Wang}{Tang and Wang}{2018}]%
        {tang2018personalized}
\bibfield{author}{\bibinfo{person}{Jiaxi Tang} {and} \bibinfo{person}{Ke
  Wang}.} \bibinfo{year}{2018}\natexlab{}.
\newblock \showarticletitle{Personalized top-n sequential recommendation via
  convolutional sequence embedding}. In \bibinfo{booktitle}{\emph{Proceedings
  of the 11th ACM International Conference on Web Search and Data Mining}}. New
  York: ACM Press, \bibinfo{pages}{565--573}.
\newblock


\bibitem[\protect\citeauthoryear{Tay, Luu, and Hui}{Tay et~al\mbox{.}}{2018}]%
        {tay2018multi}
\bibfield{author}{\bibinfo{person}{Yi Tay}, \bibinfo{person}{Anh~Tuan Luu},
  {and} \bibinfo{person}{Siu~Cheung Hui}.} \bibinfo{year}{2018}\natexlab{}.
\newblock \showarticletitle{Multi-pointer co-attention networks for
  recommendation}. In \bibinfo{booktitle}{\emph{Proceedings of the 24th ACM
  SIGKDD International Conference on Knowledge Discovery \& Data Mining}}. New
  York: ACM Press, \bibinfo{pages}{2309--2318}.
\newblock


\bibitem[\protect\citeauthoryear{Terzi, Rowe, Ferrario, and Whittle}{Terzi
  et~al\mbox{.}}{2014}]%
        {terzi2014text}
\bibfield{author}{\bibinfo{person}{Maria Terzi}, \bibinfo{person}{Matthew
  Rowe}, \bibinfo{person}{Maria-Angela Ferrario}, {and} \bibinfo{person}{Jon
  Whittle}.} \bibinfo{year}{2014}\natexlab{}.
\newblock \showarticletitle{Text-based user-knn: Measuring user similarity
  based on text reviews}. In \bibinfo{booktitle}{\emph{International Conference
  on User Modeling, Adaptation, and Personalization}}. Berlin-Heidelberg,
  Germany: Springer, \bibinfo{pages}{195--206}.
\newblock


\bibitem[\protect\citeauthoryear{Tuan and Phuong}{Tuan and Phuong}{2017}]%
        {tuan20173d}
\bibfield{author}{\bibinfo{person}{Trinh~Xuan Tuan} {and}
  \bibinfo{person}{Tu~Minh Phuong}.} \bibinfo{year}{2017}\natexlab{}.
\newblock \showarticletitle{3D convolutional networks for session-based
  recommendation with content features}. In
  \bibinfo{booktitle}{\emph{Proceedings of the 12th ACM Conference on
  Recommender Systems}}. New York: ACM Press, \bibinfo{pages}{138--146}.
\newblock


\bibitem[\protect\citeauthoryear{Vasile, Smirnova, and Conneau}{Vasile
  et~al\mbox{.}}{2016}]%
        {vasile2016meta}
\bibfield{author}{\bibinfo{person}{Flavian Vasile}, \bibinfo{person}{Elena
  Smirnova}, {and} \bibinfo{person}{Alexis Conneau}.}
  \bibinfo{year}{2016}\natexlab{}.
\newblock \showarticletitle{Meta-prod2vec: Product embeddings using
  side-information for recommendation}. In
  \bibinfo{booktitle}{\emph{Proceedings of the 11th ACM Conference on
  Recommender Systems}}. New York: ACM Press, \bibinfo{pages}{225--232}.
\newblock


\bibitem[\protect\citeauthoryear{Vaswani, Shazeer, Parmar, Uszkoreit, Jones,
  Gomez, Kaiser, and Polosukhin}{Vaswani et~al\mbox{.}}{2017}]%
        {vaswani2017attention}
\bibfield{author}{\bibinfo{person}{Ashish Vaswani}, \bibinfo{person}{Noam
  Shazeer}, \bibinfo{person}{Niki Parmar}, \bibinfo{person}{Jakob Uszkoreit},
  \bibinfo{person}{Llion Jones}, \bibinfo{person}{Aidan~N Gomez},
  \bibinfo{person}{{\L}ukasz Kaiser}, {and} \bibinfo{person}{Illia
  Polosukhin}.} \bibinfo{year}{2017}\natexlab{}.
\newblock \showarticletitle{Attention is all you need}. In
  \bibinfo{booktitle}{\emph{Advances in Neural Information Processing
  Systems}}. Long Beach, California, \bibinfo{pages}{5998--6008}.
\newblock


\bibitem[\protect\citeauthoryear{Veli{\v{c}}kovi{\'c}, Cucurull, Casanova,
  Romero, Lio, and Bengio}{Veli{\v{c}}kovi{\'c} et~al\mbox{.}}{2017}]%
        {velivckovic2017graph}
\bibfield{author}{\bibinfo{person}{Petar Veli{\v{c}}kovi{\'c}},
  \bibinfo{person}{Guillem Cucurull}, \bibinfo{person}{Arantxa Casanova},
  \bibinfo{person}{Adriana Romero}, \bibinfo{person}{Pietro Lio}, {and}
  \bibinfo{person}{On~the Properties of Neural Machine Translation:
  Encoder-Decoder~Approaches Bengio}.} \bibinfo{year}{2017}\natexlab{}.
\newblock \bibinfo{title}{Graph attention networks}.
\newblock \bibinfo{howpublished}{arXiv preprint arXiv:1710.10903}.
\newblock


\bibitem[\protect\citeauthoryear{Vincent, Larochelle, Bengio, and
  Manzagol}{Vincent et~al\mbox{.}}{2008}]%
        {vincent2008extracting}
\bibfield{author}{\bibinfo{person}{Pascal Vincent}, \bibinfo{person}{Hugo
  Larochelle}, \bibinfo{person}{Yoshua Bengio}, {and}
  \bibinfo{person}{Pierre-Antoine Manzagol}.} \bibinfo{year}{2008}\natexlab{}.
\newblock \showarticletitle{Extracting and composing robust features with
  denoising autoencoders}. In \bibinfo{booktitle}{\emph{International
  Conference on Machine Learning}}. Helsinki, Finland: JMLR.org,
  \bibinfo{pages}{1096--1103}.
\newblock


\bibitem[\protect\citeauthoryear{Vincent, Larochelle, Lajoie, Bengio, and
  Manzagol}{Vincent et~al\mbox{.}}{2010}]%
        {vincent2010stacked}
\bibfield{author}{\bibinfo{person}{Pascal Vincent}, \bibinfo{person}{Hugo
  Larochelle}, \bibinfo{person}{Isabelle Lajoie}, \bibinfo{person}{Yoshua
  Bengio}, {and} \bibinfo{person}{Pierre-Antoine Manzagol}.}
  \bibinfo{year}{2010}\natexlab{}.
\newblock \showarticletitle{Stacked denoising autoencoders: Learning useful
  representations in a deep network with a local denoising criterion}.
\newblock \bibinfo{journal}{\emph{Journal of Machine Learning Research}}
  \bibinfo{volume}{11}, \bibinfo{number}{Dec} (\bibinfo{year}{2010}),
  \bibinfo{pages}{3371--3408}.
\newblock


\bibitem[\protect\citeauthoryear{Wang, Fu, Wang, Yin, Du, and Xiong}{Wang
  et~al\mbox{.}}{2017a}]%
        {wang2017location}
\bibfield{author}{\bibinfo{person}{Hao Wang}, \bibinfo{person}{Yanmei Fu},
  \bibinfo{person}{Qinyong Wang}, \bibinfo{person}{Hongzhi Yin},
  \bibinfo{person}{Changying Du}, {and} \bibinfo{person}{Hui Xiong}.}
  \bibinfo{year}{2017}\natexlab{a}.
\newblock \showarticletitle{A location-sentiment-aware recommender system for
  both home-town and out-of-town users}. In
  \bibinfo{booktitle}{\emph{Proceedings of the 23rd ACM SIGKDD International
  Conference on Knowledge Discovery and Data Mining}}. New York: ACM Press,
  \bibinfo{pages}{1135--1143}.
\newblock


\bibitem[\protect\citeauthoryear{Wang, Wang, and Yeung}{Wang
  et~al\mbox{.}}{2015c}]%
        {wang2015collaborative}
\bibfield{author}{\bibinfo{person}{Hao Wang}, \bibinfo{person}{Naiyan Wang},
  {and} \bibinfo{person}{Dit-Yan Yeung}.} \bibinfo{year}{2015}\natexlab{c}.
\newblock \showarticletitle{Collaborative deep learning for recommender
  systems}. In \bibinfo{booktitle}{\emph{Proceedings of the 21rd ACM SIGKDD
  International Conference on Knowledge Discovery and Data Mining}}. New York:
  ACM Press, \bibinfo{pages}{1235--1244}.
\newblock


\bibitem[\protect\citeauthoryear{Wang, Zhang, Wang, Zhao, Li, Xie, and
  Guo}{Wang et~al\mbox{.}}{2018b}]%
        {wang2018ripplenet}
\bibfield{author}{\bibinfo{person}{Hongwei Wang}, \bibinfo{person}{Fuzheng
  Zhang}, \bibinfo{person}{Jialin Wang}, \bibinfo{person}{Miao Zhao},
  \bibinfo{person}{Wenjie Li}, \bibinfo{person}{Xing Xie}, {and}
  \bibinfo{person}{Minyi Guo}.} \bibinfo{year}{2018}\natexlab{b}.
\newblock \showarticletitle{RippleNet: Propagating user preferences on the
  knowledge graph for recommender systems}. In
  \bibinfo{booktitle}{\emph{Proceedings of the 27th ACM International
  Conference on Information and Knowledge Management}}. New York: ACM Press,
  \bibinfo{pages}{417--426}.
\newblock


\bibitem[\protect\citeauthoryear{Wang, Zhang, Wang, Zhao, Li, Xie, and
  Guo}{Wang et~al\mbox{.}}{019a}]%
        {wang2019exploring}
\bibfield{author}{\bibinfo{person}{Hongwei Wang}, \bibinfo{person}{Fuzheng
  Zhang}, \bibinfo{person}{Jialin Wang}, \bibinfo{person}{Miao Zhao},
  \bibinfo{person}{Wenjie Li}, \bibinfo{person}{Xing Xie}, {and}
  \bibinfo{person}{Minyi Guo}.} \bibinfo{year}{2019a}\natexlab{}.
\newblock \showarticletitle{Exploring High-Order User Preference on the
  Knowledge Graph for Recommender Systems}.
\newblock \bibinfo{journal}{\emph{ACM Transactions on Information Systems
  (TOIS)}} \bibinfo{volume}{37}, \bibinfo{number}{3} (\bibinfo{year}{2019a}),
  \bibinfo{pages}{32}.
\newblock


\bibitem[\protect\citeauthoryear{Wang, Zhang, Xie, and Guo}{Wang
  et~al\mbox{.}}{2018c}]%
        {wang2018dkn}
\bibfield{author}{\bibinfo{person}{Hongwei Wang}, \bibinfo{person}{Fuzheng
  Zhang}, \bibinfo{person}{Xing Xie}, {and} \bibinfo{person}{Minyi Guo}.}
  \bibinfo{year}{2018}\natexlab{c}.
\newblock \showarticletitle{DKN: Deep knowledge-aware network for news
  recommendation}. In \bibinfo{booktitle}{\emph{Proceedings of the 2018 World
  Wide Web Conference on World Wide Web}}. New York: ACM Press,
  \bibinfo{pages}{1835--1844}.
\newblock


\bibitem[\protect\citeauthoryear{Wang, Zhang, Zhang, Leskovec, Zhao, Li, and
  Wang}{Wang et~al\mbox{.}}{2019b}]%
        {wang2019knowledge}
\bibfield{author}{\bibinfo{person}{Hongwei Wang}, \bibinfo{person}{Fuzheng
  Zhang}, \bibinfo{person}{Mengdi Zhang}, \bibinfo{person}{Jure Leskovec},
  \bibinfo{person}{Miao Zhao}, \bibinfo{person}{Wenjie Li}, {and}
  \bibinfo{person}{Zhongyuan Wang}.} \bibinfo{year}{2019}\natexlab{b}.
\newblock \bibinfo{title}{Knowledge Graph Convolutional Networks for
  Recommender Systems with Label Smoothness Regularization}.
\newblock \bibinfo{howpublished}{arXiv preprint arXiv:1905.04413}.
\newblock


\bibitem[\protect\citeauthoryear{Wang, Zhang, Zhao, Li, Xie, and Guo}{Wang
  et~al\mbox{.}}{2019c}]%
        {wang2019multi}
\bibfield{author}{\bibinfo{person}{Hongwei Wang}, \bibinfo{person}{Fuzheng
  Zhang}, \bibinfo{person}{Miao Zhao}, \bibinfo{person}{Wenjie Li},
  \bibinfo{person}{Xing Xie}, {and} \bibinfo{person}{Minyi Guo}.}
  \bibinfo{year}{2019}\natexlab{c}.
\newblock \bibinfo{title}{Multi-Task Feature Learning for Knowledge Graph
  Enhanced Recommendation}.
\newblock \bibinfo{howpublished}{arXiv preprint arXiv:1901.08907}.
\newblock


\bibitem[\protect\citeauthoryear{Wang, Zhao, Xie, Li, and Guo}{Wang
  et~al\mbox{.}}{2019d}]%
        {wang2019knowledgegraph}
\bibfield{author}{\bibinfo{person}{Hongwei Wang}, \bibinfo{person}{Miao Zhao},
  \bibinfo{person}{Xing Xie}, \bibinfo{person}{Wenjie Li}, {and}
  \bibinfo{person}{Minyi Guo}.} \bibinfo{year}{2019}\natexlab{d}.
\newblock \bibinfo{title}{Knowledge graph convolutional networks for
  recommender systems}.
\newblock \bibinfo{howpublished}{arXiv preprint arXiv:1904.12575}.
\newblock


\bibitem[\protect\citeauthoryear{Wang, Yu, Zhang, Gong, Xu, Wang, Zhang, and
  Zhang}{Wang et~al\mbox{.}}{2017e}]%
        {wang2017irgan}
\bibfield{author}{\bibinfo{person}{Jun Wang}, \bibinfo{person}{Lantao Yu},
  \bibinfo{person}{Weinan Zhang}, \bibinfo{person}{Yu Gong},
  \bibinfo{person}{Yinghui Xu}, \bibinfo{person}{Benyou Wang},
  \bibinfo{person}{Peng Zhang}, {and} \bibinfo{person}{Dell Zhang}.}
  \bibinfo{year}{2017}\natexlab{e}.
\newblock \showarticletitle{Irgan: A minimax game for unifying generative and
  discriminative information retrieval models}. In
  \bibinfo{booktitle}{\emph{Proceedings of the 40th International ACM SIGIR
  conference on Research and Development in Information Retrieval}}. New York:
  ACM Press, \bibinfo{pages}{515--524}.
\newblock


\bibitem[\protect\citeauthoryear{Wang, Guo, Lan, Xu, Wan, and Cheng}{Wang
  et~al\mbox{.}}{2015a}]%
        {wang2015learning}
\bibfield{author}{\bibinfo{person}{Pengfei Wang}, \bibinfo{person}{Jiafeng
  Guo}, \bibinfo{person}{Yanyan Lan}, \bibinfo{person}{Jun Xu},
  \bibinfo{person}{Shengxian Wan}, {and} \bibinfo{person}{Xueqi Cheng}.}
  \bibinfo{year}{2015}\natexlab{a}.
\newblock \showarticletitle{Learning hierarchical representation model for
  nextbasket recommendation}. In \bibinfo{booktitle}{\emph{Proceedings of the
  38th International ACM SIGIR conference on Research and Development in
  Information Retrieval}}. New York: ACM Press, \bibinfo{pages}{403--412}.
\newblock


\bibitem[\protect\citeauthoryear{Wang, Mao, Wang, and Guo}{Wang
  et~al\mbox{.}}{2017b}]%
        {wang2017knowledge}
\bibfield{author}{\bibinfo{person}{Quan Wang}, \bibinfo{person}{Zhendong Mao},
  \bibinfo{person}{Bin Wang}, {and} \bibinfo{person}{Li Guo}.}
  \bibinfo{year}{2017}\natexlab{b}.
\newblock \showarticletitle{Knowledge graph embedding: A survey of approaches
  and applications}.
\newblock \bibinfo{journal}{\emph{IEEE Transactions on Knowledge and Data
  Engineering}} \bibinfo{volume}{29}, \bibinfo{number}{12}
  (\bibinfo{year}{2017}), \bibinfo{pages}{2724--2743}.
\newblock


\bibitem[\protect\citeauthoryear{Wang, Tang, Wang, and Liu}{Wang
  et~al\mbox{.}}{2015b}]%
        {wang2015exploring}
\bibfield{author}{\bibinfo{person}{Suhang Wang}, \bibinfo{person}{Jiliang
  Tang}, \bibinfo{person}{Yilin Wang}, {and} \bibinfo{person}{Huan Liu}.}
  \bibinfo{year}{2015}\natexlab{b}.
\newblock \showarticletitle{Exploring Implicit Hierarchical Structures for
  Recommender Systems}. In \bibinfo{booktitle}{\emph{International Joint
  Conference on Artificial Intelligence}}. Menlo Park: AAAI Press,
  \bibinfo{pages}{1813--1819}.
\newblock


\bibitem[\protect\citeauthoryear{Wang, Wang, Tang, Shu, Ranganath, and
  Liu}{Wang et~al\mbox{.}}{2017c}]%
        {wang2017your}
\bibfield{author}{\bibinfo{person}{Suhang Wang}, \bibinfo{person}{Yilin Wang},
  \bibinfo{person}{Jiliang Tang}, \bibinfo{person}{Kai Shu},
  \bibinfo{person}{Suhas Ranganath}, {and} \bibinfo{person}{Huan Liu}.}
  \bibinfo{year}{2017}\natexlab{c}.
\newblock \showarticletitle{What your images reveal: Exploiting visual contents
  for point-of-interest recommendation}. In
  \bibinfo{booktitle}{\emph{Proceedings of the 2017 World Wide Web Conference
  on World Wide Web}}. New York: ACM Press, \bibinfo{pages}{391--400}.
\newblock


\bibitem[\protect\citeauthoryear{Wang, Pan, Dahlmeier, and Xiao}{Wang
  et~al\mbox{.}}{2016}]%
        {wang2016recursive}
\bibfield{author}{\bibinfo{person}{Wenya Wang}, \bibinfo{person}{Sinno~Jialin
  Pan}, \bibinfo{person}{Daniel Dahlmeier}, {and} \bibinfo{person}{Xiaokui
  Xiao}.} \bibinfo{year}{2016}\natexlab{}.
\newblock \bibinfo{title}{Recursive neural conditional random fields for
  aspect-based sentiment analysis}.
\newblock \bibinfo{howpublished}{arXiv preprint arXiv:1603.06679}.
\newblock


\bibitem[\protect\citeauthoryear{Wang, He, Cao, Liu, and Chua}{Wang
  et~al\mbox{.}}{2019a}]%
        {wang2019kgat}
\bibfield{author}{\bibinfo{person}{Xiang Wang}, \bibinfo{person}{Xiangnan He},
  \bibinfo{person}{Yixin Cao}, \bibinfo{person}{Meng Liu}, {and}
  \bibinfo{person}{Tat-Seng Chua}.} \bibinfo{year}{2019}\natexlab{a}.
\newblock \bibinfo{title}{KGAT: Knowledge Graph Attention Network for
  Recommendation}.
\newblock \bibinfo{howpublished}{arXiv preprint arXiv:1905.07854}.
\newblock


\bibitem[\protect\citeauthoryear{Wang, Wang, Xu, He, Cao, and Chua}{Wang
  et~al\mbox{.}}{2018a}]%
        {wang2018explainable}
\bibfield{author}{\bibinfo{person}{Xiang Wang}, \bibinfo{person}{Dingxian
  Wang}, \bibinfo{person}{Canran Xu}, \bibinfo{person}{Xiangnan He},
  \bibinfo{person}{Yixin Cao}, {and} \bibinfo{person}{Tat-Seng Chua}.}
  \bibinfo{year}{2018}\natexlab{a}.
\newblock \bibinfo{title}{Explainable Reasoning over Knowledge Graphs for
  Recommendation}.
\newblock \bibinfo{howpublished}{arXiv preprint arXiv:1811.04540}.
\newblock


\bibitem[\protect\citeauthoryear{Wang, Xia, Tang, Wu, and Zhuang}{Wang
  et~al\mbox{.}}{2017d}]%
        {wang2017flickr}
\bibfield{author}{\bibinfo{person}{Yueyang Wang}, \bibinfo{person}{Yuanfang
  Xia}, \bibinfo{person}{Siliang Tang}, \bibinfo{person}{Fei Wu}, {and}
  \bibinfo{person}{Yueting Zhuang}.} \bibinfo{year}{2017}\natexlab{d}.
\newblock \showarticletitle{Flickr group recommendation with auxiliary
  information in heterogeneous information networks}.
\newblock \bibinfo{journal}{\emph{Multimedia Systems}} \bibinfo{volume}{23},
  \bibinfo{number}{6} (\bibinfo{year}{2017}), \bibinfo{pages}{703--712}.
\newblock


\bibitem[\protect\citeauthoryear{Weng, Xu, Li, and Nayak}{Weng
  et~al\mbox{.}}{2008}]%
        {weng2008exploiting}
\bibfield{author}{\bibinfo{person}{Li-Tung Weng}, \bibinfo{person}{Yue Xu},
  \bibinfo{person}{Yuefeng Li}, {and} \bibinfo{person}{Richi Nayak}.}
  \bibinfo{year}{2008}\natexlab{}.
\newblock \showarticletitle{Exploiting item taxonomy for solving cold-start
  problem in recommendation making}. In \bibinfo{booktitle}{\emph{2008 20th
  IEEE International Conference on Tools with Artificial Intelligence}},
  Vol.~\bibinfo{volume}{2}. Washington, DC: IEEE Computer Society Press,
  \bibinfo{pages}{113--120}.
\newblock


\bibitem[\protect\citeauthoryear{Wibowo, Siddharthan, Lin, and Masthoff}{Wibowo
  et~al\mbox{.}}{2017}]%
        {wibowo2017matrix}
\bibfield{author}{\bibinfo{person}{Agung~Toto Wibowo}, \bibinfo{person}{Advaith
  Siddharthan}, \bibinfo{person}{Chenghua Lin}, {and} \bibinfo{person}{Judith
  Masthoff}.} \bibinfo{year}{2017}\natexlab{}.
\newblock \showarticletitle{Matrix Factorization for Package Recommendations}.
  In \bibinfo{booktitle}{\emph{Proceedings of the RecSys 2017 Workshop on
  Recommendation in Complex Scenarios}}. New York: ACM Press,
  \bibinfo{pages}{1--5}.
\newblock


\bibitem[\protect\citeauthoryear{Wu, Ahmed, Beutel, and Smola}{Wu
  et~al\mbox{.}}{2017a}]%
        {wu2016joint}
\bibfield{author}{\bibinfo{person}{Chao-Yuan Wu}, \bibinfo{person}{Amr Ahmed},
  \bibinfo{person}{Alex Beutel}, {and} \bibinfo{person}{Alexander~J Smola}.}
  \bibinfo{year}{2017}\natexlab{a}.
\newblock \showarticletitle{Joint training of ratings and reviews with
  recurrent recommender networks}. In \bibinfo{booktitle}{\emph{Workshop on
  International Conference on Learning Representations}}. Toulon, France,
  \bibinfo{pages}{4--12}.
\newblock


\bibitem[\protect\citeauthoryear{Wu, Ahmed, Beutel, Smola, and Jing}{Wu
  et~al\mbox{.}}{2017b}]%
        {wu2017recurrent}
\bibfield{author}{\bibinfo{person}{Chao-Yuan Wu}, \bibinfo{person}{Amr Ahmed},
  \bibinfo{person}{Alex Beutel}, \bibinfo{person}{Alexander~J Smola}, {and}
  \bibinfo{person}{How Jing}.} \bibinfo{year}{2017}\natexlab{b}.
\newblock \showarticletitle{Recurrent recommender networks}. In
  \bibinfo{booktitle}{\emph{Proceedings of the 10th ACM International
  Conference on Web Search and Data Mining}}. New York: ACM Press,
  \bibinfo{pages}{495--503}.
\newblock


\bibitem[\protect\citeauthoryear{Xin, He, Zhang, Zhang, and Jose}{Xin
  et~al\mbox{.}}{2019}]%
        {xin2019relational}
\bibfield{author}{\bibinfo{person}{Xin Xin}, \bibinfo{person}{Xiangnan He},
  \bibinfo{person}{Yongfeng Zhang}, \bibinfo{person}{Yongdong Zhang}, {and}
  \bibinfo{person}{Joemon Jose}.} \bibinfo{year}{2019}\natexlab{}.
\newblock \bibinfo{title}{Relational Collaborative Filtering: Modeling Multiple
  Item Relations for Recommendation}.
\newblock \bibinfo{howpublished}{arXiv preprint arXiv:1904.12796}.
\newblock


\bibitem[\protect\citeauthoryear{Xiong, Chen, Huang, Schneider, and
  Carbonell}{Xiong et~al\mbox{.}}{2010}]%
        {xiong2010temporal}
\bibfield{author}{\bibinfo{person}{Liang Xiong}, \bibinfo{person}{Xi Chen},
  \bibinfo{person}{Tzu-Kuo Huang}, \bibinfo{person}{Jeff Schneider}, {and}
  \bibinfo{person}{Jaime~G Carbonell}.} \bibinfo{year}{2010}\natexlab{}.
\newblock \showarticletitle{Temporal collaborative filtering with bayesian
  probabilistic tensor factorization}. In \bibinfo{booktitle}{\emph{Proceedings
  of the 2010 Society for Industrial and Applied Mathematics International
  Conference on Data Mining}}. Columbus, Ohio: SIAM, \bibinfo{pages}{211--222}.
\newblock


\bibitem[\protect\citeauthoryear{Xu, Ba, Kiros, Cho, Courville, Salakhutdinov,
  Zemel, and Bengio}{Xu et~al\mbox{.}}{2015}]%
        {xu2015show}
\bibfield{author}{\bibinfo{person}{Kelvin Xu}, \bibinfo{person}{Jimmy Ba},
  \bibinfo{person}{Ryan Kiros}, \bibinfo{person}{Kyunghyun Cho},
  \bibinfo{person}{Aaron Courville}, \bibinfo{person}{Ruslan Salakhutdinov},
  \bibinfo{person}{Richard Zemel}, {and} \bibinfo{person}{Misc Bengio}.}
  \bibinfo{year}{2015}\natexlab{}.
\newblock \bibinfo{title}{Show, attend and tell: Neural image caption
  generation with visual attention}.
\newblock \bibinfo{howpublished}{arXiv preprint arXiv:1502.03044}.
\newblock


\bibitem[\protect\citeauthoryear{Xu, Dutta, and Ge}{Xu et~al\mbox{.}}{2018}]%
        {xu2018adjective}
\bibfield{author}{\bibinfo{person}{Xiaoying Xu}, \bibinfo{person}{Kaushik
  Dutta}, {and} \bibinfo{person}{Chunmian Ge}.}
  \bibinfo{year}{2018}\natexlab{}.
\newblock \showarticletitle{Do adjective features from user reviews address
  sparsity and transparency in recommender systems?}
\newblock \bibinfo{journal}{\emph{Electronic Commerce Research and
  Applications}}  \bibinfo{volume}{29} (\bibinfo{year}{2018}),
  \bibinfo{pages}{113--123}.
\newblock


\bibitem[\protect\citeauthoryear{Xue, Dai, Zhang, Huang, and Chen}{Xue
  et~al\mbox{.}}{2017}]%
        {xue2017deep}
\bibfield{author}{\bibinfo{person}{Hong-Jian Xue}, \bibinfo{person}{Xinyu Dai},
  \bibinfo{person}{Jianbing Zhang}, \bibinfo{person}{Shujian Huang}, {and}
  \bibinfo{person}{Jiajun Chen}.} \bibinfo{year}{2017}\natexlab{}.
\newblock \showarticletitle{Deep Matrix Factorization Models for Recommender
  Systems}. In \bibinfo{booktitle}{\emph{International Joint Conference on
  Artificial Intelligence}}. Menlo Park: AAAI Press,
  \bibinfo{pages}{3203--3209}.
\newblock


\bibitem[\protect\citeauthoryear{Yan, Xu, Zhang, Zhang, Yang, and Lin}{Yan
  et~al\mbox{.}}{2007}]%
        {yan2007graph}
\bibfield{author}{\bibinfo{person}{Shuicheng Yan}, \bibinfo{person}{Dong Xu},
  \bibinfo{person}{Benyu Zhang}, \bibinfo{person}{Hong-Jiang Zhang},
  \bibinfo{person}{Qiang Yang}, {and} \bibinfo{person}{Stephen Lin}.}
  \bibinfo{year}{2007}\natexlab{}.
\newblock \showarticletitle{Graph embedding and extensions: A general framework
  for dimensionality reduction}.
\newblock \bibinfo{journal}{\emph{IEEE Transactions on Pattern Analysis \&
  Machine Intelligence}} \bibinfo{volume}{1}, \bibinfo{number}{1}
  (\bibinfo{year}{2007}), \bibinfo{pages}{40--51}.
\newblock


\bibitem[\protect\citeauthoryear{Yang, Lei, Liu, and Liu}{Yang
  et~al\mbox{.}}{2013a}]%
        {liusocial}
\bibfield{author}{\bibinfo{person}{Bo Yang}, \bibinfo{person}{Yu Lei},
  \bibinfo{person}{Dayou Liu}, {and} \bibinfo{person}{Jiming Liu}.}
  \bibinfo{year}{2013}\natexlab{a}.
\newblock \showarticletitle{Social Collaborative Filtering by Trust}. In
  \bibinfo{booktitle}{\emph{International Joint Conference on Artificial
  Intelligence}}. Menlo Park: AAAI Press, \bibinfo{pages}{1633--1647}.
\newblock


\bibitem[\protect\citeauthoryear{Yang, Yih, He, Gao, and Deng}{Yang
  et~al\mbox{.}}{2014}]%
        {yang2014embedding}
\bibfield{author}{\bibinfo{person}{Bishan Yang}, \bibinfo{person}{Wen-tau Yih},
  \bibinfo{person}{Xiaodong He}, \bibinfo{person}{Jianfeng Gao}, {and}
  \bibinfo{person}{Li Deng}.} \bibinfo{year}{2014}\natexlab{}.
\newblock \bibinfo{title}{Embedding entities and relations for learning and
  inference in knowledge bases}.
\newblock \bibinfo{howpublished}{arXiv preprint arXiv:1412.6575}.
\newblock


\bibitem[\protect\citeauthoryear{Yang, Bai, Zhang, Yuan, and Han}{Yang
  et~al\mbox{.}}{2017}]%
        {yang2017bridging}
\bibfield{author}{\bibinfo{person}{Carl Yang}, \bibinfo{person}{Lanxiao Bai},
  \bibinfo{person}{Chao Zhang}, \bibinfo{person}{Quan Yuan}, {and}
  \bibinfo{person}{Jiawei Han}.} \bibinfo{year}{2017}\natexlab{}.
\newblock \showarticletitle{Bridging collaborative filtering and
  semi-supervised learning: a neural approach for poi recommendation}. In
  \bibinfo{booktitle}{\emph{Proceedings of the 23rd ACM SIGKDD International
  Conference on Knowledge Discovery and Data Mining}}. New York: ACM Press,
  \bibinfo{pages}{1245--1254}.
\newblock


\bibitem[\protect\citeauthoryear{Yang, Zhang, Yu, and Wang}{Yang
  et~al\mbox{.}}{2013b}]%
        {yang2013sentiment}
\bibfield{author}{\bibinfo{person}{Dingqi Yang}, \bibinfo{person}{Daqing
  Zhang}, \bibinfo{person}{Zhiyong Yu}, {and} \bibinfo{person}{Zhu Wang}.}
  \bibinfo{year}{2013}\natexlab{b}.
\newblock \showarticletitle{A sentiment-enhanced personalized location
  recommendation system}. In \bibinfo{booktitle}{\emph{Proceedings of the 24th
  ACM Conference on Hypertext and Social Media}}. New York: ACM Press,
  \bibinfo{pages}{119--128}.
\newblock


\bibitem[\protect\citeauthoryear{Yang, Drake, Damianou, and Maarek}{Yang
  et~al\mbox{.}}{2018}]%
        {yang2018leveraging}
\bibfield{author}{\bibinfo{person}{Jie Yang}, \bibinfo{person}{Thomas Drake},
  \bibinfo{person}{Andreas Damianou}, {and} \bibinfo{person}{Yoelle Maarek}.}
  \bibinfo{year}{2018}\natexlab{}.
\newblock \showarticletitle{Leveraging crowdsourcing data for deep active
  learning an application: Learning intents in alexa}. In
  \bibinfo{booktitle}{\emph{Proceedings of the 2018 World Wide Web Conference
  on World Wide Web}}. New York: ACM Press, \bibinfo{pages}{23--32}.
\newblock


\bibitem[\protect\citeauthoryear{Yang, Sun, Bozzon, and Zhang}{Yang
  et~al\mbox{.}}{2016a}]%
        {yang2016learning}
\bibfield{author}{\bibinfo{person}{Jie Yang}, \bibinfo{person}{Zhu Sun},
  \bibinfo{person}{Alessandro Bozzon}, {and} \bibinfo{person}{Jie Zhang}.}
  \bibinfo{year}{2016}\natexlab{a}.
\newblock \showarticletitle{Learning hierarchical feature influence for
  recommendation by recursive regularization}. In
  \bibinfo{booktitle}{\emph{Proceedings of the 11th ACM Conference on
  Recommender Systems}}. New York: ACM Press, \bibinfo{pages}{51--58}.
\newblock


\bibitem[\protect\citeauthoryear{Yang, Steck, and Liu}{Yang
  et~al\mbox{.}}{2012}]%
        {yang2012circle}
\bibfield{author}{\bibinfo{person}{Xiwang Yang}, \bibinfo{person}{Harald
  Steck}, {and} \bibinfo{person}{Yong Liu}.} \bibinfo{year}{2012}\natexlab{}.
\newblock \showarticletitle{Circle-based recommendation in online social
  networks}. In \bibinfo{booktitle}{\emph{Proceedings of the 18th ACM SIGKDD
  International Conference on Knowledge Discovery and Data Mining}}. New York:
  ACM Press, \bibinfo{pages}{1267--1275}.
\newblock


\bibitem[\protect\citeauthoryear{Yang, Yang, Dyer, He, Smola, and Hovy}{Yang
  et~al\mbox{.}}{2016b}]%
        {yang2016hierarchical}
\bibfield{author}{\bibinfo{person}{Zichao Yang}, \bibinfo{person}{Diyi Yang},
  \bibinfo{person}{Chris Dyer}, \bibinfo{person}{Xiaodong He},
  \bibinfo{person}{Alex Smola}, {and} \bibinfo{person}{Eduard Hovy}.}
  \bibinfo{year}{2016}\natexlab{b}.
\newblock \showarticletitle{Hierarchical attention networks for document
  classification}. In \bibinfo{booktitle}{\emph{Proceedings of the 2016
  Conference of the North American Chapter of the Association for Computational
  Linguistics: Human Language Technologies}}. \bibinfo{pages}{1480--1489}.
\newblock


\bibitem[\protect\citeauthoryear{Yao, Zhang, Huang, and Bi}{Yao
  et~al\mbox{.}}{2017}]%
        {yao2017serm}
\bibfield{author}{\bibinfo{person}{Di Yao}, \bibinfo{person}{Chao Zhang},
  \bibinfo{person}{Jianhui Huang}, {and} \bibinfo{person}{Jingping Bi}.}
  \bibinfo{year}{2017}\natexlab{}.
\newblock \showarticletitle{SERM: A recurrent model for next location
  prediction in semantic trajectories}. In
  \bibinfo{booktitle}{\emph{Proceedings of the 26th ACM International
  Conference on Information and Knowledge Management}}. New York: ACM Press,
  \bibinfo{pages}{2411--2414}.
\newblock


\bibitem[\protect\citeauthoryear{Ye, Yin, and Lee}{Ye et~al\mbox{.}}{2010}]%
        {ye2010location}
\bibfield{author}{\bibinfo{person}{Mao Ye}, \bibinfo{person}{Peifeng Yin},
  {and} \bibinfo{person}{Wang-Chien Lee}.} \bibinfo{year}{2010}\natexlab{}.
\newblock \showarticletitle{Location recommendation for location-based social
  networks}. In \bibinfo{booktitle}{\emph{Proceedings of the 18th SIGSPATIAL
  International Conference on Advances in Geographic Information Systems}}. New
  York: ACM Press, \bibinfo{pages}{458--461}.
\newblock


\bibitem[\protect\citeauthoryear{Yin, Sun, Cui, Hu, and Chen}{Yin
  et~al\mbox{.}}{2013}]%
        {yin2013lcars}
\bibfield{author}{\bibinfo{person}{Hongzhi Yin}, \bibinfo{person}{Yizhou Sun},
  \bibinfo{person}{Bin Cui}, \bibinfo{person}{Zhiting Hu}, {and}
  \bibinfo{person}{Ling Chen}.} \bibinfo{year}{2013}\natexlab{}.
\newblock \showarticletitle{LCARS: a location-content-aware recommender
  system}. In \bibinfo{booktitle}{\emph{Proceedings of the 19th ACM SIGKDD
  International Conference on Knowledge Discovery and Data Mining}}. New York:
  ACM Press, \bibinfo{pages}{221--229}.
\newblock


\bibitem[\protect\citeauthoryear{Yin, Wang, Wang, Chen, and Zhou}{Yin
  et~al\mbox{.}}{2017}]%
        {yin2017spatial}
\bibfield{author}{\bibinfo{person}{Hongzhi Yin}, \bibinfo{person}{Weiqing
  Wang}, \bibinfo{person}{Hao Wang}, \bibinfo{person}{Ling Chen}, {and}
  \bibinfo{person}{Xiaofang Zhou}.} \bibinfo{year}{2017}\natexlab{}.
\newblock \showarticletitle{Spatial-aware hierarchical collaborative deep
  learning for POI recommendation}.
\newblock \bibinfo{journal}{\emph{IEEE Transactions on Knowledge and Data
  Engineering}} \bibinfo{volume}{29}, \bibinfo{number}{11}
  (\bibinfo{year}{2017}), \bibinfo{pages}{2537--2551}.
\newblock


\bibitem[\protect\citeauthoryear{Yu, Liu, Wu, Wang, and Tan}{Yu
  et~al\mbox{.}}{2016}]%
        {yu2016dynamic}
\bibfield{author}{\bibinfo{person}{Feng Yu}, \bibinfo{person}{Qiang Liu},
  \bibinfo{person}{Shu Wu}, \bibinfo{person}{Liang Wang}, {and}
  \bibinfo{person}{Tieniu Tan}.} \bibinfo{year}{2016}\natexlab{}.
\newblock \showarticletitle{A dynamic recurrent model for next basket
  recommendation}. In \bibinfo{booktitle}{\emph{Proceedings of the 39th
  International ACM SIGIR conference on Research and Development in Information
  Retrieval}}. New York: ACM Press, \bibinfo{pages}{729--732}.
\newblock


\bibitem[\protect\citeauthoryear{Yu, Zhang, He, Chen, Xiong, and Qin}{Yu
  et~al\mbox{.}}{2018}]%
        {yu2018aesthetic}
\bibfield{author}{\bibinfo{person}{Wenhui Yu}, \bibinfo{person}{Huidi Zhang},
  \bibinfo{person}{Xiangnan He}, \bibinfo{person}{Xu Chen}, \bibinfo{person}{Li
  Xiong}, {and} \bibinfo{person}{Zheng Qin}.} \bibinfo{year}{2018}\natexlab{}.
\newblock \showarticletitle{Aesthetic-based clothing recommendation}. In
  \bibinfo{booktitle}{\emph{Proceedings of the 2017 World Wide Web Conference
  on World Wide Web}}. New York: ACM Press, \bibinfo{pages}{649--658}.
\newblock


\bibitem[\protect\citeauthoryear{Yu, Ren, Gu, Sun, and Han}{Yu
  et~al\mbox{.}}{2013a}]%
        {yu2013collaborative}
\bibfield{author}{\bibinfo{person}{Xiao Yu}, \bibinfo{person}{Xiang Ren},
  \bibinfo{person}{Quanquan Gu}, \bibinfo{person}{Yizhou Sun}, {and}
  \bibinfo{person}{Jiawei Han}.} \bibinfo{year}{2013}\natexlab{a}.
\newblock \showarticletitle{Collaborative filtering with entity similarity
  regularization in heterogeneous information networks}.
\newblock \bibinfo{journal}{\emph{The 1st International Joint Conference on
  Artificial Intelligence Workshop on Heterogeneous Information Network
  Analysis}}  \bibinfo{volume}{27} (\bibinfo{year}{2013}).
\newblock


\bibitem[\protect\citeauthoryear{Yu, Ren, Sun, Gu, Sturt, Khandelwal, Norick,
  and Han}{Yu et~al\mbox{.}}{2014}]%
        {yu2014personalized}
\bibfield{author}{\bibinfo{person}{Xiao Yu}, \bibinfo{person}{Xiang Ren},
  \bibinfo{person}{Yizhou Sun}, \bibinfo{person}{Quanquan Gu},
  \bibinfo{person}{Bradley Sturt}, \bibinfo{person}{Urvashi Khandelwal},
  \bibinfo{person}{Brandon Norick}, {and} \bibinfo{person}{Jiawei Han}.}
  \bibinfo{year}{2014}\natexlab{}.
\newblock \showarticletitle{Personalized entity recommendation: A heterogeneous
  information network approach}. In \bibinfo{booktitle}{\emph{Proceedings of
  the 7th ACM International Conference on Web Search and Data Mining}}. New
  York: ACM Press, \bibinfo{pages}{283--292}.
\newblock


\bibitem[\protect\citeauthoryear{Yu, Ren, Sun, Sturt, Khandelwal, Gu, Norick,
  and Han}{Yu et~al\mbox{.}}{2013b}]%
        {yu2013recommendation}
\bibfield{author}{\bibinfo{person}{Xiao Yu}, \bibinfo{person}{Xiang Ren},
  \bibinfo{person}{Yizhou Sun}, \bibinfo{person}{Bradley Sturt},
  \bibinfo{person}{Urvashi Khandelwal}, \bibinfo{person}{Quanquan Gu},
  \bibinfo{person}{Brandon Norick}, {and} \bibinfo{person}{Jiawei Han}.}
  \bibinfo{year}{2013}\natexlab{b}.
\newblock \showarticletitle{Recommendation in heterogeneous information
  networks with implicit user feedback}. In
  \bibinfo{booktitle}{\emph{Proceedings of the 8th ACM Conference on
  Recommender Systems}}. New York: ACM Press, \bibinfo{pages}{347--350}.
\newblock


\bibitem[\protect\citeauthoryear{Zhang, Yuan, Lian, Xie, and Ma}{Zhang
  et~al\mbox{.}}{2016}]%
        {zhang2016collaborative}
\bibfield{author}{\bibinfo{person}{Fuzheng Zhang},
  \bibinfo{person}{Nicholas~Jing Yuan}, \bibinfo{person}{Defu Lian},
  \bibinfo{person}{Xing Xie}, {and} \bibinfo{person}{Wei-Ying Ma}.}
  \bibinfo{year}{2016}\natexlab{}.
\newblock \showarticletitle{Collaborative knowledge base embedding for
  recommender systems}. In \bibinfo{booktitle}{\emph{Proceedings of the 16th
  ACM SIGKDD International Conference on Knowledge Discovery and Data Mining}}.
  New York: ACM Press, \bibinfo{pages}{353--362}.
\newblock


\bibitem[\protect\citeauthoryear{Zhang and Chow}{Zhang and Chow}{2015}]%
        {zhang2015geosoca}
\bibfield{author}{\bibinfo{person}{Jia-Dong Zhang} {and}
  \bibinfo{person}{Chi-Yin Chow}.} \bibinfo{year}{2015}\natexlab{}.
\newblock \showarticletitle{GeoSoCa: Exploiting geographical, social and
  categorical correlations for point-of-interest recommendations}. In
  \bibinfo{booktitle}{\emph{Proceedings of the 38th International ACM SIGIR
  Conference on Research and Development in Information Retrieval}}. New York:
  ACM Press, \bibinfo{pages}{443--452}.
\newblock


\bibitem[\protect\citeauthoryear{Zhang, Tay, Yao, and Sun}{Zhang
  et~al\mbox{.}}{2018}]%
        {zhang2018next}
\bibfield{author}{\bibinfo{person}{Shuai Zhang}, \bibinfo{person}{Yi Tay},
  \bibinfo{person}{Lina Yao}, {and} \bibinfo{person}{Aixin Sun}.}
  \bibinfo{year}{2018}\natexlab{}.
\newblock \bibinfo{title}{Next Item Recommendation with Self-Attention}.
\newblock \bibinfo{howpublished}{arXiv preprint arXiv:1808.06414}.
\newblock


\bibitem[\protect\citeauthoryear{Zhang, Wang, Ford, and Makedon}{Zhang
  et~al\mbox{.}}{2006}]%
        {zhang2006learning}
\bibfield{author}{\bibinfo{person}{Sheng Zhang}, \bibinfo{person}{Weihong
  Wang}, \bibinfo{person}{James Ford}, {and} \bibinfo{person}{Fillia Makedon}.}
  \bibinfo{year}{2006}\natexlab{}.
\newblock \showarticletitle{Learning from incomplete ratings using non-negative
  matrix factorization}. In \bibinfo{booktitle}{\emph{Proceedings of the 2006
  SIAM International Conference on Data Mining}}. SIAM,
  \bibinfo{pages}{549--553}.
\newblock


\bibitem[\protect\citeauthoryear{Zhang, Yao, and Sun}{Zhang
  et~al\mbox{.}}{2017c}]%
        {zhang2017deep}
\bibfield{author}{\bibinfo{person}{Shuai Zhang}, \bibinfo{person}{Lina Yao},
  {and} \bibinfo{person}{Aixin Sun}.} \bibinfo{year}{2017}\natexlab{c}.
\newblock \bibinfo{title}{Deep learning based recommender system: A survey and
  new perspectives}.
\newblock \bibinfo{howpublished}{arXiv preprint arXiv:1707.07435}.
\newblock


\bibitem[\protect\citeauthoryear{Zhang, Yao, Sun, and Tay}{Zhang
  et~al\mbox{.}}{2019}]%
        {zhang2019deep}
\bibfield{author}{\bibinfo{person}{Shuai Zhang}, \bibinfo{person}{Lina Yao},
  \bibinfo{person}{Aixin Sun}, {and} \bibinfo{person}{Yi Tay}.}
  \bibinfo{year}{2019}\natexlab{}.
\newblock \showarticletitle{Deep learning based recommender system: A survey
  and new perspectives}.
\newblock \bibinfo{journal}{\emph{ACM Computing Surveys (CSUR)}}
  \bibinfo{volume}{52}, \bibinfo{number}{1} (\bibinfo{year}{2019}),
  \bibinfo{pages}{5}.
\newblock


\bibitem[\protect\citeauthoryear{Zhang, Ai, Chen, and Croft}{Zhang
  et~al\mbox{.}}{2017a}]%
        {zhang2017joint}
\bibfield{author}{\bibinfo{person}{Yongfeng Zhang}, \bibinfo{person}{Qingyao
  Ai}, \bibinfo{person}{Xu Chen}, {and} \bibinfo{person}{W~Bruce Croft}.}
  \bibinfo{year}{2017}\natexlab{a}.
\newblock \showarticletitle{Joint representation learning for top-n
  recommendation with heterogeneous information sources}. In
  \bibinfo{booktitle}{\emph{Proceedings of the 26th ACM International
  Conference on Information and Knowledge Management}}. New York: ACM Press,
  \bibinfo{pages}{1449--1458}.
\newblock


\bibitem[\protect\citeauthoryear{Zhang, Lai, Zhang, Zhang, Liu, and Ma}{Zhang
  et~al\mbox{.}}{2014a}]%
        {zhang2014explicit}
\bibfield{author}{\bibinfo{person}{Yongfeng Zhang}, \bibinfo{person}{Guokun
  Lai}, \bibinfo{person}{Min Zhang}, \bibinfo{person}{Yi Zhang},
  \bibinfo{person}{Yiqun Liu}, {and} \bibinfo{person}{Shaoping Ma}.}
  \bibinfo{year}{2014}\natexlab{a}.
\newblock \showarticletitle{Explicit factor models for explainable
  recommendation based on phrase-level sentiment analysis}. In
  \bibinfo{booktitle}{\emph{Proceedings of the 37th International ACM SIGIR
  conference on Research and Development in Information Retrieval}}. New York:
  ACM Press, \bibinfo{pages}{83--92}.
\newblock


\bibitem[\protect\citeauthoryear{Zhang, Zhang, Zhang, Liu, and Ma}{Zhang
  et~al\mbox{.}}{2014b}]%
        {zhang2014users}
\bibfield{author}{\bibinfo{person}{Yongfeng Zhang}, \bibinfo{person}{Haochen
  Zhang}, \bibinfo{person}{Min Zhang}, \bibinfo{person}{Yiqun Liu}, {and}
  \bibinfo{person}{Shaoping Ma}.} \bibinfo{year}{2014}\natexlab{b}.
\newblock \showarticletitle{Do users rate or review?: Boost phrase-level
  sentiment labeling with review-level sentiment classification}. In
  \bibinfo{booktitle}{\emph{Proceedings of the 37th International ACM SIGIR
  Conference on Research \& Development in Information Retrieval}}. New York:
  ACM Press, \bibinfo{pages}{1027--1030}.
\newblock


\bibitem[\protect\citeauthoryear{Zhang, Li, Wu, Sun, Ye, and Luo}{Zhang
  et~al\mbox{.}}{2017b}]%
        {zhang2017next}
\bibfield{author}{\bibinfo{person}{Zhiqian Zhang}, \bibinfo{person}{Chenliang
  Li}, \bibinfo{person}{Zhiyong Wu}, \bibinfo{person}{Aixin Sun},
  \bibinfo{person}{Dengpan Ye}, {and} \bibinfo{person}{Xiangyang Luo}.}
  \bibinfo{year}{2017}\natexlab{b}.
\newblock \bibinfo{title}{Next: a neural network framework for next POI
  recommendation}.
\newblock \bibinfo{howpublished}{arXiv preprint arXiv:1704.04576}.
\newblock


\bibitem[\protect\citeauthoryear{Zhao, Zhang, and Wang}{Zhao
  et~al\mbox{.}}{2013}]%
        {zhao2013interactive}
\bibfield{author}{\bibinfo{person}{Xiaoxue Zhao}, \bibinfo{person}{Weinan
  Zhang}, {and} \bibinfo{person}{Jun Wang}.} \bibinfo{year}{2013}\natexlab{}.
\newblock \showarticletitle{Interactive collaborative filtering}. In
  \bibinfo{booktitle}{\emph{Proceedings of the 22nd ACM International
  Conference on Information and Knowledge Management}}. New York: ACM Press,
  \bibinfo{pages}{1411--1420}.
\newblock


\bibitem[\protect\citeauthoryear{Zhao, Wang, Wan, and Lai}{Zhao
  et~al\mbox{.}}{2017}]%
        {zhao2017recommendation}
\bibfield{author}{\bibinfo{person}{Zhi-Lin Zhao}, \bibinfo{person}{Chang-Dong
  Wang}, \bibinfo{person}{Yuan-Yu Wan}, {and} \bibinfo{person}{Jian-Huang
  Lai}.} \bibinfo{year}{2017}\natexlab{}.
\newblock \showarticletitle{Recommendation in feature space sphere}.
\newblock \bibinfo{journal}{\emph{Electronic Commerce Research and
  Applications}}  \bibinfo{volume}{26} (\bibinfo{year}{2017}),
  \bibinfo{pages}{109--118}.
\newblock


\bibitem[\protect\citeauthoryear{Zheng, Liu, Shi, Zhuang, Li, and Wu}{Zheng
  et~al\mbox{.}}{2017a}]%
        {zheng2017recommendation}
\bibfield{author}{\bibinfo{person}{Jing Zheng}, \bibinfo{person}{Jian Liu},
  \bibinfo{person}{Chuan Shi}, \bibinfo{person}{Fuzhen Zhuang},
  \bibinfo{person}{Jingzhi Li}, {and} \bibinfo{person}{Bin Wu}.}
  \bibinfo{year}{2017}\natexlab{a}.
\newblock \showarticletitle{Recommendation in heterogeneous information network
  via dual similarity regularization}.
\newblock \bibinfo{journal}{\emph{International Journal of Data Science and
  Analytics}} \bibinfo{volume}{3}, \bibinfo{number}{1} (\bibinfo{year}{2017}),
  \bibinfo{pages}{35--48}.
\newblock


\bibitem[\protect\citeauthoryear{Zheng, Noroozi, and Yu}{Zheng
  et~al\mbox{.}}{2017b}]%
        {zheng2017joint}
\bibfield{author}{\bibinfo{person}{Lei Zheng}, \bibinfo{person}{Vahid Noroozi},
  {and} \bibinfo{person}{Philip~S Yu}.} \bibinfo{year}{2017}\natexlab{b}.
\newblock \showarticletitle{Joint deep modeling of users and items using
  reviews for recommendation}. In \bibinfo{booktitle}{\emph{Proceedings of the
  10th ACM International Conference on Web Search and Data Mining}}. New York:
  ACM Press, \bibinfo{pages}{425--434}.
\newblock


\bibitem[\protect\citeauthoryear{Zhou, Albatal, and Gurrin}{Zhou
  et~al\mbox{.}}{2016}]%
        {zhou2016applying}
\bibfield{author}{\bibinfo{person}{Jiang Zhou}, \bibinfo{person}{Rami Albatal},
  {and} \bibinfo{person}{Cathal Gurrin}.} \bibinfo{year}{2016}\natexlab{}.
\newblock \showarticletitle{Applying visual user interest profiles for
  recommendation and personalisation}. In
  \bibinfo{booktitle}{\emph{International Conference on Multimedia Modeling}}.
  Berlin-Heidelberg, Germany: Springer, \bibinfo{pages}{361--366}.
\newblock


\bibitem[\protect\citeauthoryear{Ziegler, Lausen, and Schmidt-Thieme}{Ziegler
  et~al\mbox{.}}{2004}]%
        {ziegler2004taxonomy}
\bibfield{author}{\bibinfo{person}{Cai-Nicolas Ziegler}, \bibinfo{person}{Georg
  Lausen}, {and} \bibinfo{person}{Lars Schmidt-Thieme}.}
  \bibinfo{year}{2004}\natexlab{}.
\newblock \showarticletitle{Taxonomy-driven computation of product
  recommendations}. In \bibinfo{booktitle}{\emph{Proceedings of the 13th ACM
  International Conference on Information \& Knowledge Management}}. New York:
  ACM Press, \bibinfo{pages}{406--415}.
\newblock


\end{thebibliography}
